\documentclass[11pt,document,nofootinbib,superscriptaddress,onecolumn,preprintnumbers,balancelastpage]{article}
\pdfoutput=1
\usepackage{jheppub}
\usepackage{graphicx}
\usepackage{epstopdf}
\usepackage{dcolumn}
\usepackage{bm}
\usepackage{hyperref}
\usepackage{booktabs}
\usepackage[dvipsnames]{xcolor}
\usepackage{amsmath}
\usepackage{cancel}
\usepackage{xpatch}
\usepackage{mathtools}
\usepackage{bbold}
\usepackage{booktabs, multirow} 
\usepackage{soul}
\usepackage{tikz}
\usepackage{graphicx} 
\usepackage{physics}

\usepackage{lipsum}

\def\ellipse{\raisebox{-1pt}{\scalebox{1.35}[0.75]{$\circ$}}}
\def\dellipse{\ooalign{\raise-.25pt\hbox{\ellipse}\cr\ellipse}}

\begin{document}

\title{Primordial Black Holes from Axion Domain Wall Collapse}
\author[a]{David I. Dunsky}
\author[a,b]{and Marius Kongsore}

\affiliation[a]{Center for Cosmology and Particle Physics, Department of Physics,
New York University, New York, NY 10003, USA}
\affiliation[b]{Kavli Institute for Theoretical Physics, University of California, Santa Barbara, CA 93106, USA}
\emailAdd{ddunsky@nyu.edu}
\emailAdd{mkongsore@nyu.edu}

\abstract{The QCD axion can solve the Strong CP Problem and  be the dark matter of our universe. If the PQ symmetry breaking scale associated with the axion is below the inflationary reheating temperature, axion strings and domain walls populate the universe. Most of these strings and walls decay away into axion dark matter, but a small subset of the walls will be self-enclosed surfaces that are not attached to any strings. These enclosed walls can collapse in on themselves, compressing a large amount of energy into a small volume and potentially forming primordial black holes (PBHs). We study the number density and dynamics of these self-enclosed walls, taking into account their size distribution, Hubble expansion, asphericities, and all stages of domain wall dynamics using a combination of semi-analytic and numerical approaches. We find that axion models with a high axion decay constant $f_a$, such as those of interest in early matter-dominated cosmologies, yield a PBH abundance potentially observable by future gravitational lensing surveys. We note that the formalism developed here is also useful for predicting relic PBH abundances in other models that exhibit unstable domain walls.}

\date{\today}
\maketitle 

\section{Introduction}

The QCD axion (henceforth, just \textit{the axion}\footnote{Not to be confused with \textit{axion-like particles} (ALPs), which we consider in Appendix~\ref{sec:ALPS}.}) is a highly motivated dark matter (DM) candidate~\cite{PRESKILL1983127,ABBOTT1983133,DINE1983137} and may solve the Strong CP Problem in the Standard Model (SM). The axion is a pseudoscalar particle which arises naturally as a pseudo-Nambu-Goldstone boson of a broken global $U(1)_\mathrm{PQ}$ symmetry~\cite{PhysRevLett.38.1440,PhysRevD.16.1791,PhysRevLett.40.223,shifman1980can,PhysRevLett.40.279,Dine:1981rt,kim1979weak,zhitnitskii1980possible}. This PQ symmetry breaks at the PQ energy scale $f_a$ due to a complex scalar field $\Phi$ acquiring a vacuum expectation value. Two distinct axion cosmologies can occur: If the PQ scale lies above the inflationary reheating scale, the axion field configuration would have been homogeneous across the universe at early times~\cite{PRESKILL1983127,ABBOTT1983133,DINE1983137,Sikivie_2008}. However, if the PQ scale lies below the inflationary reheating scale, the axion field configuration would have varied from horizon to horizon~\cite{PhysRevLett.48.1156,Sikivie_2008}.

In this second scenario, the winding of the complex PQ field leads to the production of topological line defects known as cosmic strings~\cite{Kibble:1976sj,PhysRevLett.48.1867}. These strings dynamically evolve until the universe cools to a temperature of $T_{\rm QCD}\approx 10^2$ MeV, at which point the thermal axion mass grows larger than the Hubble scale. This creates dynamical brane-like structures known as domain walls~\cite{PhysRevLett.48.1156}. Most domain walls are appended to cosmic strings, but a few are entirely self-enclosed. These self-enclosed domain walls are the focus of this paper.

In a theory with only one classical axion field vacuum (often referred to as a theory where the \textit{domain wall number} $N_{\mathrm{DW}}$ is equal to one), the string-wall network is not topologically stable and can decay~\cite{PhysRevLett.48.1156}. For self-enclosed domain walls, this manifests itself as an inward collapse: the wall surface rapidly accelerates inwards, converting its rest mass energy into kinetic energy, and compressing the energetic wall surface into a smaller and smaller volume. At this stage, one of two things happens: The wall may break apart and begin oscillating, radiating away axion particles until the wall completely vanishes. Alternatively, if the wall is massive enough, sufficient energy may be forced towards the center of mass of the wall so that a primordial black hole (PBH) forms~\cite{Widrow:1989fe}. In this paper, we study this PBH formation mechanism and calculate the PBH abundance predicted by axion models.

A wide range of cosmological models predict the formation of PBHs in the early universe (see~refs.~\cite{,Carr:2021bzv,Bird:2022wvk,escriva2023primordial} and references therein). Depending on the model, PBHs can take on a wide range of masses, and they may make up some or all of the DM of the universe. PBHs are an especially intriguing DM candidate in part because of their observability. If they have masses below $10^{-16}\;\mathrm{M}_{\odot}$, they emit Hawking radiation in the form of gamma rays, which has led to constraints from gamma ray telescopes~\cite{Hawking:1975vcx,Carr_2010,Korwar_2023,Keith_2022}. Heavier PBHs in the mass range $10^{-10}\;\mathrm{M}_{\odot}$ to $10^{10}\;\mathrm{M}_{\odot}$ have strong gravitational fields that deflect light from background stars as they move across the sky via gravitational lensing. This has allowed multiple telescopes across the electromagnetic spectrum to constrain this mass range~\cite{PhysRevD.101.063005,PhysRevLett.111.181302,Griest_2014,Niikura_2019,Tisserand_2007,Oguri_2018,Zumalac_rregui_2018,Wilkinson_2001,Mroz:2024mse}, and near-future and current missions will constrain these PBHs even further~\cite{Chen_2023,derocco2023rogue,derocco2023new}.

In this work, we present the first detailed study of PBH formation from self-enclosed domain walls collapsing within an $N_{\rm DW}=1$ axion cosmology where the PQ phase transition occurs after inflationary reheating. We carefully account for the number density and size distribution of self-enclosed domain walls around the QCD phase transition, and study all stages of dynamical wall evolution, starting with the walls expanding with the scale factor of the universe, through their subhorizon collapse, and ending with their maximal compression. Moreover, we consider wall formation in both a standard cosmology and cosmologies with early matter domination. These latter cosmologies are consistent with $f_a\gtrsim 10^{11}$ GeV~\cite{turner1983coherent,DINE1983137,steinhardt1983saving,Kawasaki:1995vt,Visinelli_2010}, which is preferred for PBH production. Note that since we are simply considering the QCD axion, a majority of the present-day DM abundance will naturally consist of axion particles created via the misalignment mechanism and defect radiation. PBHs will necessarily make up a sub-component of the DM since only a fraction of the axion field energy density exists in self-enclosed domain walls when the string-wall network collapses.

In related work, refs.~\cite{Ferrer_2019,Gelmini_2023} discussed the collapse of axion domain walls into PBHs for axion models where the number of degenerate vacuua $N_{\rm DW}$ is greater than one. In these models, the temperature at which domain walls form and the temperature at which string-walls network begin to radiate and contract can be separated by several orders of magnitude, allowing for more efficient PBH production (see also refs.~\cite{Gelmini_2023_2,gouttenoire2023domain,gouttenoire2023primordial,ferreira2024collapsing,Kitajima:2023cek}, which emphasize gravitational wave production in similar scenarios). However, the stability of string-wall networks in these models also presents a problem: the domain walls overclose the universe~\cite{zel1974cosmological, PhysRevLett.48.1156}. This problem can be overcome by adding extra modifications to the axion potential to be compatible with current cosmological constraints~\cite{Hiramatsu_2013}. Ref.~\cite{Ge:2023rrq}  proposed a model in which axion strings form \textit{during} inflation, and experience a specific number of e-foldings such that they re-enter the horizon right before BBN, allowing closed domain walls to be orders of magnitude larger than the cosmological horizon at the QCD phase transition. Refs.~\cite{Vachaspati:2017hjw,Ge:2019ihf} discuss the potential of forming PBHs from axion domain wall collapse, but do not make a concrete prediction for the fraction of DM composed of black holes $f_{\rm PBH}$. In our work, we calculate the number density of enclosed walls and the effects of asphericities carefully and hence believe our manuscript to be the first to provide a reliable estimate of the present day PBH energy density $\Omega_{\mathrm{PBH}}$ in a fully post-inflationary scenario.

In Sec.~\ref{sec:cosmology_overview}, we briefly review axion cosmology, strings, and domain walls. We also discuss early matter domination. In Sec.~\ref{sec:closed_dw_abundance}, we calculate the statistical properties of enclosed walls such as their size and number density using the methods of percolation theory (effectively, the Kibble mechanism~\cite{Kibble:1976sj}) and random fields. In Sec.~\ref{sec:SupHorizonExpansion}, we study the dynamics of domain walls before they enter the cosmic horizon. In Sec.~\ref{sec:SubHorizonCollapse}, we study the evolution and compressibility of domain walls as they enter the horizon and collapse in isolation. In Sec.~\ref{sec:PBHConditions}, we discuss how efficiently energy must be compressed in order for a domain wall to form a PBH, and we discuss how angular momentum and collisions with other defects suppress PBH formation in part of the axion parameter space. In Sec.~\ref{sec:f_pbh}, we compute $\Omega_\mathrm{PBH}$ and $f_\mathrm{PBH}$, and we discuss these results in the context of current and future gravitational lensing surveys, which may be able to probe the PBH masses and abundances we predict. Finally, in Sec.~\ref{sec:conclusions}, we provide a summary and concluding remarks.

\section{Overview of Axion Cosmology and Topological Defects}
\label{sec:cosmology_overview}
Axion topological defects can be understood from the symmetries of the axion Lagrangian. At high temperatures in the early universe, the UV physics of the axion is captured by the Lagrangian of a complex scalar field $\Phi$~\cite{shifman1980can,PhysRevLett.40.279,Dine:1981rt,kim1979weak,zhitnitskii1980possible}
\begin{align}
    \mathcal{L}_{\rm UV} \supset |\partial_\mu \Phi|^2 - V_{\rm PQ}(\Phi)  \,.
    \label{eq:axion_lagrangian_UV}
\end{align}
The potential $V_{\rm PQ}$ is invariant under a global phase rotation of $\Phi$ (and a simultaneous phase rotation of the fields $\Phi$ couples to) and therefore respects a $U(1)$ symmetry~\cite{PhysRevLett.38.1440,PhysRevD.16.1791}. 
As the universe cools, this $U(1)$ symmetry is spontaneously broken and $\Phi$ acquires a vacuum expectation value $\langle \Phi \rangle = f_{\rm PQ}/\sqrt{2} e^{i a/f_{\rm PQ}}$,  with the axion, $a$, defined as the angular degree of freedom of $\Phi$ and $f_{\rm PQ}$ the PQ breaking scale.\footnote{If $\Phi$ couples to other complex scalar fields that also acquire vacuum expectation values, such as in DFSZ models~\cite{Srednicki:1985xd,shifman1980can,zhitnitskii1980possible,Dine:1981rt}, then $a$ becomes a linear combination of the angular degrees of freedom of those fields.}

The breaking of the $U(1)$ symmetry implies a non-trivial vacuum, which results in the production of topological strings at the temperature $T \approx f_{\rm PQ}$~\cite{vilenkin2000cosmic}. These axion strings are lines in physical space where the expectation value of $\Phi$ remains zero~\cite{Kibble:1976sj}. Outside the string core of size $\delta_s \sim m_\Phi^{-1} \sim f_{\rm PQ}^{-1}$, $\Phi$ rapidly approaches the true vacuum $|\langle \Phi \rangle| = f_{\rm PQ}/\sqrt{2}$, and $a$ winds from $0$ to $2\pi \times f_{\rm PQ}$ around the string~\cite{vilenkin2000cosmic}. The energy per unit length of the axion string, $\mu$, is roughly the energy density in the string core times the cross-section of the string core, $\mu \approx V_{\rm PQ}(\Phi = 0) \delta_s^2 \sim f_{\rm PQ}^4 \times f_{\rm PQ}^{-2} = f_{\rm PQ}^2$. A more realistic value is $\mu \approx \pi f_{\rm PQ}^2 \ln(f_{\rm PQ} L)$, where the log arises from the integration of the $U(1)$ potential roughly from the string core to the string separation length $L$~\cite{PhysRevLett.48.1867}.

Initially, the  abundance of strings formed at the PQ phase transition is set by the correlation length of the $\Phi$ field, as described by the Kibble-Zurek mechanism ~\cite{Kibble:1976sj,KIBBLE1980183,zurek1985cosmological,Zurek:1993ek}. The tangled axion strings self-intersect and approach an attractor solution with a density of approximately one string per horizon~\cite{Kibble:1976sj,Vilenkin:1984ib} as discussed in Sec.~\ref{sec:wallFormationTimes}. 
The distance between strings sets the correlation length of the axion field~\cite{Kibble:1976sj,Brandenberger:1993by} which in turn sets the typical scale of the axion domain walls that form much later around the QCD phase transition. 

The origin of axion domain walls can be seen by integrating out the heavy radial mode of the $\Phi$ field and considering the potential generated for $a$ by strong dynamics near the QCD phase transition, 
\begin{equation}
    \mathcal{L_\text{IR}}=\frac{1}{2}\partial_\mu a\:\partial^\mu a- V_{\rm QCD}(a), \quad V_{\rm QCD} \simeq \frac{m_a^2(T)f_{\rm {PQ}}^2}{N_{\rm DW}^2}\bigg[1-\cos\bigg(N_{\text{DW}}\frac{a}{f_{\rm PQ}}\bigg)\bigg],
\label{eq:axion_lagrangian_IR}
\end{equation}
where $f_a \equiv f_{\rm PQ}/N_{\rm DW}$ is the axion decay constant, 
$N_\text{DW}$ is the domain wall number, $T$ is the temperature of the SM gluons, and $m_a(T)$ is the temperature-dependent axion mass~\cite{PhysRevLett.48.1867,huang1985structure}. 

At high temperatures far above the QCD phase transition, $m_a(T)$ is small so that $V_{\rm QCD}$ is negligible. However, as the temperature of the universe drops and approaches $T_{\rm QCD} \approx \Lambda_{\rm QCD}$, $m_a(T)$ grows dramatically, and the  explicit breaking of the $U(1)_{\rm PQ}$ symmetry by $V_{\rm QCD}$ becomes important. The resultant cosine potential possesses $N_{\rm DW}$ unique minima for axion field values $\theta \equiv a/f_{\rm PQ}$ between $0$ and $2\pi$ (the minima at $\theta = 0$ and $2\pi$ must be identified). Since the axion field varies by this same amount as one winds around the axion string in physical space, any curve enclosing an axion string traces out these $N_{\rm DW}$ unique vacuua. This explicit breaking of the $U(1)_{\rm PQ}$ symmetry down to $Z_{N_{\rm DW}}$ results in the production of $N_{\rm DW}$ domain walls emanating from each string.
These domain walls are surfaces in physical space where $N_{\rm DW}\theta = \pi$ and where the axion field sharply interpolates between two adjacent minima in the potential.\footnote{For $N_{\rm DW} =1$, the wall is the axion field configuration that interpolates between $\theta = 0$ and $2\pi$.}
The physical thickness of the wall, $\delta$, is set by the inverse curvature of cosine potential, $m_a(T)^{-1}$, while the energy per unit area of the wall, $\sigma$, is set by the energy density in the wall surface times the thickness of the wall, which is given parametrically by $\sigma \approx V_{\rm QCD}(N_{\rm DW} \theta \approx \pi) \delta \sim m_a^2(T) f_a^2 \times \delta \sim m_a(T) f_a^2$. A precise derivation of $\sigma$ gives $\sigma = 8 m_a(T) f_a^2$~\cite{coleman1977classical,PhysRevLett.48.1867,huang1985structure}.\footnote{The axion potential is not exactly Sine-Gordon due to axion-pion mixing as pointed out by Sikivie~\cite{huang1985structure}.}
 
The $N_{\rm DW}$ domain walls which radially `fan' out from each axion string are initially thick in the early universe when the horizon size is smaller than their thickness, $\delta$. However, the walls become dynamical and `form' when $\delta$ becomes smaller than the horizon, $t$. Since the wall thickness $\delta \equiv m_a(T)^{-1}$ is a function of temperature, with $m_a(T)$ rapidly approaching its zero temperature value near the QCD phase transition, $\delta$ typically becomes less than $t$ around this time too. The temperature-dependent wall thickness $\delta \equiv m_a^{-1}$ is given by
\begin{align}
\label{eq:axion_mass_temp}
    \delta &\equiv m_a(T)^{-1} \simeq  m_a(T = 0)^{-1}
     \times
  \begin{dcases}
    \left(\frac{T}{T_{\text{QCD}}}\right)^{n/2} \qquad & T\gtrsim {T_\text{QCD}}, 
    \\
    1 \qquad  & T\lesssim T_{\text{QCD}} .
  \end{dcases}
\end{align}
For the QCD axion, instanton and lattice computations suggest $n \approx 7-8$ and $T_{\rm QCD} \simeq 100$ MeV~\cite{RevModPhys.53.43,PhysRevD.82.123508,Borsanyi:2016ksw}. To parameterize the uncertainty in the power $n$ as well as extend the results of this paper to other axion-like-particles (ALPs), we keep $n$ and $T_{\rm QCD}$ as free parameters for the remainder of this paper unless explicitly referring to the QCD axion. As discussed in Secs.~\ref{sec:SubHorizonCollapse} and \ref{sec:PBHConditions}, the conditions for PBH formation of collapsing axion walls do not depend explicitly on the value of $T_{\rm QCD}$ or $m_a(T=0)$, just ratios of $T/T_{\rm QCD}$ or $m_a(T)/m_a(T=0)$ up to mild factors of the relativistic degrees of freedom in the thermal bath, $g_*$. For ALPs, we define  $T_{\rm QCD} \equiv \sqrt{m_a(T =0)f_a}$. For the QCD axion, $m_a(T = 0)$ can be precisely computed in chiral perturbation theory in terms of the up and down quark masses, $m_u$ and $m_d$, as well as the pion mass, $m_{\pi}$, and decay constant,
$f_\pi$~\cite{PhysRevLett.40.223,hook2023tasi}
\begin{align}
    m_a(T =0) \simeq 6 \, \mu {\rm eV} \, \left(\frac{f_a}{10^{12} \, \rm GeV}\right)^{-1}  \qquad \text{(QCD axion)} \,.
\end{align}

For the remainder of this paper, we focus on the case where $N_{\rm DW} = 1$ which are axion models that possess one unique vacuum, such as the KSVZ axion~\cite{shifman1980can,kim1979weak}. These domain walls are surfaces in space where the axion field is $\theta = \pi$ and separate regions of space where $\theta$ goes to $0$ and $2 \pi$ on either side of the wall (see Figs.~\ref{fig:sim_plot},~\ref{fig:evolutionPanel}). Axion theories where $N_{\rm DW} = 1$ are not topologically stable since no residual discrete symmetry remains after $U(1)_{\rm PQ}$ is broken by QCD. The instability of a string-wall system is physically manifested by the walls pulling the attached string network apart into wall-bounded string membranes, which quickly decay into axions. Theories where $N_{\rm DW} > 1$, such as the DFSZ axion, possess more than one unique vacuum and are topologically stable due to the residual $Z_{N_{\rm DW}}$ symmetry. A universe with topologically stable walls present today is incompatible with our present-day cosmology since the walls will come to dominate the energy density of the universe~\cite{zel1974cosmological,PhysRevLett.48.1156}. It is possible to avoid this `domain wall problem' if inflation occurs after the axion strings form or if there exists additional explicit $U(1)_{\rm PQ}$ breaking which can lift the degeneracy between the different vacuua, thereby generating a pressure difference on the walls and causing the string-wall system to collapse. See ref.~\cite{Ferrer_2019} for formation of PBHs in such cosmologies.

We consider wall formation in either radiation-dominated (RD) or matter-dominated (MD) eras to allow for consistent cosmologies across a broad range of $f_a$: for example, for sufficiently large $f_a$, axions may be overproduced as DM from the misalignment mechanism or from axion strings bounded by walls in a RD era. In an early MD era, however, the entropy generated during reheating can dilute this background abundance of axions to the observed DM in our universe today. An early MD era has been motivated in several different contexts, e.g. long-lived relics that decay out of equilibrium ~\cite{steinhardt1983saving,de_Carlos_1993}, and the potential cosmological implications of early MD on axions have been widely studied (see refs.~\cite{turner1983coherent,steinhardt1983saving,Kawasaki:1995vt,Visinelli_2010,Nelson_2018}).

In the following section, we discuss the abundance of self-enclosed walls -- that is, walls that are not attached to any axion string, which can form by chance due to $\theta$ randomly varying on distance scales of order the string separation length $L$. These enclosed walls collapse under their own self tension and can form PBHs as discussed in Secs.~\ref{sec:SupHorizonExpansion}-\ref{sec:PBHConditions}.

\begin{figure}[!htb]
    \centering
    \includegraphics[width=\textwidth]{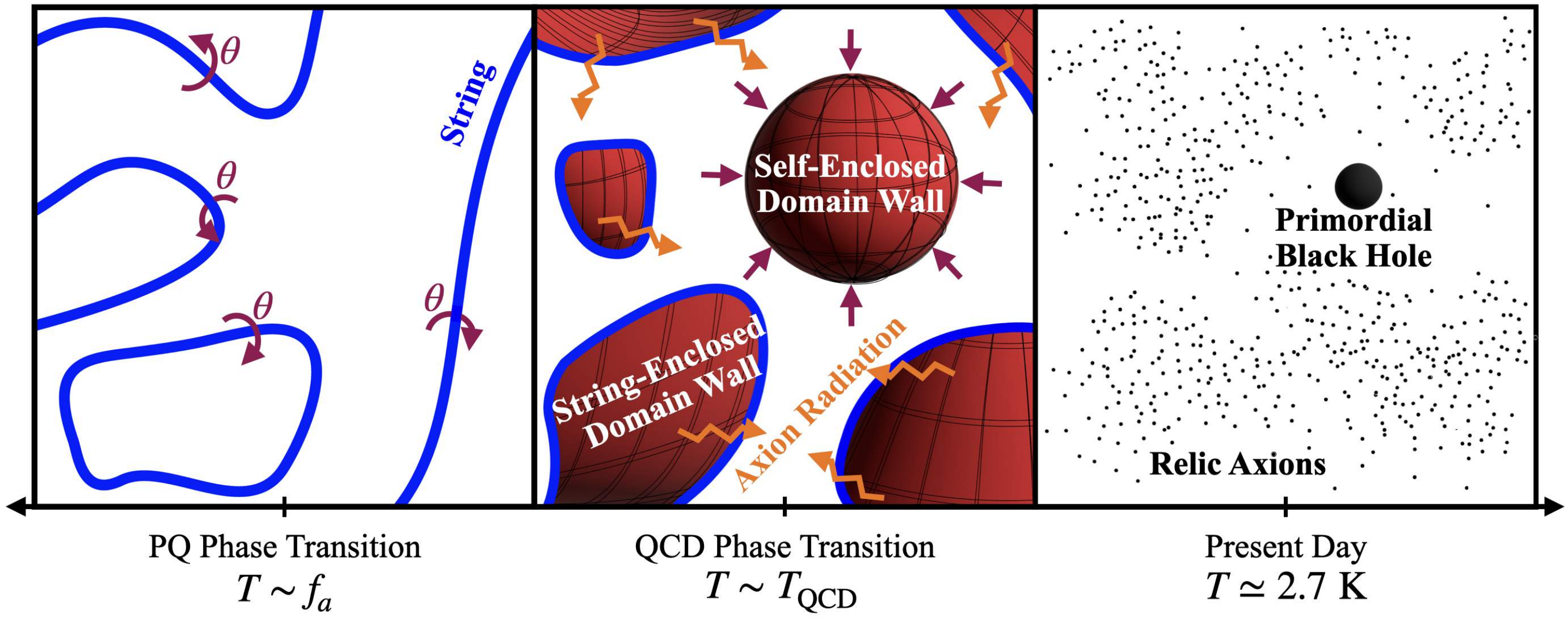}
    \vspace*{-7.5mm}
    \caption{An overview of the distinct cosmological phases of PBH formation from axion domain wall collapse. \textit{Left}: The axion string network after the PQ Phase Transition. Strings are indicated in blue. Most strings are infinite in extent, but some are closed loops~\cite{PhysRevD.30.2036,Gorghetto_2018}. The curved purple arrows indicate the $2\pi$ winding of the axion field $\theta$ around the string as discussed in Sec.~\ref{sec:cosmology_overview}. \textit{Center}: The axion string-wall network shortly after the QCD Phase Transition, where domain walls form. Note that the strings, still indicated in blue, are now attached to domain walls. Importantly, there also exist domain walls which are not attached to strings, but which are rather completely self-enclosed as discussed in Sec. \ref{sec:closed_dw_abundance}, and here indicated by a spheroid in the top middle of the panel. The purple arrows indicate that the self-enclosed domain wall eventually collapses in on itself, as discussed in Secs. \ref{sec:SupHorizonExpansion}-\ref{sec:SubHorizonCollapse}, while the orange arrows indicate axion emission. This emission arises when domain walls pull apart the infinite string-wall network, creating violent oscillations and causing the entire network to radiate away. \textit{Right}: The present-day universe. The small black dots indicate axions created from defect radiation and the misalignment mechanism.\protect\footnotemark $\;$The large black sphere indicates a PBH created from the collapse of the self-enclosed axion domain wall as discussed in Secs.~\ref{sec:PBHConditions}-\ref{sec:f_pbh}. Note that the present-day DM abundance consists of both axion particles and leftover PBHs from domain wall collapse.}
    \label{fig:overview}
\end{figure}

\section{Closed Domain Wall Abundance}
\label{sec:closed_dw_abundance}
In this section, we discuss the number density and size distribution of self-enclosed domain walls that form around the QCD phase transition. First, we explain how the lack of correlation of the axion field on distances greater than the string separation, $L$, gives rise to enclosed walls.
In Secs. \ref{sec:wall_formation} and \ref{sec:cont_lim}, we perform a series of Monte Carlo numerical simulations of both discrete and continuous uniform random fields (with support on $[0, 2\pi)$) to mimic the incoherence of the axion field in different patches of the universe and use these results to identify enclosed walls and calculate their number density and size distribution.
\subsection{Enclosed Domain Wall Number Density}
\label{sec:wall_formation}
Around the QCD phase transition, every axion string forms the boundary of a domain wall as discussed in Sec.~\ref{sec:cosmology_overview}. However, it is also possible to form  domain walls which are not attached to any string; these walls are \textit{closed} surfaces of constant $\theta = \pi$. The formation probability and size of enclosed walls arise due to the loss of correlation of the axion field  $\theta = a/f_a$ on distance scales greater than the string separation, $L$. That is, because of the randomness of $\theta$ on scales greater than $L$, it is possible, by chance, to find a region of the universe where $\theta < \pi$ surrounded entirely by a region of the universe where $\theta > \pi$ or vice versa. An enclosed domain wall is the boundary interpolating these two regions of the universe. 
\footnotetext{Here, we represent axion particles as black dots, but because typical axions have sub-eV masses, axion DM is highly wave-like.}

The probability of finding a rare patch of the universe with this axion field configuration can be understood using the methods of percolation theory~\cite{STAUFFER19791,Sahini2023ApplicationsOP}, which is closely related to the Kibble-Zurek mechanism~\cite{Kibble:1976sj,zurek1985cosmological}. The application of percolation theory to estimate the distribution of both walls bounded by strings and enclosed walls was first put forth by Sikivie \cite{sikivie1983elementary,Chang:1998tb} (see also~\cite{PhysRevD.30.2036}). We first examine a discretized version of Sikivie's algorithm to provide intuition for the formation of enclosed walls before extending the algorithm to the continuum limit, which provides both a cross-check to the discrete case and more information on the geometry of closed walls.

The discretized percolation theory approach can be understood by the following algorithm: partition the universe into a series of lattice sites where the separation between each lattice center is the correlation length of the axion field, the typical size over which $\theta$ varies randomly from patch to patch in the universe. For the axion field $\theta$, the correlation length is the string separation distance $L$ (which is roughly the horizon size). Next, randomly assign to each site the label $\theta_1, \theta_2$ or  $\theta_3$ corresponding to whether the mean $\theta$ within that patch of the universe is between $[\pi, \pi/3)$, $[\pi/3, 5\pi/3)$, or $[5\pi/3, \pi)$, respectively. We assign $\theta_{1,2,3}$ to a uniform interval between $0$ and $2\pi$ because the probability distribution of $\theta$ in a given correlation volume (lattice volume) of the universe is uniformly distributed before the axion starts oscillating near the QCD phase transition. As discussed in Sec.~\ref{sec:wallFormationTimes}, the assignment does not have to be uniformly chosen on the interval $[0, 2 \pi)$ after the axion starts oscillating.

\begin{figure}
\centering
    \includegraphics[width=.30\textwidth]{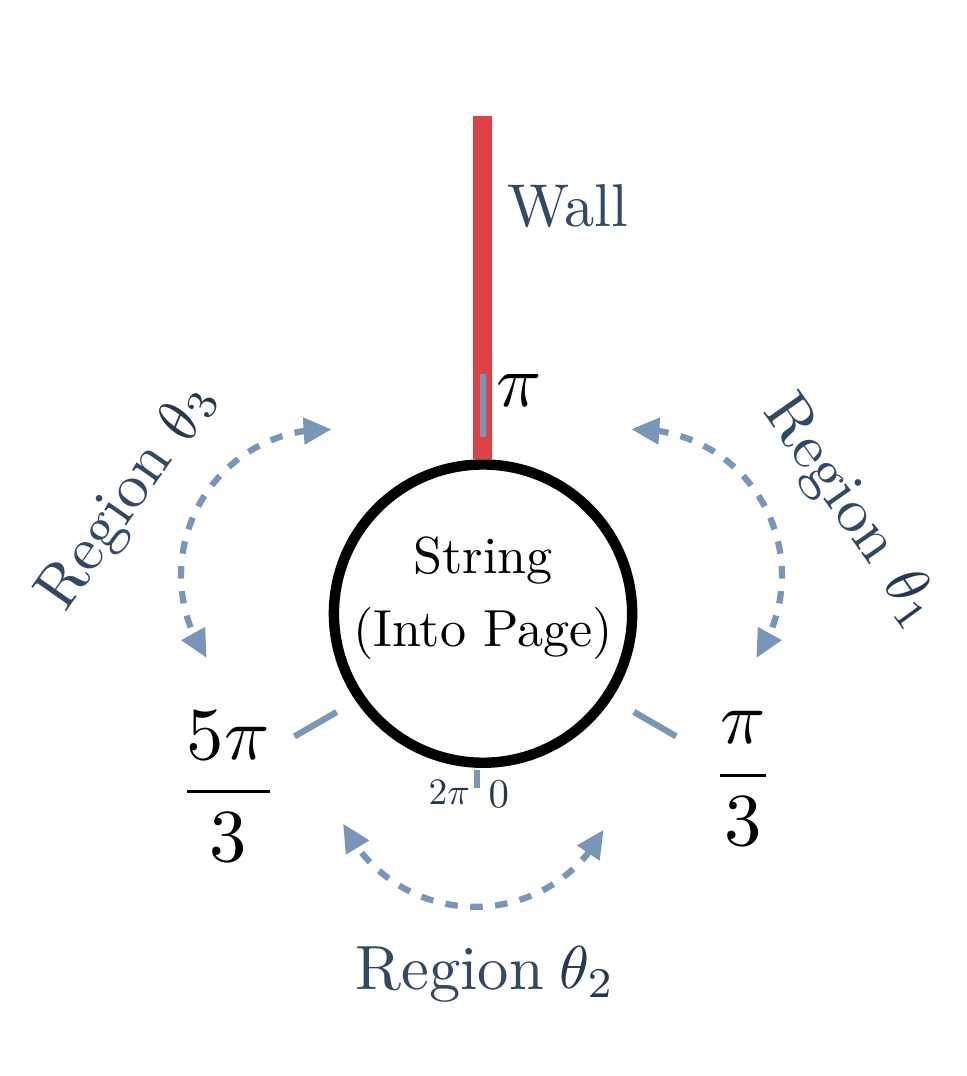}
    \hfill
    \includegraphics[width=.65\textwidth]{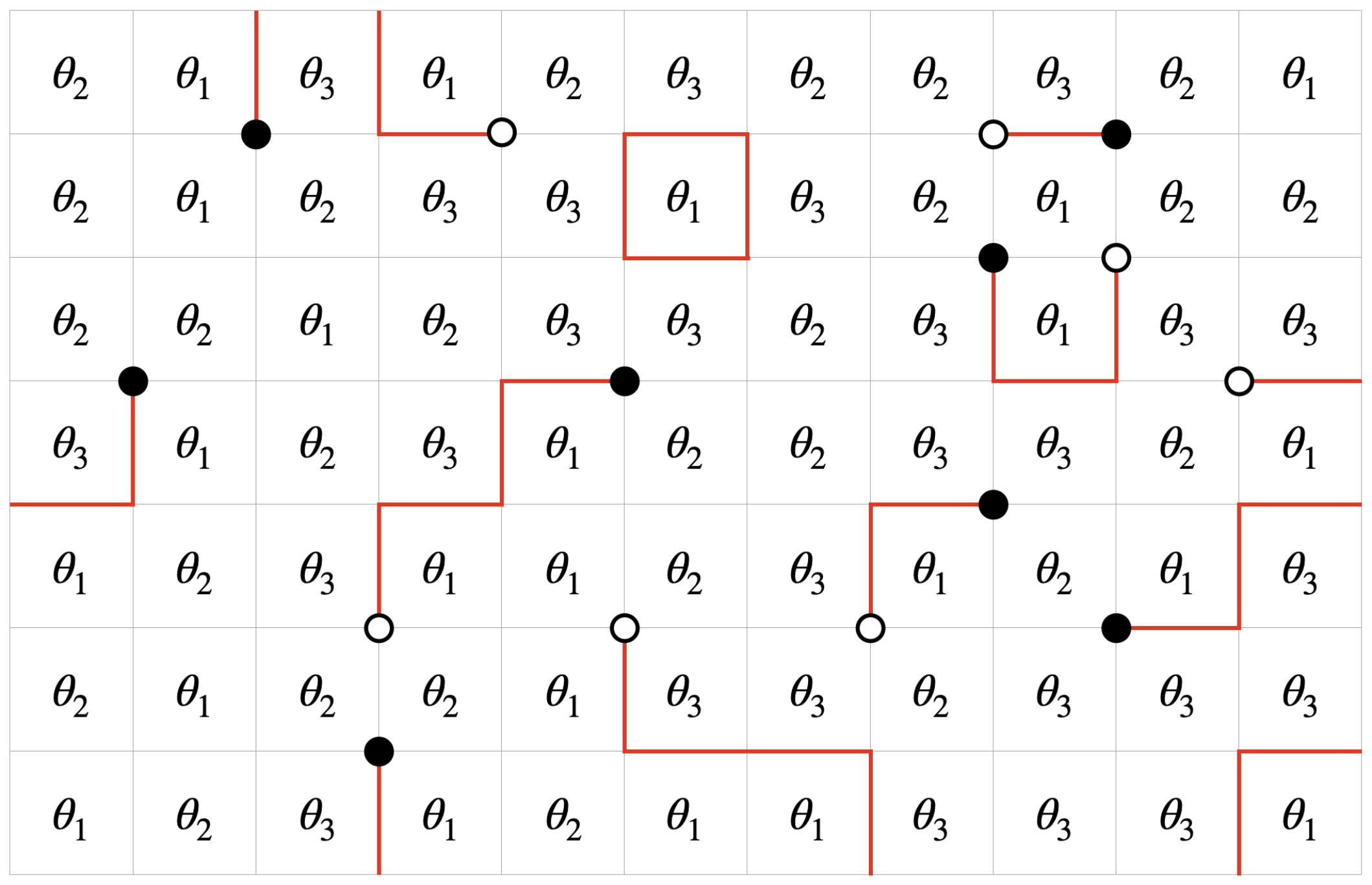}     
     \caption{An overview of the lattice defect simulation. \textit{Left}: The spatial region around a string. 
     In our lattice simulation, we divide the full $2\pi$ of the $U(1)$ circle into three equal circle segments $\theta\in\{\theta_1,\theta_2,\theta_3\}$ so that the arrangement $\theta_1\rightarrow \theta_2\rightarrow \theta_3$ corresponds to winding around a string in physical space. A string pointing into (out of) the page corresponds to a clockwise  (counterclockwise) winding of $\theta$. The red line emanating from the string represents a domain wall which separates $\theta_1$ and $\theta_3$ regions. \textit{Right}: An $x$-$y$ cross section of the defect lattice simulation. 
     The size of each lattice site
     is physically interpreted as the average string separation $L$. Each site is independently assigned a random phase $\theta_i$. From this random assignment, it is possible to identify the strings and walls: The unfilled (filled) black dots correspond to strings pointing into (out of) the page. The red lines indicate domain walls since $\theta$ must cross $\pi$ when traversing adjacent regions of $\theta_1$ and $\theta_3$. Walls that end on strings cannot be completely self-enclosed. The square closed wall shown in top middle of this cross section can correspond to a fully self-enclosed domain wall, although it can also be attached to a string further into or out of the page. The combinatorics of this simple lattice simulations allows us to estimate the number density of closed domain walls of different sizes at the QCD phase transition.}
    \label{fig:dw_sim_sketch}
\end{figure}

A planar cross-section of such a random lattice is shown in the right panel of Fig.~\ref{fig:dw_sim_sketch}. Adjacent $\theta_1$ and $\theta_3$ lattice sites are separated by a domain wall since $\theta$ crosses $\pi$ between them as indicated by the red lines on lattice boundaries. Similarly, a configuration of adjacent lattice sites that change monotonically from $\theta_1 \rightarrow \theta_2 \rightarrow \theta_3$ in a clockwise (counterclockwise) orientation indicate that they wind around an axion string going into (out of) their shared vertex, as shown by the white (black) circles.

As can be seen from the right panel of Fig.~\ref{fig:dw_sim_sketch}, every string is attached to a domain wall but there exists the rare enclosed wall as shown by the lattice site whose entire boundary is red. Most enclosed walls have an extent of order the string separation $L$, though enclosed walls that span more than one lattice site exist too, but are rarer. It is expected that these enclosed walls with size much larger than $L$ are exponentially suppressed in abundance due to the decreasing probability of finding such a patch of the universe without any strings ~\cite{sikivie1983elementary,Chang:1998tb}. Indeed, an ansatz in percolation theory for the number density of clusters of a given size (in our case, closed domain walls of a given volume) is~\cite{STAUFFER19791,PhysRevD.30.2036} 
\begin{equation}
\label{eq:percolation_scaling}
    n_{\rm c DW}(V) =A V^{-B}\exp\big[-C V^{2/3}],
\end{equation}
where $V$ is the volume enclosed by the domain wall (equivalent to the number of lattice sites it comprises), and $A$, $B$ and $C$ are constants to be determined numerically. 

\begin{figure}
    \centering
    \includegraphics[width=\textwidth,height=\textheight,keepaspectratio]{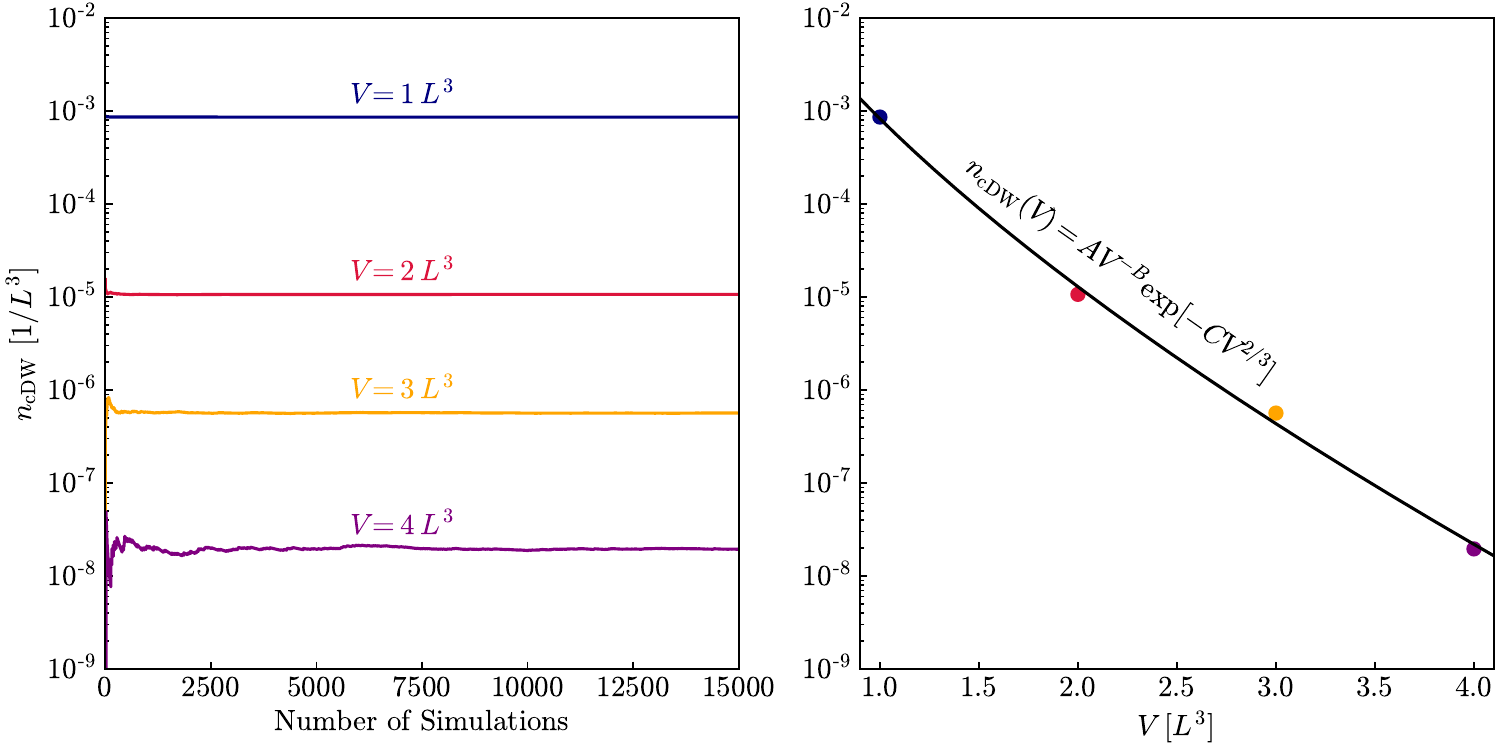}
    \caption{Results from numerical string-wall network simulations. \textit{Left}: The average number of closed domain walls of specified size per lattice volume $V$ (interpreted physically as the cubed string separation length $L^3$) as a function of the number of simulations performed. The blue, red, yellow, and purple curve correspond to volume one, two, three, and four lattice volumes, respectively. Note that all four closed domain wall densities have converged. \textit{Right}: The average number of closed domain walls of specified size per lattice volume $n_{\text{cDW}}$ as a function of the enclosed volume $V$. The colored points correspond to values obtained from simulations for a given size wall. The solid black line shows the best fit  of Eq.~\eqref{eq:percolation_scaling} to the simulation results. 
    }
    \label{fig:dw_sim_results}
\end{figure}

Using this algorithm, we run a series of simulations on a cubic lattice consisting of $25^3$ lattice sites, corresponding to roughly $25^3$ Hubble volumes. To mitigate boundary effects, we record only walls and strings two lattice sites away from the boundary of the lattice. Ensuring convergence requires realizing the random axion field phases on the lattice many times. We realize it 15,000 times, as illustrated in the left panel of Fig.~\ref{fig:dw_sim_results}. The right panel of Fig.~\ref{fig:dw_sim_results} shows the number density of enclosed walls as a function of their volume (number of lattice sites they enclose) in units of number per $L^3$. The results for $n_{\rm cDW}$ give good agreement with Eq.~\eqref{eq:percolation_scaling}, with the fit
\begin{equation}
    A=0.512,\hspace{1in} B=0.552,\hspace{1in} C=6.430 \, ,
\label{eq:percolation_parameters}
\end{equation}
which is shown by the black line in the right panel of Fig.~\ref{fig:dw_sim_results}.

Since the mean curvature radius, $R$, of the walls scales with volume as $V \propto R^3$, Eqns.~\eqref{eq:percolation_scaling}-\eqref{eq:percolation_parameters} suggest that the number density distribution of enclosed walls scales as $dn_{\rm c DW}/d \ln R  \propto -(R^3)^{2/3} = - R^2$~\cite{PhysRevD.30.2036}. 
In the following section we confirm this using the continuum approach. 
\subsection{Domain Wall Shape Distribution}
\label{sec:cont_lim}
The simulations outlined in Sec.~\ref{sec:wall_formation} are a course-grained snapshot of the actual smooth $\theta$ field configuration of the universe with the separation between each lattice site equal to the correlation length, $L$, of the axion field at that instant of time. In essence, the lattice simulations in the previous section corresponds to a smooth random field $\theta(\mathbf{x})$ with two-point function
\begin{align}
    \langle\theta(\mathbf{x}_1),\theta(\mathbf{x}_2)\rangle_{\rm Lattice} =
    \begin{cases}
        1 \quad |\mathbf{x}_1 - \mathbf{x}_2| \leq L \, ,
        \\
        0 \quad |\mathbf{x}_1 - \mathbf{x}_2| > L \, .
    \end{cases}
    \label{eq:two_point_func_discrete}
\end{align}
That is, $\theta$ is entirely correlated within one lattice cell and entirely uncorrelated outside. The real universe is of course not discrete and the axion field can take continuous values anywhere from $0$ to $2\pi$. 

A more physical continuum limit of the lattice simulations entails generating a continuous uniform random field $\theta(\mathbf{x})$ with support from $0$ to $2\pi$ and with a \textit{continuous} correlation function that is highly correlated for scales $|\mathbf{x}_1 - \mathbf{x}_2| \ll L$ and uncorrelated for $|\mathbf{x}_1 - \mathbf{x}_2| \gg L$, analogous to Eq.~\eqref{eq:two_point_func_discrete}. Such a random field can be generated first by generating a Gaussian random field with the desired correlation function and then taking its phase, which itself is a uniform random field. Generating a Gaussian random field with an arbitrary correlation function (equivalently, power spectrum) is computationally efficient, and we follow the approach of~\cite{Pen:1997up,Prunet:2008fv} which involves generating white noise in real space (which has unity correlation everywhere), fast-Fourier transforming to momentum-space, multiplying by the desired power spectrum (the Fourier transform of $ \langle\theta(\mathbf{x}_1),\theta(\mathbf{x}_2)\rangle$), and inverse fast-Fourier transforming back to real-space.  
It is possible to use the same correlation function of Eq.~\eqref{eq:two_point_func_discrete} in this method, but because of its sharp derivative at $|\mathbf{x}_1 - \mathbf{x}_2| = L$, its Fourier series is oscillatory and somewhat difficult to sample accurately for the fast-Fourier transforms.  
Instead, we take a correlation function
\begin{equation}
    \langle\theta(\mathbf{x}_1),\theta(\mathbf{x}_2)\rangle_{\rm Continuum}=e^{-|\mathbf{x}_1 - \mathbf{x}_2|^2/L^2} \, ,
    \label{eq:two_point_func} 
\end{equation}
to model the continuous fluctuations of the axion field. 
\begin{figure}
    \centering    
    \includegraphics[width=.475\textwidth]{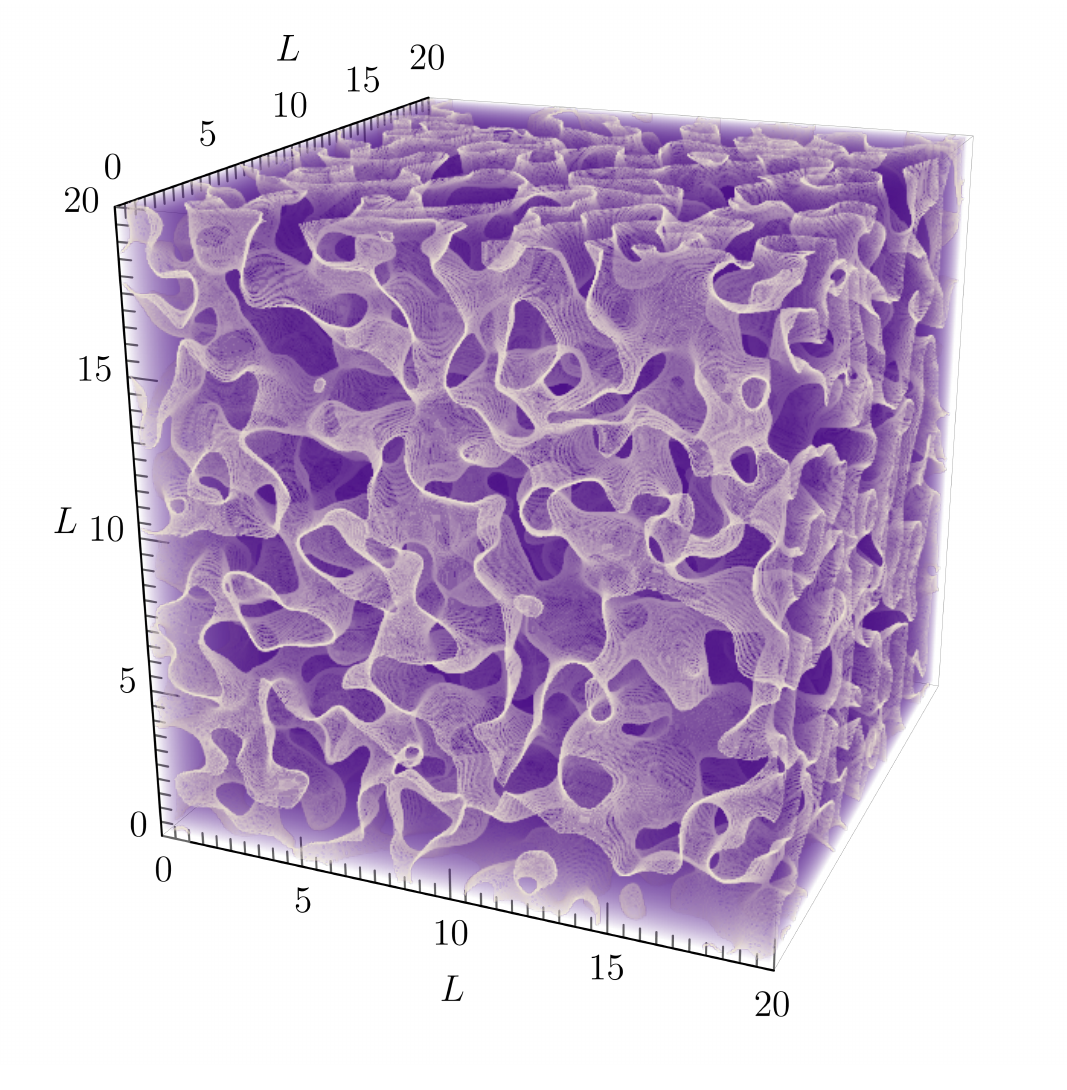}
    \hfill
    \includegraphics[width=.475\textwidth]{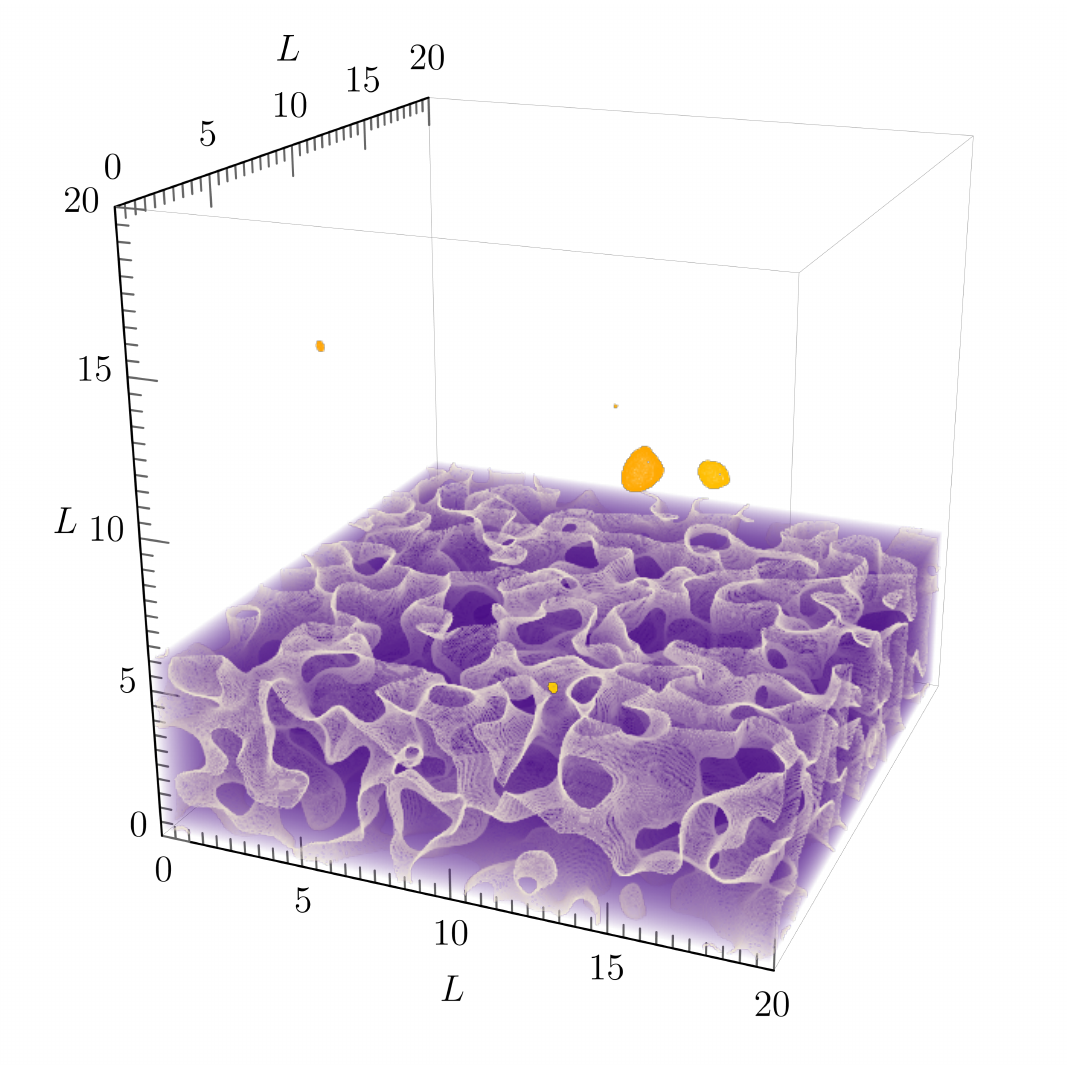}
    \caption{\textit{Left}: A contour map showing the surface of constant $\theta = \pi$ from a continuous uniform random field  (domain $0 \leq \theta < 2\pi$) generated using the two-point function Eq.~\eqref{eq:two_point_func}. The field is generated randomly in a box with size $20^3$ string correlation volumes $L^3$ and represents a statistical configuration of the axion field in the universe near $T_\mathrm{QCD}$. Domain walls can be seen as the purple mesh separating regions of space. \textit{Right}: Same as left, but with the infinite string-wall network (non-self-enclosed domain walls) removed for $z>5L$ and with self-enclosed domain walls highlighted in gold. We use this continuum uniform random field approach to verify that the number density of self-enclosed domain walls agrees with the discrete lattice simulation. We also extract the shape distribution of self-enclosed walls from this random field in order to accurately account for domain wall asphericities.}
     \label{fig:mriPlot}
\end{figure}

Figure \ref{fig:mriPlot} shows the surface of points where $\theta = \pi$ from one simulation run of a generated 3D uniform random field with correlation function given by Eq.~\eqref{eq:two_point_func} on a box of dimensions $(20L)^3$. There is one large infinite string-wall network, as shown by the purple membrane (walls) which end on strings. For distances much greater than the correlation length, $L$, the walls appear Brownian-like due to the lack of correlation in the axion field. As cross-sectional slices of the image are removed, two enclosed walls embedded in the string-wall network come into view as shown by the gold spheroidal objects in the right panel. To gather statistics on the number density distribution and shapes of these enclosed walls and to compare with the lattice results of Sec.~\ref{sec:wall_formation}, we generate 15000 realizations of the axion random field like Fig.~\ref{fig:mriPlot}, corresponding to $15000 \times (20L)^3 = 1.2 \times 10^8 L^3$ total correlation volumes (roughly Hubble volumes), and record the total number of enclosed walls observed and their following properties: volume, and the mean, maximum, and minimum distance from the wall center of mass to its surface (centroid distance). 

\begin{figure}
    \centering
    \includegraphics[width=1\textwidth]{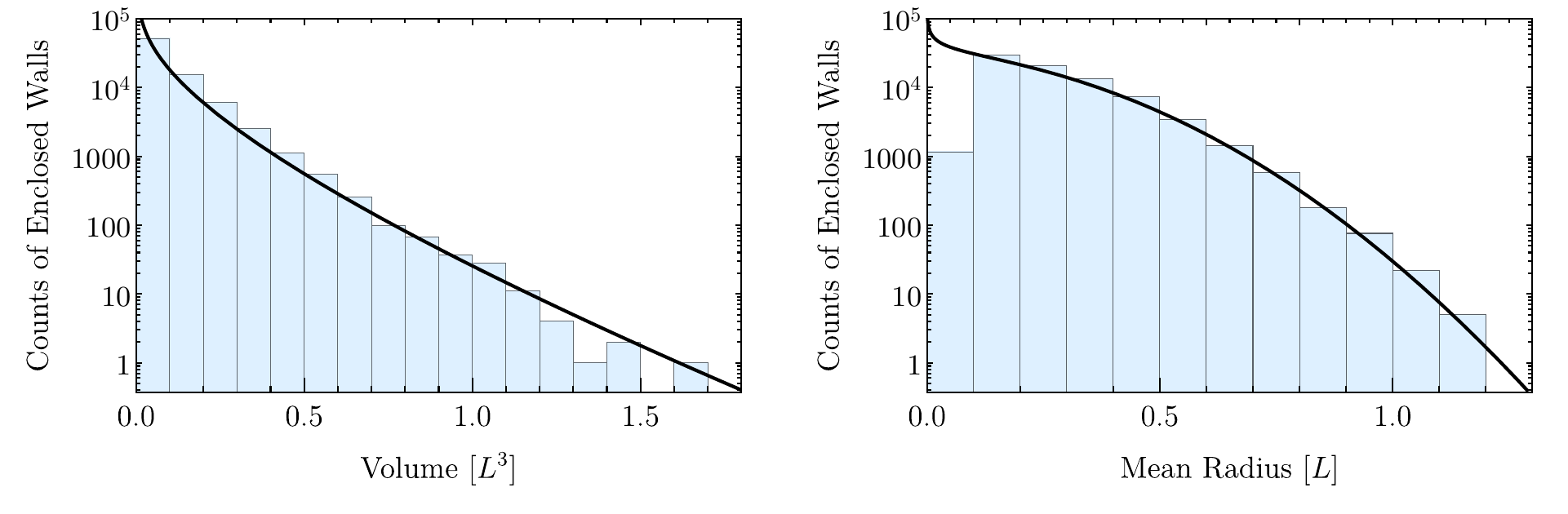}     
    \caption{\textit{Left}: The number of self-enclosed domain walls found using the continuous random field realizations with two-point function given by Eq.~\eqref{eq:two_point_func}, here shown as a function of enclosed volume. The histogram shows the data obtained from our random field realizations, while the solid black line shows the best fit analytic function, given by Eq.~\eqref{eq:continuumf(V)}. We note that there is strong agreement between the fit obtained using the lattice or continuum approaches, illustrating that the number density is governed by a simple scaling dictated by percolation theory. \textit{Right}: Same as left, but as a function of the mean wall radius. The black line shows the best fit analytic solution, given by Eq.~\eqref{eq:continuumf(R)}. Note that the dip on the leftmost side of the histogram is due to the limited resolution of the continuous random field realizations.}
\label{fig:nenclosed_dist}
\end{figure}

Tallying over all simulation runs yields nearly $8 \times 10^4$ enclosed walls. The histograms of Fig.~\ref{fig:nenclosed_dist} show how these enclosed walls are distributed as a function of their volume and mean centroid distance, which we define as their mean radius, $R$. As expected, larger walls are much rarer than smaller walls -- in fact, exponentially so as the lattice approach indicated. The black curves overlaying each histogram show an analytic fit of the enclosed wall count in each bin. Dividing by the total simulation volume and bin width gives the number density distributions of walls, which are well described by
\begin{align}
    \label{eq:continuumf(V)}
    \frac{dn_{\rm c DW}}{dV} &\approx 1.5 \times 10^{-3} \, V^{-.55} \exp\left[-7 V^{0.8} \right] \, ,
    \\
    \label{eq:continuumf(R)}
    \frac{dn_{\rm c DW}}{dR} &\approx 1.5 \times 10^{-3} \, R^{-.25} \exp \left[-6.4 R^{2} \right] \, ,
\end{align}
where $n_{\rm c DW}$, $V$ and $R$ are in units of $L^{-3}$, $L^3$ and $L$, respectively. 
We have checked that these distribution functions are independent of the histogram bin width except when the bin width is so small that there are large gaps between bins. Moreover, the above fits are of course, not unique, but the variety of alternative parameterizations do not greatly differ until extrapolating far beyond the domain of the histograms of Fig.~\ref{fig:nenclosed_dist}. Since most of the relevant parameter space for observable PBHs will not be in this regime, the final results of this paper are quite insensitive to these alternative parameterization. With these caveats in mind, the form of Eqns. \eqref{eq:continuumf(V)} and \eqref{eq:continuumf(R)} nevertheless is in good agreement with the lattice results of Sec.~\ref{sec:wall_formation}, and independently confirm the expectation that $d \ln n_{\rm c DW}/dR \propto -R^2$. Moreover, integrating Eq.~\eqref{eq:continuumf(V)} or \eqref{eq:continuumf(R)} gives the total number density of enclosed walls of $n_{\rm c, DW} \simeq 1.0 \times 10^{-3}$ per correlation volume $L^3$, which again agrees with the lattice results (see Fig.~\ref{fig:dw_sim_results}). 

\begin{figure}
    \centering
    \includegraphics[width=1\textwidth]{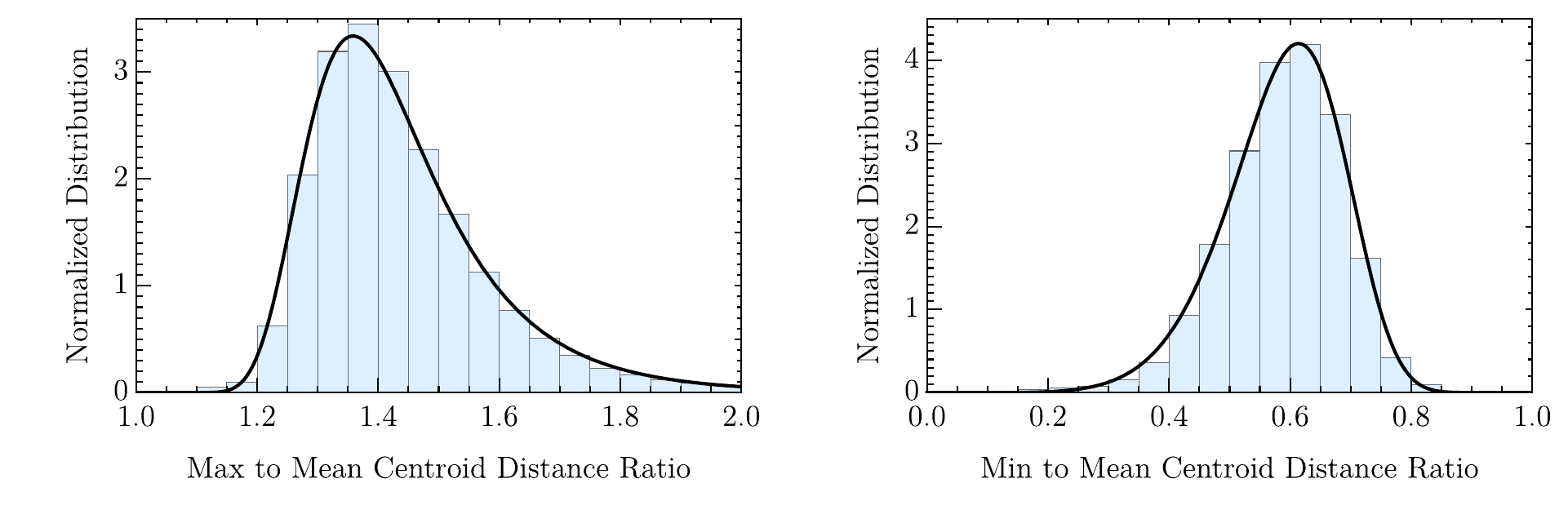}     
    \caption{\textit{Left}: The normalized distribution of the max to mean centroid distance ratio for self-enclosed domain walls generated using a continous uniform random field with a two-point function given by Eq.~\eqref{eq:two_point_func}. The histogram shows the data obtained from our random field realizations, while the solid black line shows the best fit extreme value function, given by Eq.~\eqref{eq:maxtomeanAxis}. \textit{Right}: Same as left, but showing the min to mean centroid distance ratio and best fit, given by Eq.~\eqref{eq:mintomeanAxis}.}
\label{fig:ellipticity_dist}
\end{figure}

Last, another important result from the continuum simulations is that walls are, as expected, not perfectly spherical. Fig.~\ref{fig:ellipticity_dist} shows the distribution of the ratio of the max to mean centroid distance and the min to mean centroid distance of enclosed walls. These ratios are a proxy for how ellipsoidal the walls are. Note, the asphericity of walls hinder PBH formation due to asphericity growth during collapse as we discuss in Sec.~\ref{sec:SubHorizonCollapse}.

Most walls have moderate ellipticities, with their largest (shortest) axis about a factor of $1.4$ ($0.6$) greater (smaller) than their mean radius. Perfectly spherical walls are rare. Although we show in Sec.~\ref{subsec:asphericities} that the domain wall abundance is not overly sensitive to the exact distribution of ellipticities -- as long as most are not perfectly spherical -- we find that the distributions are well described by extreme-value Frechet and Weibull distributions ~\cite{frechetDist,weibullDist} parameterized by the following probability density functions
\begin{alignat}{3}
    \label{eq:maxtomeanAxis}
    f_\mathrm{Max/Mean}(x) &\propto
    x^{-\alpha-1}\exp\left[-(x/\beta)^{-\alpha} \right] \quad &&\alpha =12.38,\; \;  &&\beta=1.37,\\
    \label{eq:mintomeanAxis}
     f_\mathrm{Min/Mean}(x) &\propto x^{\alpha-1}\exp\left[-(x/\beta)^\alpha \right] \quad &&\alpha = 7.34,\; \; &&\beta=0.63,
\end{alignat}
as shown by the black contours overlaying the histograms in Fig.~\ref{fig:ellipticity_dist}

In the following sections, we return to these distributions, Eqns.\,\eqref{eq:continuumf(R)}-\eqref{eq:mintomeanAxis}, when computing the abundance of enclosed walls that form PBHs.
\subsection{Wall Formation Times}
\label{sec:wallFormationTimes}
The percolation theory approach of Sec.~\ref{sec:wall_formation} makes manifest that the formation and size of enclosed walls is just a geometric problem. Where is the physics? At what point in time is the geometric argument valid?

The physics is encoded in the correlation length $L$ which sets the scale of the distribution functions. To \textit{map} $L$ to a \textit{physical} length scale requires knowledge of the physics that controls the spatial deviations of the axion field at a given point in time. One can apply the geometric arguments above at \textit{any time} there exists separated domains which are correlated over a given region $L$
\footnote{We thank Robert Brandenberger for emphasizing this point.}. Indeed, the mean curvature of an axion topological defect \textit{is} effectively the correlation scale of the axion field~\cite{Brandenberger:1993by}.
\begin{figure}
    \centering    
    \includegraphics[width=1.0\textwidth]{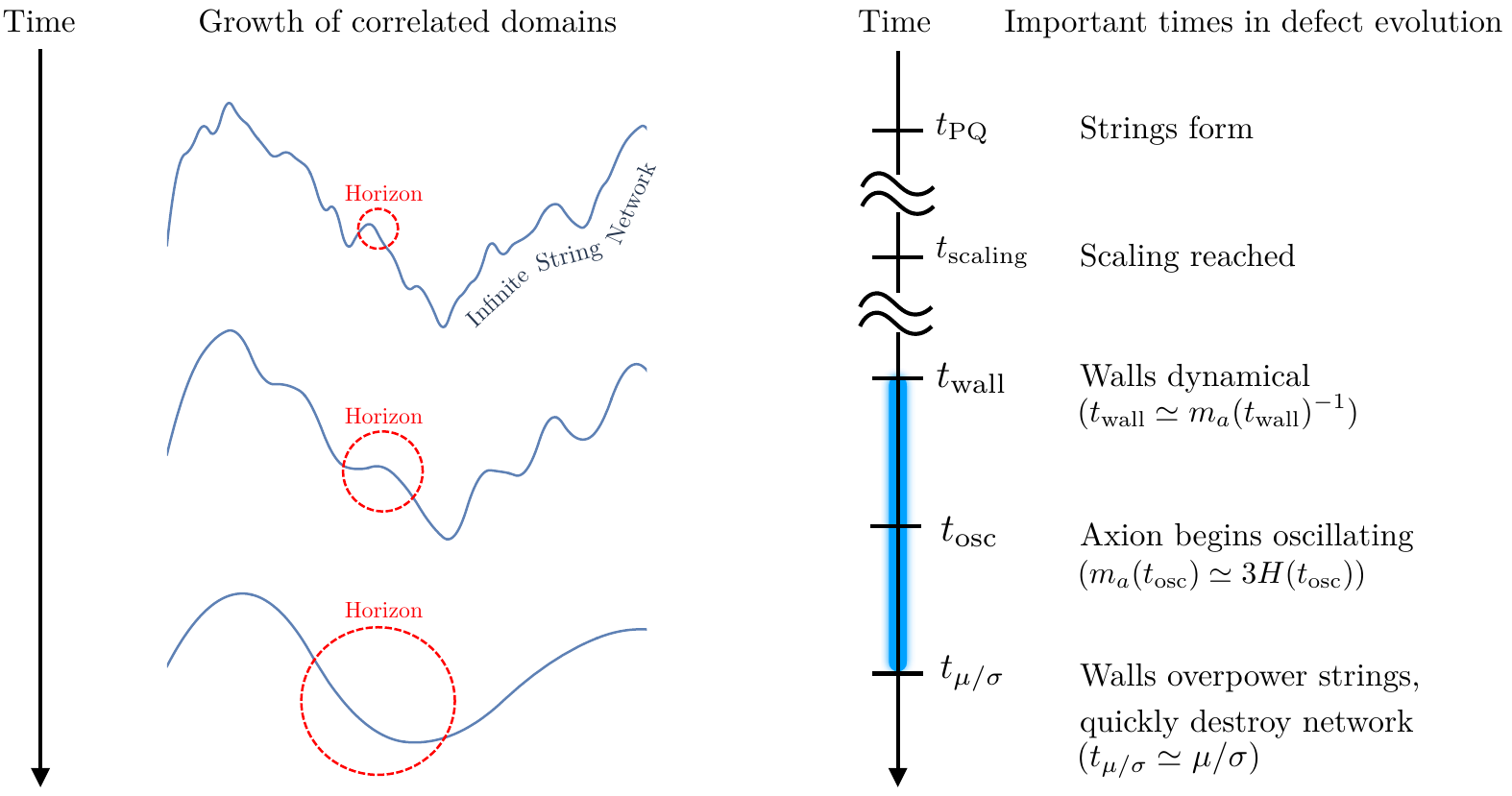}
\caption{The effect of time-evolution on the string-wall network. Strings form at $t_\mathrm{PQ}$ being highly Brownian on scales larger than the correlation length of $\Phi$. Strings eventually reach scaling at $t_\mathrm{scaling}$, at which time the string correlation length $L$ is equal to the cosmic horizon size (times potential log corrections).  Small scale structure on the strings continually damps as scaling ensures that the axion field remains largely homogeneous on scales of order the cosmic horizon size. Later, at $t_\mathrm{wall}$, domain walls begin to form since the domain wall thickness $\delta$ decreases below the horizon size. At $t_\mathrm{osc}$, the misalignment production of DM is most significant. Finally, at $t_{\mu/\sigma}$, the entire string-wall network collapses as the wall tension overpowers the string tension, meaning domain wall production ceases. The entire domain wall forming time interval is indicated by the blue band.}
\label{fig:correlation_evolution}
\end{figure}

Thus, by identifying the mean string curvature radius with the size of the correlated domain of the axion field, we can find the density of newly formed enclosed walls at different points in time in the evolution of the universe. Physically, walls are continually being produced throughout cosmic time, with their characteristic size being larger at later times. However, walls and the small scale structure of strings are also continually being destroyed, which is why the correlation length of the axion field grows with time. For example, the left panel of Fig.~\ref{fig:correlation_evolution} shows a pictorial representation of the infinite axion string network in the scaling regime, which is an attractor solution of defect network evolution where the average number of strings per horizon, $\xi$, is maintained around one  ~\cite{Kibble:1976sj,KIBBLE1980183}. At a given point in time, the infinite string has mean curvature of size $L = t/\sqrt{\xi} \simeq t$ (red circle) --  definitionally corresponding to the axion field correlation length. The superhorizon perturbations of the string are frozen by causality (Hubble friction~\cite{vilenkin2000cosmic}), while the subhorizon deviations are dynamic. The relativistic motion of the string on subhorizon  scales facilitates axion emission and string intercommutation which damps small scale axion structure and increases the size of correlated regions of the axion field. As a result, the axion field is effectively `remixed' every Hubble time ~\cite{PhysRevD.30.2036,Vilenkin:1984ib}, allowing the percolation method to be applied again with a new (and larger) interpretation of the correlation length. 

A self-enclosing wall can form at any time $t_0$ between the creation and destruction of the string-wall network. The right panel of Fig.~\ref{fig:correlation_evolution} shows some key times in the evolution history of the entire axion topological network beginning with the time time of string formation, $t_{\rm PQ}$. After the network reaches scaling at $t_{\rm scaling}$, the correlation length of the axion field is $L = t/\sqrt{\xi} \simeq t$. One can of course identify walls during this early period by denoting them as surfaces of constant $\theta = \pi$. However, the wall thickness $m_a^{-1}$ is much larger than the horizon so that the `walls' are not dynamical at this point; defining a wall tension $\sigma$ is not meaningful nor do the `walls' collapse. 

Since enclosed walls do not collapse until they become dynamical, we focus on formation times after this occurs. The walls become dynamical at time $t_\mathrm{wall}$ satisfying
\begin{align}
    t_{\rm wall} \simeq m_a(t_{\rm wall})^{-1} \qquad \text{(Walls dynamical)} \,. 
    \label{eq:twall}
\end{align}
Soon afterwards, the axion field begins oscillating towards its minimum at time $t_{\rm osc}$. This occurs when $3H(t_{\rm osc}) \simeq m_a(t_{\rm osc})$, or equivalently, at time
\begin{align}
    t_{\rm osc} \simeq  3k \, m_a(t_{\rm osc})^{-1}, 
    \qquad
    k = \begin{dcases}
        \frac{1}{2} \quad \text{RD era}
        \\
        \frac{2}{3} \quad \text{MD era}
    \end{dcases} 
    \qquad \text{(Axion oscillates)} \, .
    \label{eq:tosc}
\end{align}
Here, $k$ is the exponent in the scale factor vs time relation $a(t) \propto t^k$. Note that after $t_{\rm osc}$, the axion field is still correlated within a region of size $L$. It is just not uniform between $0$ and $2 \pi$ since the axion field oscillates about its minimum at $\langle \theta \rangle = 0$. This can be simply captured in the percolation theory approach by reducing the $\theta_2$ region (see Fig.~\ref{fig:dw_sim_sketch}) at the rate at which the amplitude of the axion field falls towards its minimum in time. 

Past $t_{\rm osc}$, the string-wall network continues to evolve in the scaling regime until the dynamics of the wall overpower that of the string. This change happens when the tension force from a string segment, $F_{\rm string} = \mu$, becomes weaker than the force of the domain wall of curvature radius $t$ attached to it, $F_{\rm wall} \approx \sigma t$~\cite{everett1982left,Vilenkin:1984ib,dunsky2021gravitational}. This occurs at time
\begin{align}
    t_{\mu / \sigma} \simeq \mu/\sigma  \simeq \frac{\pi}{4} \ln(f_a/T_{\rm QCD}) \, m_a(t_{\rm \mu/\sigma})^{-1} \qquad \text{(Walls overpower strings)}  \, .
    \label{eq:tmusigma}
\end{align}
After time $t_{\mu/ \sigma}$, the walls pull the string network apart and the entire system quickly decays into axions and gravitational waves (see Fig.~12 of ref.~\cite{dunsky2021gravitational} for how quickly walls overpower their string boundary once their curvature radius grows beyond $\mu/\sigma$). Consequently, the latest time an enclosed wall can form is shortly after $t_{\mu /\sigma}$. The PBH abundance will be dominated at the latest possible time of enclosed wall formation, which is approximately $t_{\mu / \sigma}$. This is because larger walls form later, and the larger the wall, the easier it is to form a PBH upon collapse as discussed in Sec.~\ref{sec:PBHConditions}.

Note that if the axion mass is still increasing at $t_{\rm wall}$, then the times $t_{\rm wall} \simeq t_{\rm osc} \approx t_{\mu/\sigma}$ are all comparable. This is generally the case if the QCD axion begins oscillating in a RD era. Conversely, if the axion mass is already at its zero temperature mass at $t_{\rm wall}$, then $t_{\rm wall} \approx t_{\rm osc} \approx .035 (\ln(f_a/T_{\rm QCD})/36)^{-1} t_{\mu/\sigma}$. This is generally the case if the QCD axion begins oscillating in a MD era and $f_a \gtrsim 10^{15}$ GeV or for ALPs with a (at late times) temperature-independent mass.

Finally, we note that recent numerical work
~\cite{PhysRevLett.82.4578,PhysRevD.60.103511,Yamaguchi_2003,Hiramatsu_2011,PhysRevD.85.105020} confirms this picture of scaling with $\xi \rightarrow 1$ independent of initial conditions.
\footnote{Friction between the axion strings and the background plasma is ignored in these references as well as in the methods simulating the full Euler-Lagrange equation of motion. Friction can dramatically increase the time it takes for strings to reach scaling, though they are expected to still reach scaling before $T_{\rm QCD}$~\cite{PhysRevD.53.R575,Dimopoulos_1998,Fukuda_2021}.}
However, refs.~\cite{Kawasaki:2018bzv,Gorghetto_2018,Buschmann:2019icd,Vaquero_2019,Gorghetto_2021,Buschmann_2022,Saikawa:2024bta,Kim:2024wku}, which fully simulate the field-theoretic Euler-Lagrange equations of motion for strings, find evidence of logarithmic violations to this scaling. In these simulations, $\xi$ still has an attractor solution (meaning the simulations converge to the same value of $\xi$ independently of the initial conditions used), but once the attractor is reached, $\xi$ does not stay fixed, but increases logarithmically with cosmic time (see however ~\cite{Hindmarsh:2021zkt}). Simulations find $\xi$ logarithmically grows in cosmic time up to $\xi \sim 1.2 \pm 0.2$ but cannot tell if this trend continues indefinitely due to lack of computational resources; if extrapolating this trend in the growth of $\xi$ through orders of magnitude in time all the way to the QCD phase transition, $\xi$ could be as large as $\sim 15$ at the time $t_{\rm wall}$. 

In the remaining sections, we assume the traditional value of $\xi\simeq 1$, which does not rely on an (as of the present) uncertain scaling violation coefficient. This choice is likely more robust in an early matter-dominated cosmology, such as the one we are primarily interested in, where logarithmic scaling violations are expected to be suppressed~\cite{Visinelli_2010} if they exist. However, we discuss how a different $\xi$ at the QCD phase transition would affect the present-day PBH abundance in Sec.~\ref{sec:conclusions} and in Appendix~\ref{sec:ALPS} where we consider ALPs in a radiation-dominated cosmology. In the following two sections, we consider the dynamics of a single enclosed wall that forms at some arbitrary time $t_0$, with $t_{\rm wall} \leq t_0 \leq t_{\mu/\sigma}$, as shown by the blue strip in the right panel of Fig.~\ref{fig:correlation_evolution}. Sec.~\ref{sec:SupHorizonExpansion} deals with the initial stage of superhorizon expansion  while Sec.~\ref{sec:SubHorizonCollapse} deals with subhorizon collapse.
\section{Superhorizon Expansion of Domain Walls}
\label{sec:SupHorizonExpansion}
In this section, we describe the dynamics of self-enclosed domain walls starting from when the walls form up until the moment they enter the cosmic horizon. During this stage, the rest mass of the wall increases in two distinct ways: First, the Hubble expansion of the universe stretches the wall, giving it a larger surface area. Second, the wall tension grows due to the axion mass continuing to `turn on' as the universe cools. Both mechanisms increase the probability of PBH formation. We numerically simulate the full Euler-Lagrange equations of motion of an enclosed wall in a Friedmann-Lemaitre-Robertson-Walker (FLRW) background to accurately account for these phenomena. First, we describe how these simulations are set up and executed. We then use the simulations to extract the peak radius of the wall at horizon entry and the time it takes for the wall to enter the horizon and to collapse. These horizon entry characteristics are crucial in determining whether a PBH can form, as discussed in Secs.~\ref{sec:SubHorizonCollapse}-\ref{sec:PBHConditions}. Moreover, we use this information in Sec.~\ref{sec:f_pbh} to calculate the final energy stored in a wall and hence the mass of the resultant PBH. Finally, while the analysis in this section assumes the walls are perfectly spherical during superhorizon expansion for simplicity, we also simulate and discuss the superhorizon evolution of aspherical walls in Appendix~\ref{app:superhorizon_asphericities}. Asphericities are most important during the final stage of \textit{subhorizon} collapse, as discussed in Sec.~\ref{subsec:asphericities}.

\subsection{Simulation Setup}
In Minkowski spacetime, it is energetically favorable for a self-enclosed domain wall to shrink. Even in the absence of an energy gap between the vacuum inside and outside of the domain wall, the intrinsic surface tension, $\sigma$, causes a curved wall surface to accelerate inward, leading it to rapidly approach the speed of light.

However, as discussed in Sec.~\ref{sec:closed_dw_abundance}, self-enclosed domain walls larger than the cosmic horizon can form for which  the non-flatness of an FLRW spacetime is important. Instead of collapsing immediately, superhorizon walls stretch with the scale factor, $a(t)$, of the universe. This only occurs for a finite amount of time due to the horizon size growing at a faster rate than the scale factor. For a wall that forms at time $t_0$ with radius $R_0 \gtrsim t_0$, one can equate the horizon size $t$ with the increasingly stretched radius of the wall, $R(t) = R_0 a(t)/a(t_0)$, to estimate the size of the wall at horizon re-entry, $R_{\rm RE} \approx R_0 (R_0/t_0)^{2 (3)}$ in a RD (MD) era. To make this relation more exact, we study domain wall growth numerically.

As soon as the domain wall re-enters the horizon, it is briefly at rest with a radius $R_\mathrm{RE}$,\footnote{For aspherical walls, there is not single time at which the wall is at rest, but there is a short time interval in which the entire wall goes from being completely superhorizon to collapsing inwards.}
so that its total energy at horizon re-entry is
\begin{equation}
    E_\mathrm{RE} \simeq 4\pi\sigma R_\mathrm{RE}^2 \approx 32 \pi m_a(t_\mathrm{RE}) f_a^2 R_\mathrm{RE}^2,
\label{eq:spherical_wall_energy}
\end{equation}
where $\sigma \simeq 8 m_a(T) f_a^2$ is the tension of the domain wall. Eq.~\eqref{eq:spherical_wall_energy} shows that the total domain wall energy scales with the square of the wall radius, meaning up until horizon re-entry $E(t)\propto a^2(t)m_a(t)$. Eq.~\eqref{eq:spherical_wall_energy} also accounts for $m_a$ potentially not having reached its zero temperature value at $t=t_\mathrm{RE}$. It is preferable for PBH formation for the domain walls to take a long time to collapse so that $m_a$ is as large as possible per Eq.~\eqref{eq:axion_mass_temp}.

We directly simulate the enclosed wall from formation until collapse. This is done by numerically solving the Euler-Lagrange equations associated with the IR axion Lagrangian, given by Eq.~$\eqref{eq:axion_lagrangian_IR}$, in an FLRW background. Using $\theta(x)\equiv a(x)/f_a$, the classical equation of motion for the axion field is
\begin{equation}
\label{eq:axion_hubble_eom}
    \left[\partial_t^2+3\frac{\dot{a}}{a}\partial_t-\frac{1}{a^2}\nabla^2\right]\theta(x)+m_a^2(T)\sin(\theta)=0,
\end{equation}
where $a = a(t)$ denotes the scale factor. The Hubble scale $H$ is related to the SM plasma temperature via
\begin{equation}
\label{eq:hubble_temp}
    H(T)=
        \begin{dcases}
        \sqrt{\frac{8\pi^3 g_*(T)}{90}}\frac{T^2}{M_\mathrm{Pl}} \hfill & \mathrm{RD\; Era},
        \\
         \sqrt{\frac{8\pi^3 g_*^2(T)}{90\; g_*(T_\mathrm{RH})}}\frac{T^4}{M_\mathrm{Pl}T_\mathrm{RH}^2} \hfill & \text{MD Era (Non-Adiabatic)},
    \end{dcases}
\end{equation}
where $g_*$ is the effective number of relativistic degrees of freedom in the SM plasma, $M_\text{Pl}\simeq1.22\times10^{19}$ GeV is the Planck mass, and $T_\mathrm{RH}$ is the reheat temperature in the early matter-dominated scenario~\cite{Kolb:1990vq}.

\begin{figure}
    \centering    \includegraphics[width=1.0\textwidth,height=1.0\textheight,keepaspectratio]{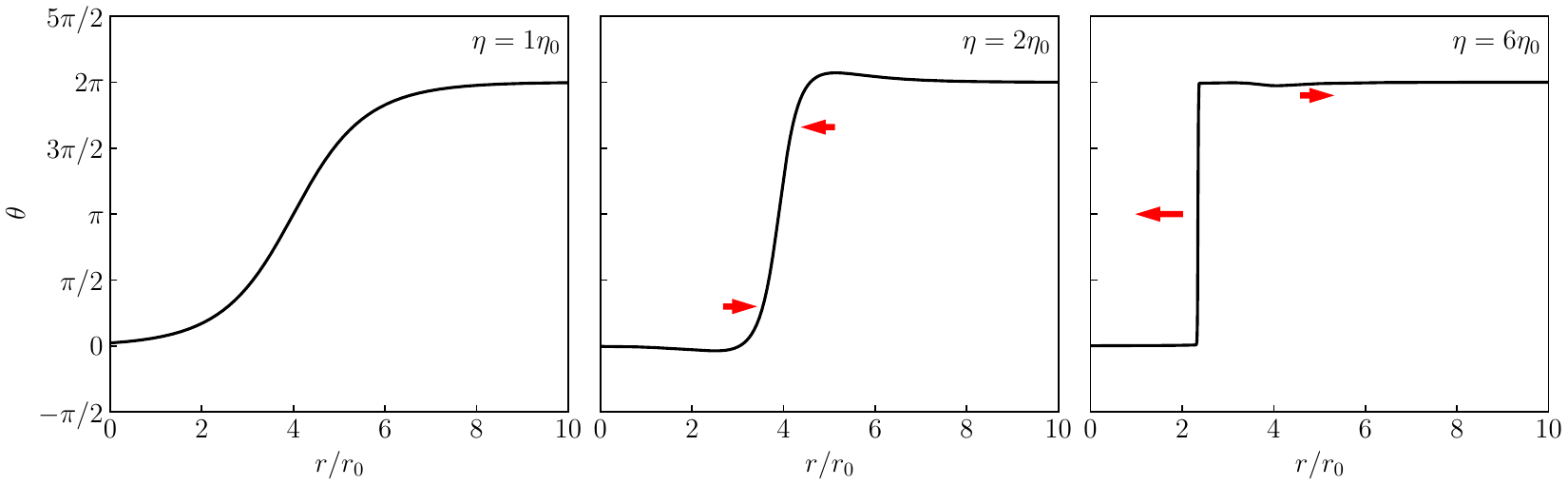}
\caption{The axion field configuration $\theta=a/f_a$ for a spherical domain wall of initial radius $R_0 = 4t_0$ as a function of the comoving radial coordinate $r/r_0$ at three conformal time slices. Here, the domain wall is assumed to have formed at $t=t_\mathrm{wall}$. \textit{Left}: The initial axion field configuration at $\eta=\eta_0$, corresponding to the moment the axion mass and Hubble scale cross. The initial field configuration is given by Eq.~\eqref{eq:dw_init_field_config} and the wall is taken to be completely at rest. \textit{Middle}: The axion field configuration at $\eta=2\eta_0$. Note that the wall appears considerably thinner due to the increase in the thermal axion mass, athough this visual effect is further enhanced due to the choice of comoving coordinates (see the $(\eta/\eta_0)^2$ factor multiplying $\tilde{m}_a^2$ in Eq.~\eqref{eq:axion_hubble_eom}). The red arrows indicate the motion of the field configuration, with the wall continuing to become thinner, but the wall otherwise being at rest in comoving coordinates as it continue to expand with the scale factor. \textit{Right}: The axion field configuration at $\eta=6\eta_0$. Here, the wall has again become dramatically thinner and has entered the cosmic horizon, causing rapid inward movement. The wall has also started radiating axions, which can be seen as outward moving ripples in the field.}
\label{fig:sim_plot}
\end{figure}

We transform the above equation into conformal time via the relation $d\eta = dt/a(t)$ and define $\eta_0\equiv r_0\equiv t_0$ with $t_0$ chosen to be the wall formation time arbitrarily lying somewhere in the interval $(t_\mathrm{wall},\; t_{\mu/\sigma})$ as discussed in Sec.~\ref{sec:wallFormationTimes}. Eq.~\eqref{eq:axion_hubble_eom} now reduces to the simple form
\begin{align}
\label{eq:dimless_w_hubble_eom}
    \frac{\partial^2 \theta}{\partial (\eta/\eta_0)^2} - \nabla^2_{r/r_0} \theta + 2 \mathcal{H} \frac{\partial \theta}{\partial (\eta/\eta_0)} + (\eta/\eta_0)^{2} \tilde{m}_a^2\left(\frac{\eta}{\eta_0}\right) \sin \theta  = 0 \, ,
\end{align}
where $\tilde{m}_a$ is the dimensionless axion mass, given by
\begin{equation}
    \tilde{m}_a\left(\frac{\eta}{\eta_0}\right) = m_a(T=0)\times\eta_0\times\min\bigg[\left(\frac{T_\mathrm{QCD}}{T(\eta/\eta_0)}\right)^{n/2},\;\; 1\bigg], \quad n \approx 6.68,
\end{equation}
and where $\mathcal{H}$ is given by
\begin{equation}
    \mathcal{H} = \frac{1}{R}\frac{dR}{d(\eta/\eta_0)} =     
    \begin{dcases}
    \frac{1}{(\eta/\eta_0)} \quad \;\;\text{RD Era},
    \\
    \frac{2}{(\eta/\eta_0)} \quad \;\;\text{MD Era}.
    \end{dcases}
\end{equation}
We numerically solve this equation starting from $\eta/\eta_0=1$ with initial field configuration
\begin{align}
\label{eq:dw_init_field_config}
    \frac{\partial \theta}{\partial (\eta/\eta_0)}\biggr\rvert_{\eta/\eta_0 = 1}  & = 0 \, ,
    \\
    \theta\left(\frac{\mathbf{r}}{r_0}\right)\biggr\rvert_{\eta/\eta_0 = 1}  &= 4 \arctan{\exp\big[\tilde{m}_a\left(\frac{\eta}{\eta_0}=1\right)\frac{1}{r_0}({\mathbf{r} - \mathbf{R}_{\rm 0}}})\cdot \hat{\mathbf{n}}\big] \, ,
    \label{eq:dw_arctan_config}
\end{align}
where $\mathbf{R}_{\rm 0}$ is a vector pointing to the wall surface along the direction of $\mathbf{r}$ and $\hat{\mathbf{n}}$ is an outward pointing vector normal to the domain wall at the point $\mathbf{R}_{\rm 0}$. This field configuration corresponds to a single self-enclosed domain wall. Note that it is simply a standard planar Sine-Gordon kink soliton with asymptotic limits $0$ and $2 \pi$, but deformed into a sphere~\cite{weinberg_2012}. This quasi-planar solution is a good approximation to the initial wall profile in the regime where the local radius of curvature of the wall is much greater than the wall thickness $\delta$, being the exact solution when neglecting the $2/r \partial_r$ term in $\nabla^2_{r/r_0}$ of Eq.~\eqref{eq:dimless_w_hubble_eom}. Due to the significant thinning of axion domain walls as the axion mass $m_a(T)=\delta^{-1}$ grows, and due to the initial comoving expansion of the wall, this is a good approximation for any wall of size $L$ or larger shortly after creation. Fig.~\ref{fig:sim_plot} shows three time slices of the simulated axion field configuration for a collapsing spherically symmetrical domain wall.
The axion mass turn-on has no significant effect on the total amount of time it takes for a domain wall collapse.\footnote{This can be seen in the Nambu-Goto picture, where adding a time-dependent mass term to the action admits the exact same solution to the equations of motion as the action with no time-dependent mass term.} However, the mass turn-on can be important in increasing the total energy of the wall and in decreasing its thickness, which makes the domain wall more easily compressible into a black hole, which we discuss in Sec.~\ref{sec:SubHorizonCollapse}.
\begin{figure}
    \centering
    \includegraphics[width=\textwidth,keepaspectratio]{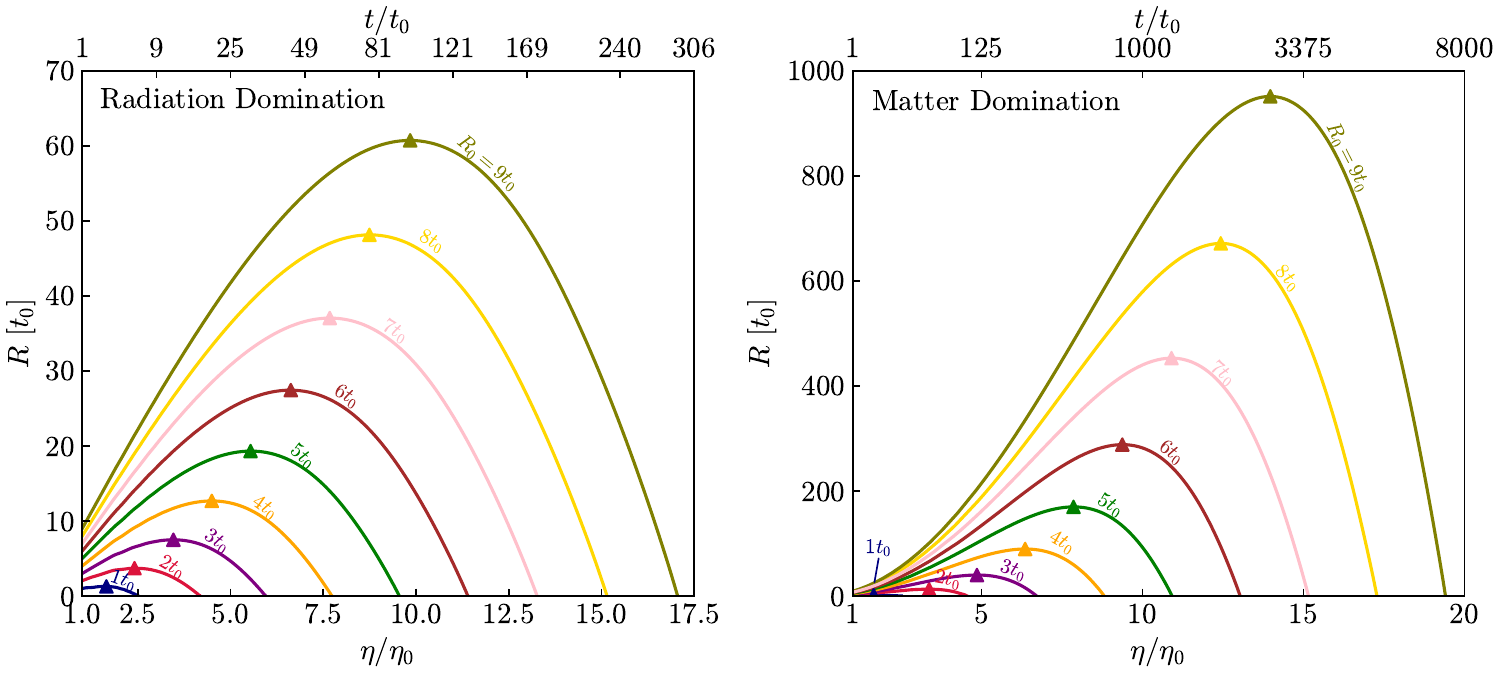}
\caption{\textit{Left}: Collapse of closed spherical domain walls assuming radiation domination. The colored curves indicate the physical radius $R$ of different closed domain walls with initial radius $R_0$ as a function of conformal time $\eta$ and physical time $t$. The triangles on each curve indicate the moment in time that the corresponding domain wall enters the horizon, which we define as when the physical velocity of the wall reverses sign and points inwards. Notice that walls with larger $R_0$ re-enter the horizon later and hence attain a larger surface area than walls with small $R_0$, giving them more energy to form black holes. \textit{Right}: Same as left, but assuming matter domination. Note here that since $a\propto t^{2/3}$ rather than $t^{1/2}$ like for radiation domination, the walls grow much larger before collapsing.}
    \label{fig:sphere_collapse_time}
\end{figure}

\subsection{Re-entry Radius and Collapse Time}
\label{sec:peak_radius_and_collapse_time}
\begin{figure}
    \centering
    \includegraphics[width=1.0\textwidth,height=1.0\textheight,keepaspectratio]{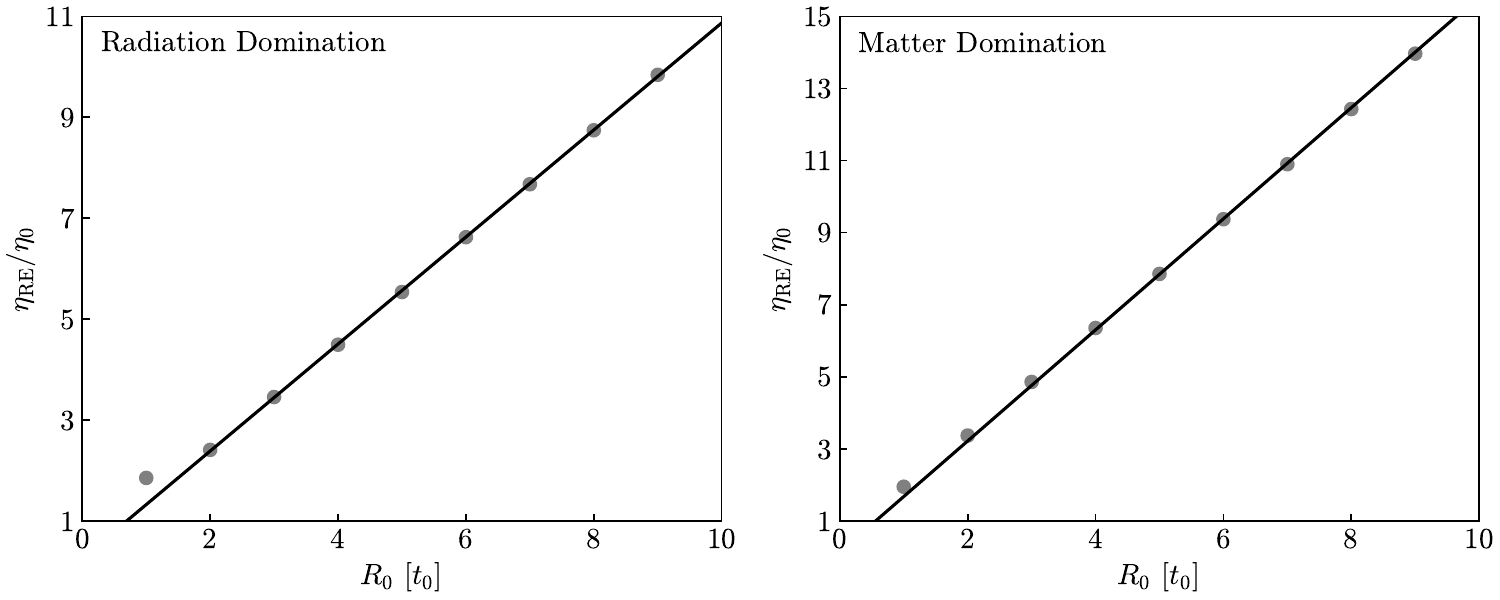}
\caption{The conformal horizon re-entry time $\eta_{\mathrm{RE}}/\eta_0$ as a function of the initial spherical wall radius $R_0$ computed from the domain wall collapse simulation shown in Figure~\ref{fig:sphere_collapse_time}. \textit{Left}: The conformal time at horizon re-entry $\eta_\mathrm{RE}/\eta_0$ assuming radiation domination. The gray points are the simulation results and the black line is a linear fit to the simulation results. Note that the relation is highly linear in conformal time, with corresponding coefficients given by Eq.~\eqref{eq:reentry_time_reln}. The $\alpha=1$ point is excluded in this fit to account for the wall already being horizon sized at formation. \textit{Right}: Same as left, but assuming matter domination. Note that in a matter-dominated cosmology, it takes longer for the walls to re-enter the horizon and ultimately collapse compared to a radiation dominated cosmology.}
    \label{fig:conformal_reentry_time_fit}
\end{figure}

\begin{figure}
    \centering
    \includegraphics[width=\textwidth,keepaspectratio]{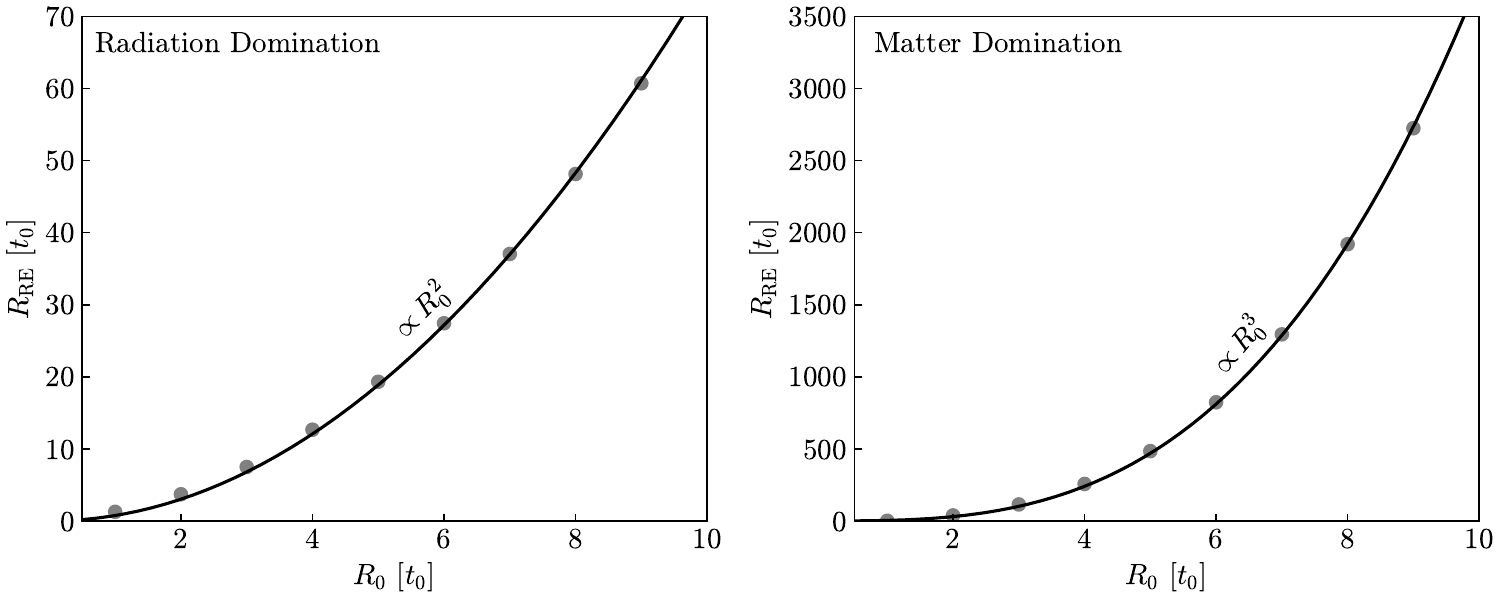}
\caption{\textit{Left}:  The physical domain wall horizon re-entry radius $R_\mathrm{RE}$ as a function of initial radius $R_0$. The gray points are the simulation results and the black line is a linear fit to the simulation results, given by Eq.~\eqref{eq:reentry_radii}. \textit{Right}: Same as left, but assuming matter domination where $a\propto t^{2/3}$, hence the larger re-entry radii.}
    \label{fig:reentry_radius_fit}
\end{figure}

With this noted, we simulate spherical domain wall collapse via Eq.~\eqref{eq:dimless_w_hubble_eom} for a range of initial wall radii $R_0$ in order to infer the relation between initial wall size, peak wall radius, and the amount of time it takes for a wall to enter the horizon. Here, we define horizon re-entry as the moment when the physical velocity of the wall reverses sign. The left (right) panel of Fig.~\ref{fig:sphere_collapse_time} shows the evolution of the physical radius $R$ of the wall as a function of time in a RD (MD) era. Note the wall stretches to a larger peak radius in a MD era due to the faster expansion rate of the universe. Similarly, Fig.~\ref{fig:conformal_reentry_time_fit} shows a linear fit to the re-entry time $\eta_{\mathrm{RE}}$ for wall collapse during both RD (left) and MD (right) eras. In conformal time $\eta$, this relation is
\begin{align}
    \frac{\eta_{\mathrm{RE}}}{\eta_0} 
    \simeq 
    \begin{dcases}
    1.06\alpha + 0.26 \quad &\text{RD Era} \, ,
    \\
    1.51\alpha + 0.38 \quad &\text{MD Era} \, .
    \end{dcases}
    \qquad \quad (\alpha \equiv R_0/t_0)
\label{eq:reentry_time_reln}
\end{align} 
The linearity of this relation is principally due to the fact that the ratio between any distance times the scale factor $a$ and the horizon size $t$ is linear in $\eta$. We also extract the re-entry radius $R_{\mathrm{RE}}$ as a function of $\alpha \equiv R_0/t_0$. It is
\begin{align}
    R_{\rm RE}  \simeq R_0\times
    \begin{dcases}
    \max\{0.75 \alpha^2,\alpha\} \quad &\text{RD Era} \, ,
    \\
    \max\{3.75 \alpha^3,\alpha\}  \quad &\text{MD Era} \, ,
    \end{dcases}
    \qquad \quad (\alpha \equiv R_0/t_0)
\label{eq:reentry_radii}
\end{align} 
where we take the maximum of the initial radius $\alpha$ and the interpolated power laws in order to account for the fact that the smallest walls ($\alpha \lesssim  1$) are already subhorizon sized upon formation. We show the power law fits to the simulation data in Fig.~\ref{fig:reentry_radius_fit}.
\subsection{Re-entry and Collapse Tension}
\label{sec:reentry_tension}
We also use simulations to account for the change in the domain wall thickness as it expands and later contracts. By Eq.~\eqref{eq:axion_mass_temp}, between formation and re-entry, the axion wall thickness changes by a factor
\begin{align}
    \frac{\delta_{\rm RE}}{\delta_0} = \frac{m_a(t_0)}{m_a(t_{\rm RE})}= \frac{\min 
    \left\{\left(\frac{T_{\rm QCD}}{T_{0}}\right)^{n/2}, 1\right\}}
    {\min 
    \left\{\left(\frac{T_{\rm QCD}}{T_{\rm RE}}\right)^{n/2}, 1\right\}}
    =
    \frac{\min 
    \left\{\left(\frac{T_{\rm QCD}}{T_{0}}\right)^{n/2}, 1\right\}}
    {\min 
    \left\{\left(\frac{T_{\rm QCD}}{T_{0}}\frac{T_{0}}{T_{\rm RE}}\right)^{n/2}, 1\right\}},
    \label{eq:wallthicknessRatio}
\end{align}
where $T_0$ ($T_{\rm RE}$) is the temperature of the universe at the wall formation (horizon re-entry) time $t_0$ ($t_{\rm RE}$). Note that $\delta_{\rm RE}/\delta_0$ lies between $(T_{\rm RE}/T_0)^{n/2} \leq 1$ and $1$, where the former case corresponds to the axion not reaching its zero temperature value by the time the wall re-enters the horizon, and the latter case corresponds to the axion mass reaching its zero-temperature value by the time the wall forms.

The wall formation temperature $T_0$ ranges from the temperature at which the wall thickness becomes smaller than the horizon, $T_{\rm wall}$, to when the string-wall network stops scaling and is destroyed, $T_{\mu/\sigma}$. As can be seen by the right-most equation of \eqref{eq:wallthicknessRatio}, the ratio of the formation temperature $T_0$ to $T_{\rm QCD}$ and to the corresponding re-entry temperature, $T_{\rm RE}$, matters significantly for $\delta_{\rm RE}/\delta_0$. When the walls form in a RD era and $f_a \lesssim 10^{17}$ GeV, the formation temperatures $T_{\rm wall} \simeq T_{\rm osc} \simeq T_{\mu /\sigma}$ are nearly identical. The value for $T_0/T_{\rm QCD} \simeq T_{\rm osc}/T_{\rm QCD}$ is given in the top line of Eq.~\eqref{eq:TOsc} as a function of $f_a$. The ratio $T_0/T_{\rm RE}$ depends only on $\alpha$ and the time-temperature relation of the cosmological era
\begin{align}
    \frac{T_0}{T_{\rm RE}} \simeq 
     \begin{dcases}
    \frac{\eta_{\rm RE}}{\eta_0} \left(\frac{g_{*,\rm RE}}{g_{*,0}}\right)^{1/3} \quad &\text{RD Era},
    \\
    \left(\frac{\eta_{\rm RE}}{\eta_0} \right)^{3/4}\left(\frac{g_{*,\rm RE}}{g_{*,0}}\right)^{1/4} \quad &\text{MD (Non-Adiabatic) Era}.
    \end{dcases}
\end{align}
where $\eta_{\rm RE}/\eta_0$ is given in Eq.~\eqref{eq:reentry_time_reln} and is a function of $\alpha$.
If the walls form in a MD era and $f_a \gtrsim 10^{15}$ GeV (as will be the case for the largest PBH signal), then the axion is already at its zero temperature value by the time the universe cools to $T_{\rm wall}$. In that case, Eq.~\eqref{eq:wallthicknessRatio} is simply unity for any $T_0$.

\begin{figure}
    \centering
    \includegraphics[width=\textwidth,keepaspectratio]{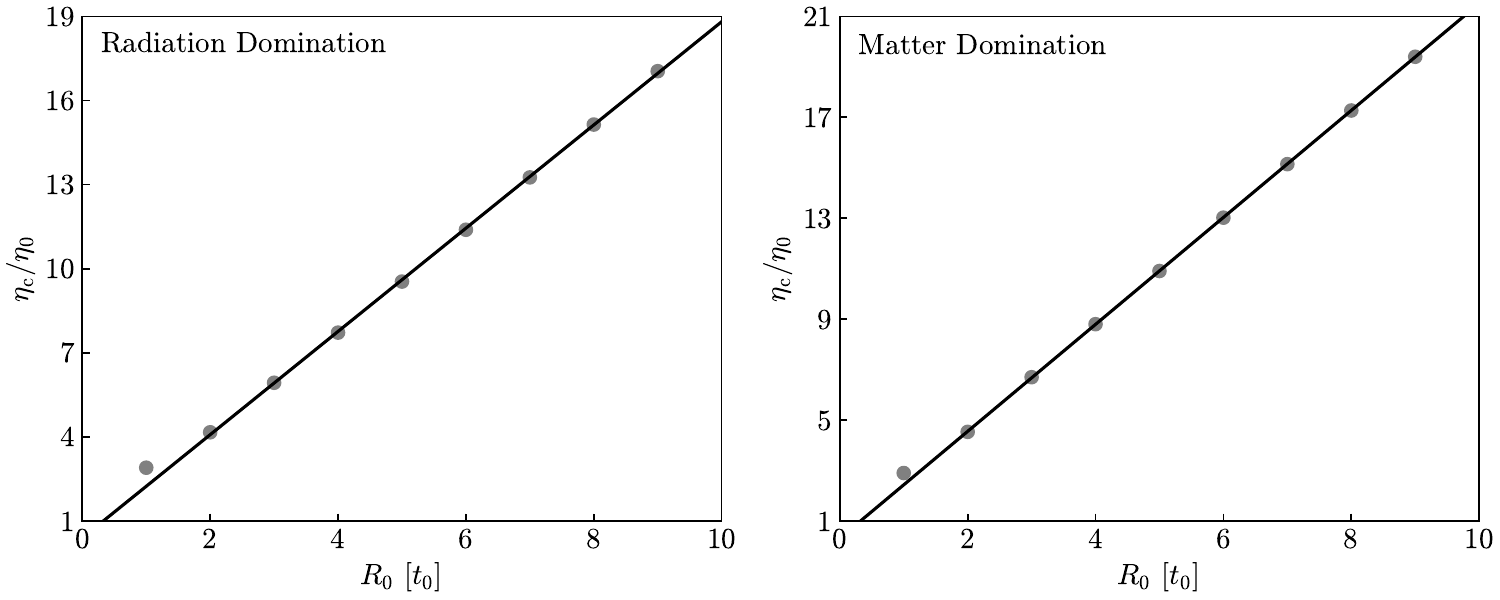}
\caption{The normalized conformal time length $\eta_c/\eta_0$ at which closed spherical domain walls have fully collapsed as a function of their initial radius $R_0$. \textit{Left}: The gray data points are collapse times obtained from our wall simulations assuming radiation domination. The black line shows the best linear fit to these data points. \textit{Right}: Same as left, but assuming matter domination.}
    \label{fig:collapse_time_fit}
\end{figure}

Finally, the total collapse time $\eta_\mathrm{c}$ of the wall can be determined from these simulations as well. Note that while we delegate studying the compression of energy into a PBH to a distinct set of simulations covered in Sec.~\ref{sec:SubHorizonCollapse}, the simulations used in this section yield an accurate prediction for the amount of time it takes for a domain wall to be maximally compressed. We show a linear fit to these collapse times in Fig.~\ref{fig:collapse_time_fit}. This linear relation is given by
\begin{equation}
    \frac{\eta_{\mathrm{c}}}{\eta_0} 
    \simeq 
    \begin{dcases}
    1.84\alpha + 0.39 \quad &\text{RD Era},
    \\
    2.12\alpha + 0.31 \quad &\text{MD Era}.
    \end{dcases}
    \qquad \quad (\alpha \equiv R_0/t_0)
    \label{eq:collapse_eta}
    \end{equation}
The results of Eqns.~\eqref{eq:reentry_time_reln}, \eqref{eq:reentry_radii} and \eqref{eq:collapse_eta} capture the superhorizon behavior of perfectly spherical walls. As (more realistic) aspherical walls expand, they deform due to the expansion of the universe. Sides of the wall with different displacement from the wall center of mass stretch by differing amounts and re-enter the horizon at different times. We run identical simulations to the spherical ones described above, but with ellipsoidal initial wall configurations and describe the results in Appendix~\ref{app:superhorizon_asphericities} to account for this. We find that prior to horizon re-entry, deformation of walls with large-scale asphericities typically amounts to a decrease in the ellipsoid parameters $\epsilon_1$ and $\epsilon_2$ of a few percent.
\section{Subhorizon Collapse of Domain Walls}
\label{sec:SubHorizonCollapse}
In this section, we study the dynamics of self-enclosed domain walls  from the moment they re-enter the cosmic horizon. We first discuss perfectly spherical walls in Sec.~\ref{subsec:sphericalWallEfficiency} before considering realistic walls with asphericities in Sec.~\ref{subsec:asphericities}. In both cases, we calculate the maximum energy within a given radius to determine how efficient walls compression is.
\subsection{Compressibility of Perfectly Spherical Walls}
\label{subsec:sphericalWallEfficiency}
At horizon re-entry, $t_{\rm RE}$, walls are at their turning point transitioning from expansion to collapse. For a perfectly spherical wall, there are only two parameters that control the dynamics of wall collapse: the wall thickness $\delta \equiv m_a^{-1}$ (in the rest frame of the wall) and the radius at horizon re-entry, $R(t_{\rm RE}) = R_{\rm RE}$. In fact, when the wall is subhorizon, Hubble friction is negligible and the equation of motion can be made scale invariant so that it depends only on the ratio $\delta/R$, even for a temperature-dependent axion mass.\footnote{This scale invariance can be seen by the change of variables $\{t, r\} \rightarrow \{\tilde{t}, \tilde{r}\} \equiv \{m_a(t) t, m_a(t) r\}$ and dropping terms $\mathcal{O}{((m_a t)^{-1})}$ which are small now that the wall is subhorizon.} Eq.~\eqref{eq:axion_hubble_eom} then becomes
\begin{align}
    \label{eq:axionEOMSubhorizon}
    (1 + \kappa)^2\frac{\partial^2 \theta }{\partial \tilde{t}^2} - \ \frac{\partial^2 \theta }{\partial \tilde{r}^2}  - \frac{2}{\tilde{r}}\frac{\partial \theta }{\partial \tilde{r}} + \sin \theta = 0 
    \qquad
    \kappa = \begin{dcases}
        \frac{n}{4} \quad \text{RD era},
        \\
        \frac{n}{3} \quad \text{MD adiabatic era},
        \\
        \frac{n}{8} \quad \text{MD non-adiabatic era},
    \end{dcases} \,
\end{align}
where $n \approx 6.68$ is the power at which $m_a^2$ increases with temperature (see Eq.~\eqref{eq:axion_mass_temp}) and $\kappa/n = -(1/2) (d\ln T/d \ln t)$ is related to the time-temperature relation at the QCD phase transition. 
\begin{figure}[!]
 \centering
    \includegraphics[width=\textwidth]{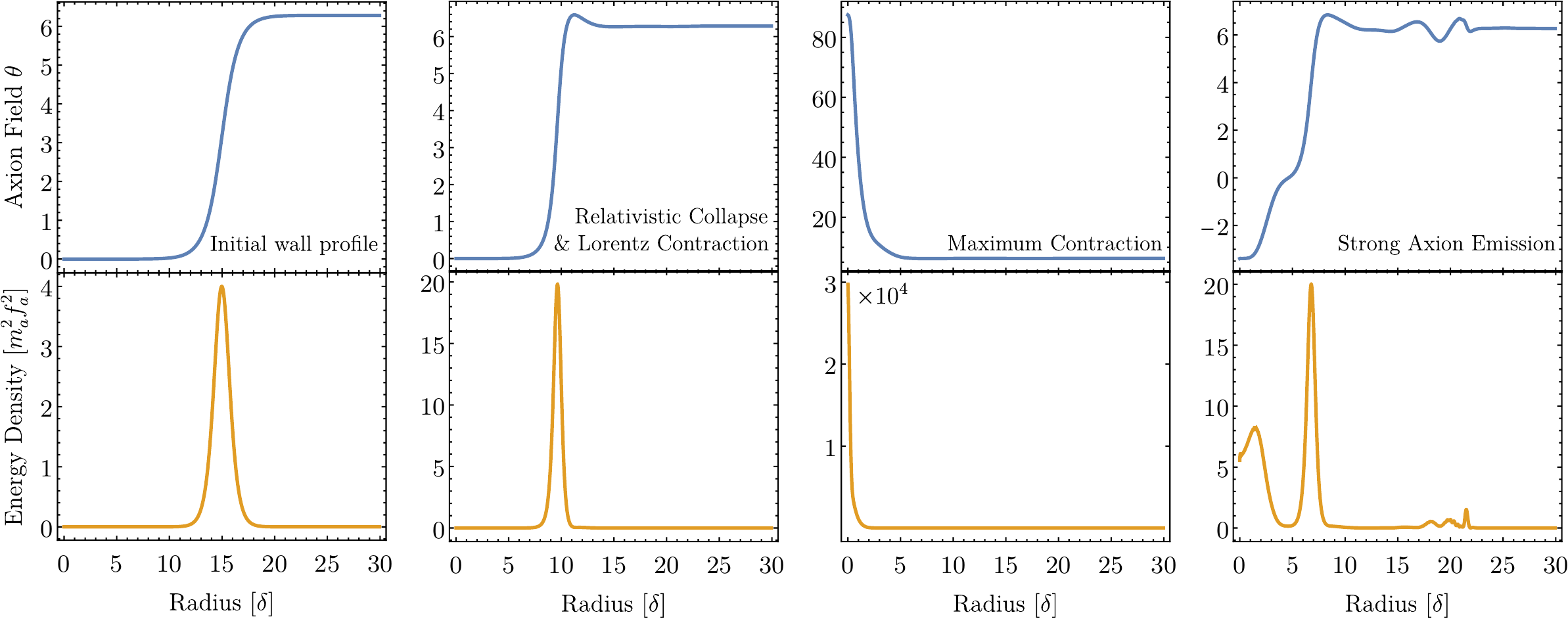}
\caption{The time evolution of a subhorizon domain wall. \textit{Top}: The axion field configuration evolving according to Eq.~\eqref{eq:axionEOMSubhorizon}, with each panel from left to right corresponding to a later time. In the first panel, the wall is at rest. In the second panel, the wall has started Lorentz contracting and moving to the left. In the third panel, the wall is maximally compressed. In the fourth panel, the wall has passed through itself, radiating a large amount of energy in the form of axions. \textit{Bottom}: The axion field energy density corresponding to the field configurations shown in the top panels. Notice that the energy density is highly localized in the domain wall surface, being maximally compressed in the third panel, but having been split into several components propagating outwards in panel four after a PBH did not successfully form.}
\label{fig:evolutionPanel}
\end{figure}

We numerically compute the evolution of the collapsing axion wall after horizon re-entry for a wide range of $\delta_{\rm RE}/R_{\rm RE}$, taking a starting axion field configuration of 
\begin{align}
    \theta(t_{\rm RE}, r)=4\tan^{-1}\exp \big[ \delta_{\rm RE}^{-1}(r - R_{\rm RE}) \big], \qquad \dot{\theta}(t_{\rm RE}, r) = 0 \, ,
    \label{eq:initConditionsSubhorizon}
\end{align}
which is the analytic solution to Eq.~\eqref{eq:axionEOMSubhorizon} for large  $R_{\rm RE}/\delta_{\rm RE}$. We use the numerical solution for $\theta(t, \mathbf{x})$ to compute the energy density of the axion field, $\rho(t, \mathbf{x})$, as
\begin{align}
     \rho(t, \mathbf{x}) = f_a^2 \left[\frac{1}{2} \dot{\theta}(t,\mathbf{x})^2 + \frac{1}{2} |\nabla \theta (t, \mathbf{x})|^2 + m_a(t)^2 (1 - \cos \theta (t, \mathbf{x}))\right]  \, ,
\end{align}
and the energy within a given enclosed radius $r_{\rm encl}$ as a function of time as
\begin{align}
    \label{eq:energyREnclosed}
    E(t, r_{\rm encl}) = \int_{|\mathbf{x}|< r_{\rm encl}} \rho(t, \mathbf{x}) \, d^3 \mathbf{x} \, .
\end{align}
Fig.~\ref{fig:evolutionPanel} shows the evolution of $\theta(t, r)$ and $\rho(t, r)$ for a fiducial $R_{\rm RE}/\delta_{\rm RE} = 15$. The evolution can be understood as follows: Initially, the wall is at rest and separates a region where $\theta = 0$ inside the wall and $\theta = 2\pi$ outside as shown by the blue contour on the top of the first panel. The thickness of the wall, which can be seen more clearly from the width of the Gaussian-like peak of the energy density, is of order $m_a^{-1}$, as shown by the orange contour on the bottom of the first panel. Next, the wall begins to collapse, converting rest mass into kinetic energy and picking up speed as shown by the second panel. At this stage of collapse, the Lorentz factor, $\gamma$, can be estimated from the Nambu-Goto effective field theory, which predicts $\gamma = (R_0/R)^2$. As the wall becomes more and more relativistic, a deviation from the arctan profile occurs with a small peak developing on the wall as shown on the top of the second panel. Here, the thin-wall limit breaks down since the curvature term, $2 \partial_{\tilde r} \theta/\tilde{r} \sim \gamma \delta_{\rm RE} \theta/R$, becomes comparable to the potential term, $\sin \theta$~\cite{Widrow:1989vj}. This occurs at the critical radius $R_* \approx \gamma_* \delta_{\rm RE} \approx R_{\rm RE} (\delta_{\rm RE}/R_{\rm RE})^{1/3}$, Lorentz factor, $\gamma_* \approx (R_{\rm RE}/\delta_{\rm RE})^{2/3}$, and Lorentz contracted wall thickness $\delta_* \approx \delta_{\rm RE}/\gamma_*$. At the final stage of collapse when $R < R_*$, the kinetic term in Eq.~\eqref{eq:axionEOMSubhorizon} is much greater than the potential ($\sin \theta$) term. The axion equation of motion effectively becomes the wave equation. The wave equation solution for $R < R_*$ is a radially inward-moving wave with a peak growing linearly with time, but with a constant width. The minimum radius that a spherical wall can be fully compressed to is thus approximately the Lorentz contracted wall thickness $\delta_* = \delta_{\rm RE} (\delta_{\rm RE}/R_{\rm RE})^{2/3}$.\footnote{Particles in the background plasma that the axion couples to can damp wall motion if they reflect off the wall~\cite{Everett:1974bs, Kibble:1976sj}. This occurs when the particle momentum is less than the inverse Lorentz contracted wall thickness $1/\delta_*$~\cite{huang1985structure, ferreira2022high, blasi2023friction,blasi2023axionic}. The $\delta_*$ of walls that dominantly end up forming PBHs is typically less than $(100 m_a)^{-1}$, so that only non-relativistic axions from misalignment are cold enough to reflect off the enclosed walls, as is the usual case for the mildly relativistic string-wall network~\cite{huang1985structure,Chang:1998tb} (see however,~\cite{Evans:1996eq}). The friction generated by the cold axion population is small, especially after $t_{\rm osc}$~\cite{huang1985structure}.}
A maximally compressed wall is shown on the third panel of Fig.~\ref{fig:evolutionPanel}. Finally, the violent collision of the wall with itself at collapse (assuming it has not formed a PBH), induces strong axion emission, as shown by the wiggles on the rightmost panel of Fig.~\ref{fig:evolutionPanel}, which travel radially outward. The remnant of the wall expands and collapses again, losing about half its energy to axion emission with each bounce.

\begin{figure}[tb]
    \centering
    \includegraphics[width=.49\textwidth]{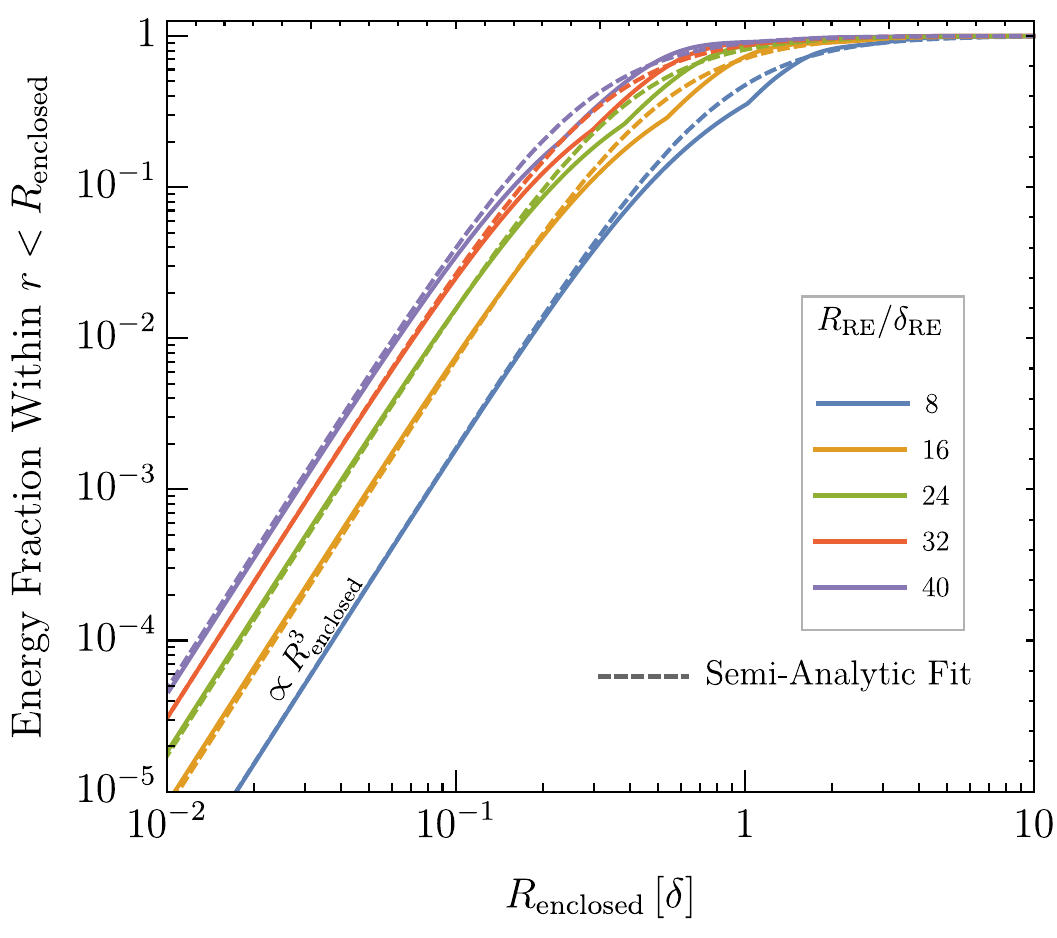} 
    \includegraphics[width=.49\textwidth]{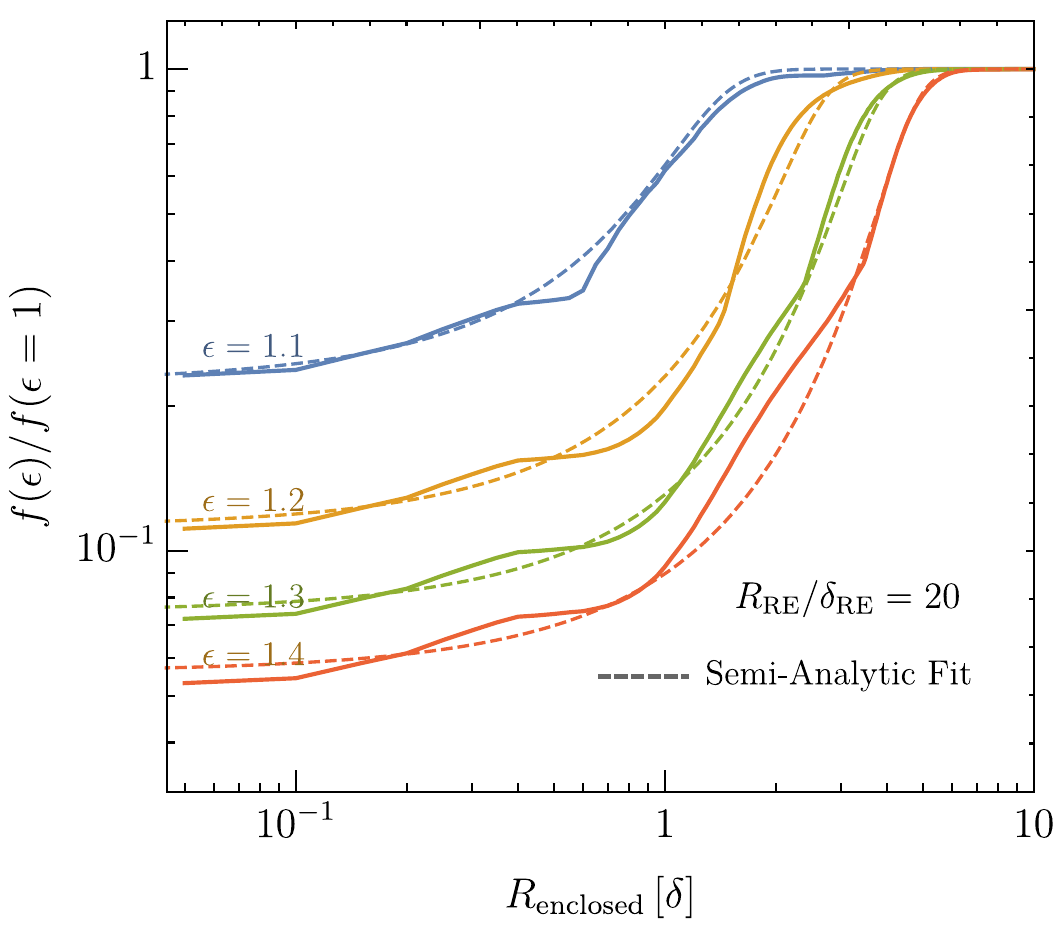} 
    \caption{Energy compression in domain wall collapse. \textit{Left}: The peak energy fraction compressed into a spherical area of radius $R_\mathrm{enclosed}$ as a function of $R_\mathrm{enclosed}$ in units wall thickness $\delta$ for spherical domain wall collapse. The solid colored lines indicate energy fractions for walls with five different re-entry radii $R_{\mathrm{RE}}$. Note, the slight kinks at $R_{\mathrm{enclosed}}\sim\delta$ are caused by the appearance of a small oscillatory correction to $\dot{\theta}$ in the final moments before collapse, where the analytic domain wall solution becomes invalid. The dashed colored lines indicate the best analytic fits to Eq.~\eqref{eq:fFractionEnclosed}. \textit{Right}: The relative suppression in energy that can be compressed into a sphere of radius $R_\mathrm{enclosed}$ as a function of $R_\mathrm{enclosed}$ assuming that the domain wall is aspherical along a single axis with ellipsoid parameter $\epsilon$. The four distinct colored lines correspond to walls with four different ellipsoid parameters. Note that the energy fraction within the sphere $f(\epsilon)$ has been normalized to the fraction in the perfectly spherical ($\epsilon=1$) case. The dashed colored lines indicate the best analytic fits to Eq.~\eqref{eq:asphericalReduction}.}
    \label{fig:spherical_efrac}
\end{figure}

The left panel of Fig.~\ref{fig:spherical_efrac} shows the largest fractional energy, $f_\circ \equiv \text{Max} \{ E(t, r_{\rm encl})/ \/E(t,\infty) \}$, that a given radius encloses for different contours of the wall re-entry radius to thickness ratio, $R_{\rm RE}/\delta_{\rm RE}$ of a spherical wall. 
Fig.~\ref{fig:spherical_efrac} demonstrates that an $\mathcal{O}(1)$ fraction of the energy in a spherical wall can be compressed into a radius of order the wall thickness. Moreover, the larger the initial wall size (larger $R_{\rm RE}/\delta_{\rm RE}$), the smaller the volume the wall can be compressed to, as shown by the contours shifting to the left with increasing $R_{\rm RE}/\delta_{\rm RE}$. The greater compressibility of larger walls is due to their greater Lorentz contraction. For a fixed $r_{\rm encl} \lesssim \delta$, the fractional energy enclosed grows as $(R_{\rm RE}/\delta_{\rm RE})^2$, which is proportional to the volume of the minimum Lorentz contracted wall thickness, $\delta_*^3$. The energy enclosed falls off rapidly ($\propto r_{\rm encl}^3)$ for $r_{\rm encl} \ll \delta_*$, since the energy density, $\rho$, of the axion field at small $r_{\rm encl}$ is roughly a sharp peak of thickness $\delta_*$, so that for $r_{\rm encl} \ll \delta_*$, $\rho$ is approximately constant. Eq.~\eqref{eq:energyREnclosed} then indicates that $E(t, r_{\rm encl}) \propto r_{\rm encl}^3$ as seen in the simulations. 

In summary, numerical simulations indicate that for a collapsing spherical wall, the maximum fraction of its total energy enclosed within a given radius, $r_{\rm encl}$, is nearly unity for $r_{\rm encl} \gg \delta_*$, but rapidly drops as $r_{\rm encl}^3$ for $r_{\rm encl} \ll \delta_*$ and is proportional to the cube of the minimum Lorentz contracted wall thickness $\delta_*^3 \propto (R_{\rm RE}/\delta_{\rm RE})^2$ for small $r_{\rm encl}$. This is captured by the semi-analytic broken power law
\begin{align}
    \label{eq:fFractionEnclosed}
    f_\circ \equiv \text{Max}\frac{E(t, r_{\rm encl})}{E(t, \infty)} &\approx \left[\left(\mathcal{A}^{-1}\left(\frac{r_{\rm encl}}{\delta}\right)^{-3} \left(\frac{R_{\rm RE}}{\delta_{\rm RE}} \right)^{-2}\right)^{2/3} + 1\right]^{-3/2} \\
    &\simeq 
    \begin{dcases}
        \mathcal{A}\left(\frac{r_{\rm encl}}{\delta}\right)^{3} \left(\frac{R_{\rm RE}}{\delta_{\rm RE}} \right)^{2} \quad & r_{\rm encl} \ll \delta_*,
        \\
        1 \quad & r_{\rm encl} \gg \delta_*,
    \end{dcases}
\end{align}
where the constant $\mathcal{A} \approx 1/34$ is calibrated by the numerical simulations and $\delta$ is the \textit{instantaneous} rest frame wall thickness. The semi-analytic fit, Eq.~\eqref{eq:fFractionEnclosed}, is shown by the dashed contours in the left panel of Fig.~\ref{fig:spherical_efrac}.
\subsection{Compressibility of Realistic Walls with Asphericities}
\label{subsec:asphericities}
\begin{figure}
    \centering
    \includegraphics[width=1.0\textwidth,height=1.0\textheight,keepaspectratio]{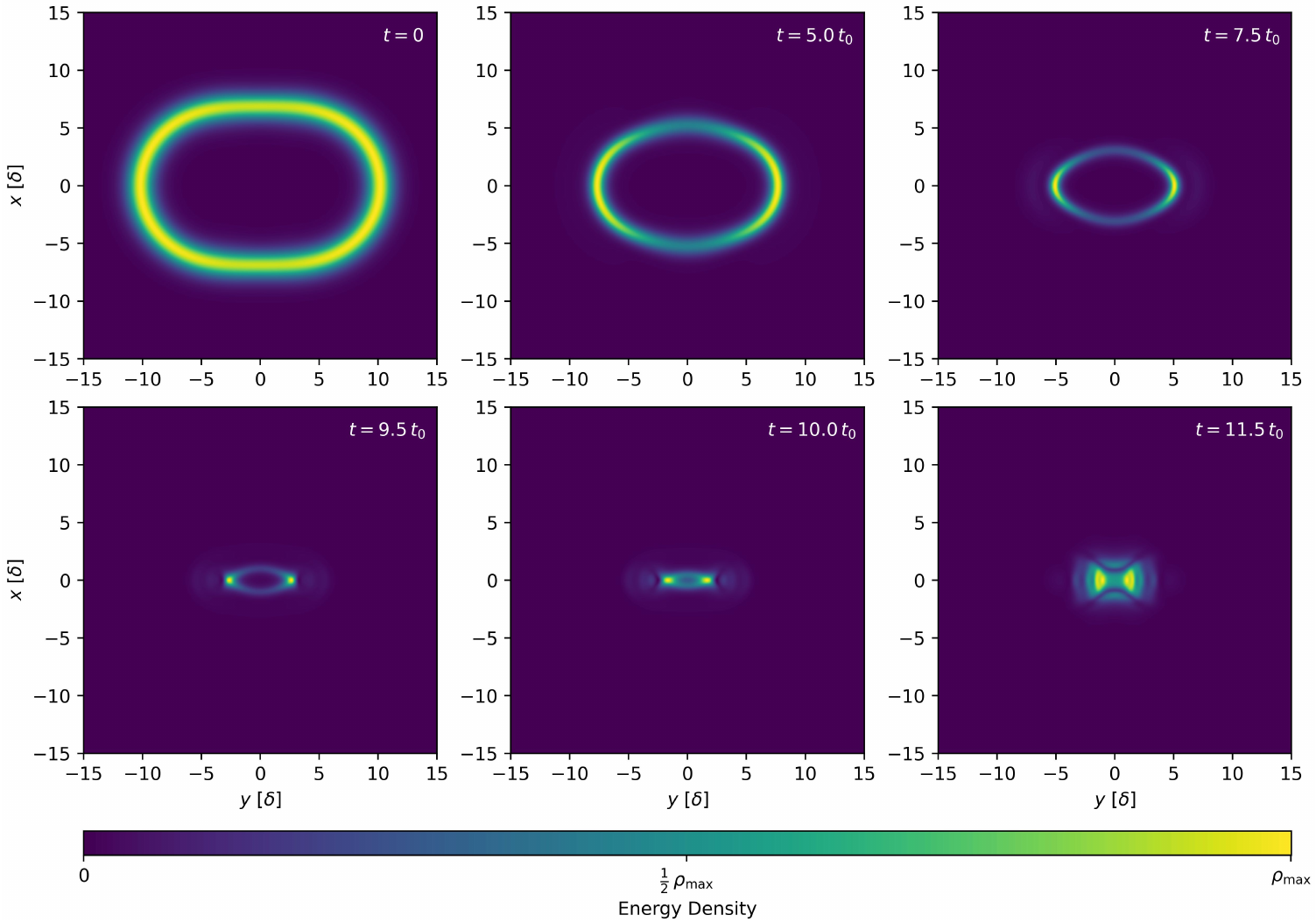}
\caption{Simulation of subhorizon aspherical wall collapse (with Hubble expansion neglected). Each panel is a contour map of the axion field energy density $\rho$ at a specific time viewed in the $z=0$ plane.\protect\footnotemark ~Note that the color shading is individually normalized in each panel. The initial field configuration used in this example is given by Eq.~\eqref{eq:aspherical_field_config} with $R_0=7\delta$ and $\epsilon_1=1.5$, $\epsilon_2=1.0$. As the wall shrinks, the energy density grows larger on parts of the wall surface where the curvature is high. The bottom middle panel shows the field configuration near maximal energy compression. The bottom right panel shows the field configuration past the point of maximal compression -- axion radiation escapes outward.}
    \label{fig:aspherica_collapse_panels}
\end{figure}

In this section, we extend the analysis of Sec.~\ref{subsec:sphericalWallEfficiency} to determine the condition for PBH formation for realistic, non-spherical walls. The effect of asphericities on collapse was first discussed by Widrow~\cite{Widrow:1989fe} who analytically found that large scale aspherical perturbations on an enclosed domain wall are slightly damped in the early phases of collapse but then grow dramatically during the final moments (see similarly Appendix B of~\cite{Deng:2017uwc}). There are two important caveats to this analysis: first, asphericities were only shown to grow to linear order in perturbation theory, that is, only valid in a regime when the  perturbations are small compared to the size of the wall. Second, the analysis of~\cite{Widrow:1989fe,Deng:2017uwc} was done in the Nambu-Goto picture which breaks down in the regime where the perturbations are expected to grow the most (at $R < R_*$). To overcome these limitations, we study the effect of asphericities numerically using the full Euler-Lagrange equation of motion for the axion field \eqref{eq:axion_hubble_eom}. While we do find asphericities grow during collapse (see Fig.~\ref{fig:aspherica_collapse_panels}), this growth is not as severe as previously expected in the literature; the ultimate effect is to moderately reduce the efficiency of PBH formation.

Large scale asphericities on the wall are more important than small scale asphericities for two reasons: First, Hubble freezes perturbations on the wall larger than the horizon while perturbations smaller than the horizon are free to move which become damped from oscillating and emitting axions or gravitational waves. After re-entering the horizon, the walls of typical curvature radius $R_{\rm RE} \approx t_{\rm RE}$ will mostly have perturbations of size $\mathcal{O}(R_{\rm RE})$. Second, from the mode analysis of~\cite{Widrow:1989fe} in the Nambu-Goto picture, it was found that the largest scale perturbations grow the most. For these reasons, we only consider large scale asphericities and approximate them by modeling the enclosed wall as an ellipsoid. 

We repeat the simulations and analysis of Sec.~\ref{subsec:sphericalWallEfficiency} with an initial wall configuration that possesses $\theta$ and $\phi$ dependence
\begin{align}
    \theta(t_{\rm RE}, r,\theta, \phi) = 4 \tan^{-1} \exp[ \delta_{\rm RE}^{-1}(r - R_{\ellipse}(\theta,\phi))]\, \quad   \dot{\theta}(t_{\rm RE}, r,\theta, \phi) = 0 \, ,
\label{eq:aspherical_field_config}
\end{align}
where $R_{\ellipse}(\theta,\phi)$ is the radial profile of the ellipsoidal wall with semi-axes $(R_{\rm RE}, {\epsilon_1} R_{\rm RE}, {\epsilon_2} R_{\rm RE} $),
\begin{align}
    R_{\ellipse}(\theta,\phi) = \frac{R_{\rm RE}}{\sqrt{(\cos^2 \phi + \epsilon_1^{-2} \sin^2 \phi) \sin^2 \theta + \epsilon_2^{-2} \cos^2 \theta}} \, .
\label{eq:ellipsoid_eqn}
\end{align}
Without loss of generality, we take $\epsilon_1 \geq 1$ and $\epsilon_2 \leq 1$ so that the curvature radius $R_{\rm RE}$ is defined as the median semi-axis at horizon re-entry. The random field simulations of Sec.~\ref{sec:closed_dw_abundance} give a typical $\epsilon_1 \sim 1.4$ an $\epsilon_2 \sim 0.6$ as shown by Fig.~\ref{fig:ellipticity_dist}. We show time slices of an aspherical wall simulation in Fig.~\ref{fig:aspherica_collapse_panels}.
\footnotetext{We note that due to the $\tan^{-1}$ wall profile being set by the radial vector rather than the orthogonal surface vector, highly aspherical walls appear slightly more pill-shaped in our simulations, with negligible effects on dynamics.}

Our goal is to numerically compute the reduction factor for the enclosed energy within a given radius, $r_{\rm encl}$, for an ellipsoidal wall of initial curvature radius $R_{\rm RE}$ and ellipsoidal parameters ($\epsilon_1, \epsilon_2$) compared to a perfectly spherical wall of the same curvature radius $R_{\rm RE}$ (i.e. with $\epsilon_1=\epsilon_2 = 1$).

The right panel of Fig.~\ref{fig:spherical_efrac} shows this reduction in enclosed energy of an ellipsoidal wall compared to a spherical wall for $R_{\rm RE}/\delta_{\rm RE} = 20$ with $\epsilon_2 = 1$ and $\epsilon_1$ allowed to vary (i.e. the walls are more like a baguette than a pancake) as shown by the colored contours. As expected, the more $\epsilon_1$ deviates from unity, the larger the suppression in enclosed energy compared to a perfectly spherical wall. Moreover, the right panel of Fig.~\ref{fig:spherical_efrac} indicates that most of the energy in an elliptical wall can still be compressed into a radius $r_{\rm encl} \gg \text{few} \times \delta$ because the relative suppression compared to a spherical wall of the same mean $R_{\rm RE}$ is negligible. However, for $r_{\rm encl} \lesssim \text{few} \times \delta$, the eccentricity of the walls grows dramatically and the suppression in the energy enclosed compared to a spherical wall is severe, becoming more suppressed, with larger $R_{\rm RE}/\delta_{\rm RE}$. By simulating a variety of $R_{\rm RE}/\delta_{\rm RE}$ and $\epsilon_1$, we find a semi-analytic function for the relative suppression of an ellipsoidal wall of eccentricity $\epsilon_1 >1$ and $\epsilon_2 = 1$ compared to a spherical wall ($\epsilon_1 = \epsilon_2 = 1$) as shown by the dashed curves in the right panel of Fig.~\ref{fig:spherical_efrac}.  This fit is given approximately by a broken power law
\begin{align}
    \label{eq:asphericalReduction}
     \frac{f(\epsilon_1, \epsilon_2 = 1)}{f_\circ} \equiv f_{\ellipse}(\epsilon_1)  
     &\approx
     \left[1 + \left(\mathcal{B}^{-1}|\epsilon_1 -1|\left(\frac{R_{\rm RE}}{\delta_{\rm RE}}\right)^{1.83} S\left(\frac{r_{\rm encl}}{\delta}, \epsilon_1\right)\right)^3 \right]^{-1/3},
\end{align}
where $\mathcal{B} \approx 5.4$ and $S$ is a sigmoid function that goes to $1$ for $r_{\rm encl}/\delta \ll 1$ and to $0$ for $r_{\rm encl}/\delta \gg 1$. Parametrically, $S$ is given by 
\begin{align}
    S\left(\frac{r_{\rm encl}}{\delta}, \epsilon_1\right) \approx \left(\frac{2}{1 + \exp\left[\mathcal{C} \frac{r_{\rm encl}}{\delta}|\epsilon_1 - 1|^{-.55}\right]} \right)^{1.5} \, ,
\end{align}
where $\mathcal{C} \approx 0.33$. The sigmoid function $S$ gives rise to a characteristic radius $r_{\rm break} > \delta$ where for $r_{\rm encl} \gtrsim r_{\rm break}$, the maximum fractional energy enclosed within a radius $r_{\rm encl}$ is effectively unity and unchanged relative to a spherical wall of the same re-entry radius $R_{\rm RE}$; conversely, for $r_{\rm encl} \ll r_{\rm break}$, the maximum fractional energy enclosed is much less than the spherical wall of the same $R_{\rm RE}$ due to the large scale asphericities on the wall which grow strongly for $r_{\rm encl} \ll r_{\rm break}$. That is, the relative suppression of an ellipsoidal wall compared to a spherical wall, Eq.~\eqref{eq:asphericalReduction} is approximately
\begin{align}
    \label{eq:approxAsphericalSuppression}
    \frac{f(\epsilon_1, \epsilon_2 = 1)}{f_\circ} 
     &\approx 
    \begin{dcases}
        1 \hfill & r_{\rm encl} \gtrsim r_{\rm break},
        \\
         \mathcal{B}\left(\epsilon_1 - 1\right)^{-1} \left(\frac{R_{\rm RE}}{\delta_{\rm RE}} \right)^{-1.83} \hfill & r_{\rm encl} \ll r_{\rm break},
    \end{dcases}
\end{align}
where
\begin{align}
    \label{eq:rbreak}
    \frac{r_{\rm break}}{\delta} \approx 5\left(\frac{\epsilon_1 - 1}{0.3}\right)^{0.55} \left(1 + \frac{1}{5}\ln \left[\left( \frac{R_{\rm RE}/\delta_{\rm RE}}{20}\right)^{1.83} \left(\frac{\epsilon_1 - 1}{0.3}\right) \right]\right) \, .
\end{align}
We emphasize that numerical limitations prohibit simulating aspherical wall dynamics for walls much larger than $R_{\rm RE}/\delta_{\rm RE} \sim 30$, which is why it is important to simulate a variety of walls to infer how the enclosed energy fraction Eq. \eqref{eq:asphericalReduction} scales with $R_{\rm RE}/\delta_{\rm RE}$ so that extrapolation to larger initial walls can be inferred. In the relevant range of wall parameters where the PBH abundance contribution is largest, the extrapolation to larger $R_{\rm RE}/\delta_{\rm RE}$ is typically a factor of $\sim 10-50$ times larger than the largest walls we can simulate, so that the true value of $r_{\rm break}$ \eqref{eq:rbreak} may be slightly different. This can mildly change the conditions for PBH formation as discussed in Sec. \ref{sec:compressionToSchwarzchild}.

The results of Eqns.~\eqref{eq:approxAsphericalSuppression} and \eqref{eq:rbreak} have important implications for the formation of PBHs. Since most walls are not perfectly spherical (see the distribution of Fig.~\ref{fig:ellipticity_dist}), and the suppression in enclosed energy is strong for $r_{\rm encl} < r_{\rm break}$, most walls only form PBHs if their Schwarzschild radius is larger than $r_{\rm break}$, which is parameterically much larger than $\delta_*$ for perfectly spherical walls. We discuss this quantitatively in the following section.

Finally, realistic walls have  $\epsilon_2 < 1$. The maximum radius the entire wall can be compressed into,  $r_{\rm break}$, does not significantly change because the semi-axis $R_{\rm RE} \epsilon_2$ is less than the other two semi-axes $R_{\rm RE}$ and $R_{\rm RE} \epsilon_1$. However, for $r_{\rm encl} \ll r_{\rm break}$, the asphericities along the semi-axis associated with $\epsilon_2$ can grow. Through simulations of collapsing walls with varying $\epsilon_1$ and $\epsilon_2$, we find that the reduction factor in the energy enclosed within a radius $r_{\rm encl}$ compared to a spherical wall of the same $R_{\rm RE}$ roughly factorizes in terms of $f_{\ellipse}(\epsilon)$ of Eq. ~\eqref{eq:asphericalReduction} for each axis,
\begin{align}
    \label{eq:eccentricityTotalRatio}
    \frac{f(\epsilon_1, \epsilon_2)}{f_\circ}
     &\approx
     f_{\ellipse}(\epsilon_1) \times f_{\ellipse}(\epsilon_2) 
     \, ,
\end{align}
though this is harder to numerically check for  $R_{\rm RE}/\delta_{\rm RE} \gg 10$ due to the computational limitations of simulating high resolution three-dimensional walls. Note for $r > r_{\rm break}$, \eqref{eq:eccentricityTotalRatio} is effectively equivalent to \eqref{eq:approxAsphericalSuppression}.

In summary, the maximum fractional energy enclosed within a radius $r_{\rm encl}$ of a collapsing wall in terms of its horizon re-entry radius $R_{\rm RE}$ and ellipsoidal parameters $\epsilon_1$, $\epsilon_2$ is given by
\begin{align}
     f \equiv \text{Max}\frac{E(t, r_{\rm encl})}{E(t, \infty)} &= f_\circ \times \frac{f(\epsilon_1,\epsilon_2)}{f_\circ},
     \label{eq:totalFractionalEnergy}
\end{align}
where $f_{\circ}$ is given by Eq.~\eqref{eq:fFractionEnclosed} and  $f(\epsilon_1, \epsilon_2)/f_{\circ}$ is given by Eq.~\eqref{eq:eccentricityTotalRatio}.
\section{Conditions for PBH Formation}
\label{sec:PBHConditions}

In this section, we determine the conditions for PBH formation of collapsing domain walls. First, we discuss the conditions for the wall to be compressed to within a Schwarzschild radius using the results of Sec.~\ref{sec:SubHorizonCollapse}. We find a minimum $R_{\rm RE}/\delta_{\rm RE}$ needed to form a PBH, which increases with decreasing $f_a$. This follows because walls with low $f_a$ must compensate by having large radii to achieve the same mass compared to a wall with larger $f_a$. Next, we carefully treat potential impediments to PBH formation, such as angular momentum which can cause some walls to form a centrifugal barrier or break apart prior to compressing to within a Schwarzschild radius, and collisions with other topological defects. We find that angular momentum is only an obstacle to PBH formation if the domain wall is very light, while collision with other defects has a mild suppressive effect across the whole wall mass range. In Appendix~\ref{app:self_gravitation}, we furthermore discuss why gravity does not impede PBH formation.

\subsection{Compression into a Schwarzschild Radius}
\label{sec:compressionToSchwarzchild}
A black hole forms when $r_{\rm encl}$ is smaller than the corresponding Schwarzschild radius, $R_s = 2 G E f$, where $E = E(t, \infty) = A_{\rm wall} \sigma$ is the total energy of the wall, $\sigma \simeq 8 m_a  f_a^2 = 8 \delta^{-1} f_a^2$ is the wall tension, $f$ is the fraction of the total energy within $r_{\rm encl}$ \eqref{eq:totalFractionalEnergy}, and $A_{\rm wall}$ is the wall surface area at horizon re-entry, given by $A_{\rm wall} = 4\pi R_{\rm RE}^2 \mathcal{I}_{\ellipse}(\epsilon_1, \epsilon_2)$, where $\mathcal{I}_{\ellipse}(\epsilon_1, \epsilon_2)/\epsilon_1 \epsilon_2 \equiv R_G(\epsilon_1^{-2}; \epsilon_2^{-2}; 1) \sim 1$ is the Carlson G elliptic integral.\footnote{The Carlson G elliptic integral is the integral giving the surface area of an ellipsoid with semi-major axes $R_{\rm RE} \epsilon_1$ and $R_{\rm RE} \epsilon_2$ relative to a sphere of radius $R_{\rm RE}$~\cite{carlsonGInt}.} 

It is helpful first to gain some intuition for the conditions imposed on the wall parameters $(R_{\rm RE}/\delta_{\rm RE}, \epsilon_1, \epsilon_2)$ to successfully collapse within a Schwarzchild radius for simpler walls or for naive assumptions about collapse before considering how realistic walls with asphericities modify these conditions for PBH formation. First, assume that the wall is perfectly spherical and can compress most of its energy to within a radius of order its rest-frame thickness, $\delta \equiv 1/m_a$; that is $f\sim 1$ for $r_{\rm encl} \sim \delta$. Setting $r_{\rm encl} = R_s$ gives the condition on the minimum size of the wall re-entry radius to its thickness
\begin{align}
    \frac{R_{\rm RE}}{\delta_{\rm RE}} = \left(\frac{M_{\rm Pl}}{f_a}\right) 
    \left(\frac{\delta}{\delta_{\rm RE}}\right)
    \left(\frac{1}{64\pi}\right)^{ \scalebox{1.01}{$\frac{1}{2}$}} 
    \qquad 
    \Big(\parbox{4.8cm}{\centering
    PBH Formation Condition: \\ Naive Assumption}\Big).
    \label{eq:PBHConditionNaive}
\end{align}
Although Eq.~\eqref{eq:PBHConditionNaive} is not correct for realistic walls, it captures some of the parameteric dependence of the wall on forming a PBH, namely that the larger $f_a$ is, the smaller the wall needs to be to collapse to within a Schwarzschild radius and form a PBH. 

\begin{figure}
    \centering
    \includegraphics[width=.75\textwidth]{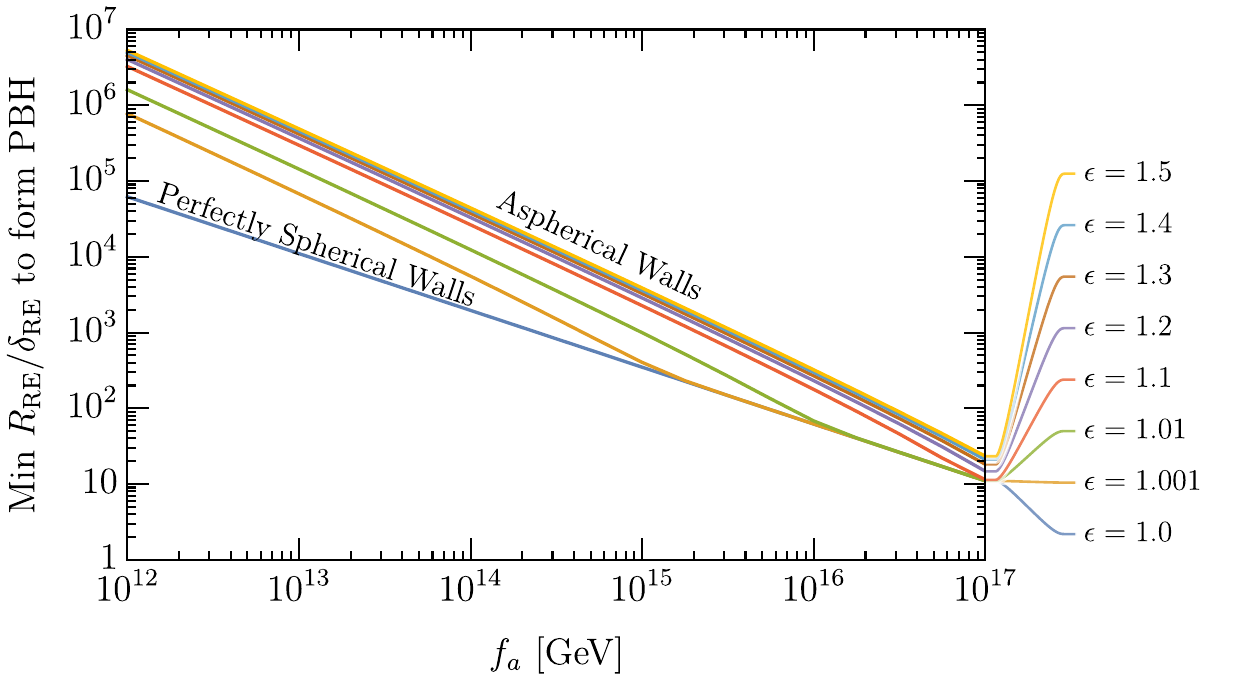}
    \caption{The minimum wall re-entry radius $R_\mathrm{RE}$ in units of the re-entry wall thickness $\delta_\mathrm{RE}$ needed to form a PBH, as a function of $f_a$. The different colored lines show the radius needed for different wall eccentricity parameters, where one axis is allowed to vary in length while the other two axes are fixed at radius $R_\mathrm{RE}$. Note that for aspherical walls, the minimal radius needed for PBH formation is substantially larger than in the spherical case (the blue line). This is due to the fact that it is substantially harder to compress energy into a small volume if the wall is aspherical. This difficulty can be overcome, however, by driving up the domain wall radius to the point where the Schwarzschild radius of the potential black hole formed is equal to the thickness of the wall.}
    \label{fig:minRREPlot}
\end{figure}

Next, drop the assumption that $f \sim 1$ for $r_{\rm encl} \sim \delta$. As discussed in Sec.~\ref{sec:SubHorizonCollapse}, because of the Lorentz contraction of the wall thickness, a perfectly spherical wall can be compressed into a radius smaller than its rest-frame thickness. Setting $r_{\rm encl} = R_s$, but with $f  = f_\circ$ of Eq.~\eqref{eq:fFractionEnclosed}, gives an improved condition on the minimum size of the wall re-entry radius to its thickness
\begin{align}
    \frac{R_{\rm RE}}{\delta_{\rm RE}} = \left(\frac{M_{\rm Pl}}{f_a}\right)^{ \scalebox{1.01}{$\frac{3}{4}$}}  \left(\frac{\delta}{\delta_{\rm RE}}\right)^{ \scalebox{1.01}{$\frac{3}{4}$}} \left(\frac{1}{64 \pi} \frac{1}{\mathcal{A}^{1/3}} \frac{1}{f(f^{-2/3} - 1)^{1/2}}\right)^{ \scalebox{1.01}{$\frac{3}{8}$}}
    \quad 
    \Big(\parbox{4.7cm}{\centering
    PBH Formation Condition: \\ Perfectly Spherical Walls}\Big).
    \label{eq:PBHConditionPerfectlySpherical}
\end{align}
Although Eq.~\eqref{eq:PBHConditionPerfectlySpherical} is similar to the naive expectation of Eq.~\eqref{eq:PBHConditionNaive}, there are a few important differences: First, the power in $M_{\rm Pl}/f_a$ is reduced from $1$ to $3/4$ which arises from the Lorentz contraction of the wall, making it easier to form a PBH by not needing a large $R_{\rm RE}/\delta_{\rm RE}$. Second, the energy fraction $f$ is a free parameter in Eq.~\eqref{eq:PBHConditionPerfectlySpherical} which can be chosen to minimize the right-hand-side of Eq.~\eqref{eq:PBHConditionPerfectlySpherical}. For $f \gtrsim 0.95$, the right-hand side of \eqref{eq:PBHConditionPerfectlySpherical} blows up because $r_{\rm encl}$ is so large that it is difficult to satisfy the black hole formation condition of $r_{\rm encl} = R_s$. Conversely, for $f \lesssim 0.05$, the right-hand side again blows up because while $r_{\rm encl}$ is small, the energy fraction within $r_{\rm encl}$ falls off so rapidly that $R_s$ decreases much faster than $r_{\rm encl}$, making it again difficult to satisfy $r_{\rm encl} = R_s$. For $f$ between these two values, the right-hand side is nearly constant and saturated near its minimum value. The blue contour labeled $\epsilon = 1.0$ in Fig.~\ref{fig:minRREPlot} shows the minimum $R_{\rm RE}/\delta_{\rm RE}$ needed for a PBH formation as a function of $f_a$ for a perfectly spherical wall (which occurs for an energy fraction $f \approx 0.54$).

Finally, drop the assumption that the wall is perfectly spherical. The condition for PBH formation is still given by $r_{\rm encl} = R_s$, but now with $f$ of Eq.~\eqref{eq:totalFractionalEnergy} for aspherical walls. Because of the complicated dependence of $f$ on $r_{\rm encl}$, it is not possible to write an exact analytic bound like Eqns. \eqref{eq:PBHConditionNaive} and \eqref{eq:PBHConditionPerfectlySpherical}, but the equation can be solved numerically for the minimum $R_{\rm RE}/\delta_{\rm RE}$ to form a PBH as shown by the $\epsilon \neq 1$ contours of Fig.~\ref{fig:minRREPlot}. The $\epsilon = 1.001$ (orange) and $\epsilon = 1.01$ (green) contours highlight that the walls must be spherical to a high degree at horizon re-entry for the perfectly spherical wall approximation, Eq.~\eqref{eq:PBHConditionPerfectlySpherical}, to hold. For $\epsilon \gtrsim 1$, aspheriticies hinder compression into $R_s$. In this regime, an approximate equation for the numerical solution of the minimum $R_{\rm RE}/\delta_{\rm RE}$ can be understood by taking $f \sim 1$ for $r_{\rm encl} \sim r_{\rm break}$ of Eq.~\eqref{eq:rbreak} due to the rapid fall-off in enclosed energy for radii smaller than $r_{\rm break}$. Solving the equation $r_{\rm encl} = R_s$ perturbatively gives the following minimum size of the wall re-entry radius to form a PBH
\begin{align}
    \frac{R_{\rm RE}}{\delta_{\rm RE}} \approx \left(\frac{M_{\rm Pl}}{f_a}\right) 
    \left(\frac{\delta}{\delta_{\rm RE}}\right) 
    \left(\frac{1}{64\pi \mathcal{I}_{\ellipse}(\epsilon_1,\epsilon_2)}\right)^{ \scalebox{1.01}{$\frac{1}{2}$}} \left(\frac{r_{\rm break}}{\delta} \right)^{ \scalebox{1.01}{$\frac{1}{2}$}}
    \qquad 
    \Big(\parbox{4.8cm}{\centering
    PBH Formation Condition: \\ Aspherical Walls}\Big).
    \label{eq:PBHConditionApprox}
\end{align}
Here, $r_{\rm break}/\delta$ is given in Eq.~\eqref{eq:rbreak} with $R_{\rm RE}/\delta_{\rm RE}$ in its natural logarithm argument evaluated at the logarithmic-independent piece of Eq.~\eqref{eq:PBHConditionApprox} (i.e. first order in perturbation theory).  
\subsection{Impediments to Collapse: Angular Momentum}
\label{sec:gravity_spin}
A simple requirement for PBH formation is that the angular momentum of the wall, $J_{\rm DW}$, must be less than the maximum angular momentum a black hole can possess (the so-called \textit{extremal limit})~\cite{Teukolsky:2014vca}
\begin{align}
    \label{eq:angularMomentumCondition}
    J_{\rm DW} < G M_{\rm DW}^2 \quad {\text{(Maximum Wall Angular Momentum)}} \, .
\end{align}
Enclosed axion domain walls may form with, or acquire, angular momentum that violate condition~\eqref{eq:angularMomentumCondition}. The intrinsic angular momentum of the wall can arise from the uncorrelated velocity of the axion field upon formation while the acquired angular momentum can arise in three main ways: (1) collisions with other defects, (2) fracturing into multiple walls during collapse, and (3) tidal gravitational forces from background density inhomogeneities. We now discuss each scenario in turn.

The angular momentum of an enclosed wall is given by
\begin{align}
\label{eq:ang_momentum_jdw}
    \mathbf{J}_{{\rm DW}} &= \oint \mathbf{R} \times d\mathbf{p} = 
    \oint  \mathbf{R} \times \mathbf{v} \,  \gamma \sigma  dA
    \\
    &= M_{\rm DW}  \bar{R} v_{\rm RMS}
    \left(\frac{\oint  \mathbf{R} \times d\mathbf{A} \gamma \mathrm{v}}{\bar{R} v_{\rm RMS} \int dA} \right) \, ,
\end{align}
where $\mathbf{R}$ is a vector pointing from the wall center of mass to a given point on the wall surface and $d\mathbf{p} = \mathbf{v} \gamma \sigma dA$ the momentum of a patch of wall of infinitesimal wall surface area $dA$, velocity $\mathbf{v}$, Lorentz factor $\gamma$, and tension $\sigma = M_{\rm DW}/\oint dA$. $\bar{R}$ and $v_{\rm RMS}$ are the mean wall radius and root-mean-square velocity of the wall surface, respectively.

In the second line, we use the fact that the wall velocity is always normal to the wall surface. Thus, $\mathrm{v}$ can be thought of as 1-dimensional random velocity field at a given $(\theta, \phi)$ coordinate on the wall surface; it may be positive  or negative depending on whether the motion is outward or inward, respectively, at a given point on the surface.

For an enclosed wall that is not significantly distorted, $\mathbf{R}$ and the surface normal vector are roughly aligned. Consequently,  the dimensionless factor in parenthesis -- which involves the perpendicular projection of these two vectors  -- is typically much smaller than unity. Moreover, it vanishes identically if $\gamma \mathrm{v}$ is constant everywhere on the wall surface since by Stokes' theorem, the integral can be transformed into an integral over the boundary of an enclosed surface, which is $0$. For any random velocity field $\mathrm{v}(\theta, \phi)$ on the wall surface -- that is, a periodic random field on the rectangle $0 \leq \theta < \pi$, $0 \leq \phi < 2\pi$ -- which is neither highly correlated nor highly uncorrelated, we find the magnitude of the term in parenthesis is of order $\sim 10^{-2}$.
\footnote{Closed string loops (including string-bounded walls) have a somewhat larger specific angular momentum~\cite{Scherrer:1990pj,Blanco-Pillado:2015ana}. The reason for this is likely geometric: only motion perpendicular to a defect is physical~\cite{vilenkin2000cosmic}. A segment of string can move anywhere in the plane transverse to its tangent vector, while a segment of wall can only only move parallel to its normal vector. The latter constraint naturally reduces the perpendicular projection of the lever arm momentum in Eq.~\eqref{eq:ang_momentum_jdw}. }
For velocity fields $\mathrm{v}$ that are either highly correlated or highly uncorrelated, which correspond to the limit where the velocity field is the same everywhere on the wall or totally random everywhere, this factor in parenthesis is smaller, as expected.

After horizon re-entry, the magnitude of the wall angular momentum compared to an extremal black hole is therefore
\begin{align}
    \label{eq:intrinsicJ}
    \frac{J_{\rm DW}}{ G M_{\rm DW}^2} \lesssim .01 v_{\rm RMS} \, \frac{R_{\rm RE}}{G M_{\rm DW}} = .02 v_{\rm RMS} \frac{R_{\rm RE}}{R_s} \, ,
\end{align}
where $R_{s}$ is the Schwarzschild radius of the PBH. 

Since the continual growth and remixing of correlated domains of the axion field in the universe -- which are the source of enclosed walls --- are generated by the motion of the infinite string-wall network, it is expected that $v_{\rm RMS}$ inherits its value from the network at the time $t_0$ when the enclosed wall forms. Numerical simulations and analytic expectations give a typical speed $v_{\infty} \sim 0.4-0.6$ of the string-wall network~\cite{Avelino:2005kn,Martins:2000cs}. By the time the wall re-enters the horizon, this speed may decrease due to the redshifting of momentum with the Hubble expansion~\cite{vilenkin2000cosmic}. Consequently, Eq.~\eqref{eq:intrinsicJ} indicates that the intrinsic angular momentum of an enclosed wall likely cannot stop a PBH from forming if the wall needs to compress by factor of $\sim$ a few hundred to fall within its Schwarzschild radius. From the minimum $R_{\rm RE}/\delta$ needed to form a PBH and the scale $r_{\rm break}/\delta \approx R_s/\delta$ into which an $\mathcal{O}(1)$ fraction of the wall energy can be compressed (Eq.~\eqref{eq:rbreak}), it can be seen that $R_{\rm RE}/R_s \lesssim 10^2$ for $f_a \gtrsim 10^{15}$ GeV, which designates the region where Eq.~\eqref{eq:angularMomentumCondition} is satisfied.

Enclosed walls that collide with the wall-bounded strings that surround them can acquire further angular momentum. Since the relative speeds between the two colliding defects is comparable to $v_{\infty}$, this additional angular momentum is similar to Eq.~\eqref{eq:intrinsicJ} when the two colliding defects have comparable masses. If the enclosed wall is more massive compared to the string-bounded wall it collides with, which is typically the case, then the angular momentum transfer can be suppressed further by the ratio of their masses. If the wall is highly distorted and collides with itself during collapse, the wall can break apart into multiple enclosed walls. But by the same logic as before, each piece will have some angular momentum comparable to Eq.~\eqref{eq:intrinsicJ} . Moreover, the shape of the enclosed wall must be highly non-spherical for this fracturing to occur, which will not be the case unless the initial wall size is far greater than the horizon at formation, at which point, the surface is more Brownian-like. However, the walls that eventually become the majority of PBHs form with  sizes only mildly greater than the horizon (i.e. $\alpha \equiv R_0/t_0 \sim 1-2$). Walls with $\alpha \gg 1$ are anyway far too suppressed in abundance.

Finally, if the enclosed wall is treated as a rigid object, then the wall can acquire angular momentum from gravitational tidal forces if it is aspherical and possesses a quadrupole moment~\cite{Eroshenko_2021}. The angular momentum acquired by a rigid elliptical wall of mass $M_{\rm DW} = M_{\rm BH}\sim 4 \pi \sigma t_{\rm RE}^2$ from such tidal forces corresponds to a spin parameter~\cite{Eroshenko_2021}
\begin{align}
    \label{eq:JTidal}
    \frac{J_{\rm DW}}{G M_{\rm DW}^2} \sim \left(\frac{M_{\rm DW}}{10^{-5} M_{\odot}} \right)^{-1}\left( \frac{t_{\rm RE}}{t_{\rm QCD}} \right) \left( \frac{\Sigma}{5 \times 10^{-5}} \right) \, ,
\end{align}
where $\Sigma$ is the variance in the primordial background density perturbations and $t_{\rm QCD}$ is the time of the QCD phase transition which is the typical timescale at which the walls re-enter the horizon and collapse.   
Eq.~\eqref{eq:JTidal} indicates that QCD axion domain walls with mass $M_{\rm DW} \gtrsim 10^{-5} M_{\odot}$ and moderate eccentricities are sub-extremal and able to form PBHs if Eq.~\eqref{eq:angularMomentumCondition} is satisfied. This is typically the case for any $f_a \gtrsim 10^{14}$ GeV as discussed in Sec. \ref{sec:f_pbh}.
\subsection{Impediments to Collapse: Collisions with Neighboring Defects}
Collisions with the surrounding string-bounded walls can also damage or destroy the enclosed walls via axion emission or wall puncturing which, if the resultant holes are larger than $R_{\mu/\sigma} \approx \mu/\sigma$, will expand and `eat' the wall~\cite{everett1982left,PhysRevLett.48.1156,barr1987some,1986LIACo..26..173S}. How often does this occur?

Consider an enclosed wall of initial mean physical radius $R_0 = \alpha t_0$ during superhorizon expansion. In the comoving frame, the wall radius is stationary at fixed coordinate radius $r = R_0$ and the comoving flux of surrounding defects impinging on the enclosed wall is
\begin{align}
    \Phi = \int n v \cos \theta f(v) d^3 v = \frac{1}{4} n \bar{v}  \, ,
\end{align}
where $\bar{v} = \int v f(v) d^3 v \approx v_{\infty} \times a(t_0)/a(t)$ is the mean \textit{coordinate} velocity of the enclosed wall relative to the surrounding defects 
\footnote{This follows because the coordinate velocity is $d \mathbf{x}/dt = v_{\rm pec} a(t_0)/a(t)$ where $v_{\rm pec}$ is the peculiar velocity of the surrounding defects which is the speed of the infinite string-wall network $v_{\infty} \sim 0.5$.} 
and $n =  t^{-3} \times (a(t)/a(t_0))^3$ is the comoving number density of the other defects in the medium. The rate of collisions is then
\begin{align}
    \Gamma_{\rm coll} \approx 4\pi r^2 \Phi \approx \pi \alpha^2 t_0^2 \frac{1}{t^3}\left(\frac{t}{t_0} \right)^{3k} v_{\infty} \left(\frac{t_0}{t}\right)^{k} \, ,
    \label{eq:rateOfCollision}
\end{align}
where $k$ is again the exponent in the relation $a(t) \propto t^k$ and is equal to $1/2 \, (2/3)$ in a RD (MD) era. It follows from \eqref{eq:rateOfCollision} that the probability of an enclosed wall \textit{not} colliding with another defect after a time $t$ is
\begin{align}
    \label{eq:probNoCollision}
    P(\text{Not colliding after time $t$}) \equiv P_{\cancel{\rm coll}} &= \exp\left(-\int_{t_0}^t \Gamma_{\rm{coll}}(t') dt' \right) \nonumber
    \\
    &= \exp \left[- \frac{\pi \alpha^2 v_{\infty}}{2 - 2k} \left( 1 - \left(\frac{t_0}{t}\right)^{2 - 2 k}\right) \right] \, .
\end{align}
Eq.~\eqref{eq:probNoCollision} demonstrates that the majority of large walls collide with other defects in the network, leading to their destruction.
The factor of $\alpha^2$ in Eq.~\eqref{eq:probNoCollision} can be understood as a geometric factor in the sense that the larger the total surface area of the wall, the greater its cross-sectional area. The asymptotic behavior of $P_{\cancel{\rm coll}}$ to the constant $\exp[-\pi \alpha^2 v_{\infty}/(2-2k)]$ can be understood by the increase in physical separation of defects with the expansion of the universe, which reduces the rate at which enclosed walls collide with other defects.

Note that Eq.~\eqref{eq:probNoCollision} is valid while the wall is superhorizon and while the infinite string-wall network remains in scaling. After $t_{\mu/\sigma}$, the infinite string-wall network collapses relativistically so that the network decays within roughly a Hubble time $t_{\mu/\sigma}$~\cite{Ryden:1989vj}. Thus Eq.~\eqref{eq:probNoCollision} should be evaluated at the minimum of $t \approx t_{\rm RE}$ and $t \approx 2 t_{\mu/\sigma}$. For example, for walls that form at time near network collapse $ t_{\mu/\sigma}$, $P_{\cancel{\rm coll}}(\alpha) \approx \exp[-1.6  \alpha^2 v_{\infty}]$ in a RD or MD era.
\section{PBH Abundance and Detection Prospects}
\label{sec:f_pbh}
In this section, we combine the results for the number density of enclosed walls  (Sec.~\ref{sec:closed_dw_abundance}) with their superhorizon growth (Sec.~\ref{sec:SupHorizonExpansion}), subhorizon collapse (Sec.~\ref{sec:SubHorizonCollapse}), and conditions for PBH formation (Sec.~\ref{sec:PBHConditions}) to determine the relic abundance of PBHs $f_\mathrm{PBH}$ as a function of $f_a$, as well as the mass associated with the PBHs produced. We compare the computed abundances with the reach of current and future gravitational lensing surveys. We focus on the early MD cosmological scenario, which yields an especially favorable $f_\mathrm{PBH}$, but share analogous results for axion-like particles in RD cosmologies in Appendix~\ref{sec:ALPS}.

The total energy density in PBHs at the moment of production $\rho_\mathrm{PBH}$ is parameterized by the integral
\begin{equation}
    \rho_\mathrm{PBH}=\int d\alpha\;d\epsilon_1\;d\epsilon_2 \;M_\mathrm{PBH}(\alpha,\epsilon_1,\epsilon_2) \frac{dn_\mathrm{PBH}(\alpha,\epsilon_1,\epsilon_2)}{d\alpha}f_{\rm max/mean}(\epsilon_1)f_{\rm min/mean}(\epsilon_2),
\end{equation}
where $f_{\rm max/mean}(\epsilon_1)$ and $f_{\rm min/mean}(\epsilon_2)$ are the normalized eccentricity probability distribution functions given by Eq.~\eqref{eq:maxtomeanAxis} and Eq.~\eqref{eq:mintomeanAxis}, respectively, $M_\mathrm{PBH}(\alpha,\epsilon_1,\epsilon_2)$ is the mass of a PBH formed from a closed domain wall of radius $\alpha=R_0/t_0$ and eccentricity parameters $\epsilon_1$ and $\epsilon_2$. $dn_\mathrm{PBH}/d\alpha$ is the number density distribution of PBHs, which we construct as
\begin{equation}
    \frac{dn_\mathrm{PBH}(\alpha)}{d\alpha}\equiv \frac{dn_\mathrm{cDW}(\alpha)}{d\alpha} \times P_{\cancel{\rm{coll}}}(\alpha)\times \Theta\left[\alpha-\alpha_\mathrm{min}\right],
\end{equation}
where $dn_\mathrm{PBH}(\alpha)/d\alpha$ is the self-enclosed domain wall number density given by Eq.~\eqref{eq:continuumf(R)}, $P_{\cancel{\rm{coll}}}$ is the probability a self-enclosed wall does not collide with a neighboring defect, given by Eq.~\eqref{eq:probNoCollision}. $\Theta$ is a Heaviside function with $\alpha_\mathrm{min}$ being the solution to Eq.~\eqref{eq:PBHConditionApprox} taking into account superhorizon expansion of Eq.~\eqref{eq:reentry_radii}. For ease of computation, we also decompose $M_\mathrm{PBH}$ in the following way
\begin{align}
    M_{\rm PBH}(\alpha, \epsilon_1, \epsilon_2)  &\simeq 4\pi R_{\rm RE}^2(\alpha) \mathcal{I}_{\ellipse}(\epsilon_1,\epsilon_2) \sigma f(\alpha,\epsilon_1,\epsilon_2)
    \\
                &\simeq 32 \pi R_\mathrm{RE}^2(\alpha) \mathcal{I}_{\ellipse}(\epsilon_1,\epsilon_2) f_a^2 m_a(t_c) f(\alpha,\epsilon_1,\epsilon_2).
    \label{eq:MPBH}
\end{align}
Here, $f(\alpha,\epsilon_1,\epsilon_2) \sim 1$  is the fraction of the total wall energy that compresses to within a Schwarzschild radius as given in Eq.~\eqref{eq:totalFractionalEnergy}, $R_\mathrm{RE}$ is the wall re-entry radius as given in Eq.~\eqref{eq:reentry_radii}, $t_c$ is the collapse time given by Eq.~\eqref{eq:reentry_time_reln}. We take the latest possible $t_0 \simeq t_{\mu/\sigma}$ which is the moment that the string-wall network scaling relation ends and the entire string-wall network begins to contract, given by Eq.~\eqref{eq:tmusigma}. $\mathcal{I}_{\ellipse}$ is the surface area of an ellipsoidal domain wall with semi-major and semi-minor axis lengths $R_\mathrm{RE}\epsilon_1$ and $R_\mathrm{RE}\epsilon_2$, respectively, normalized to the corresponding spherical surface area. We evaluate this integral numerically.

With $\rho_\mathrm{PBH}$ calculated, the final step is to compute the total fraction of DM consisting of PBHs $f_\mathrm{PBH}$. In QCD axion models, the relic axion particle energy density $\Omega_\mathrm{a}$ (not including PBHs) has two components: the misalignment energy density $\Omega_\mathrm{mis}$ and the relic density of axions produced from defect radiation (primarily strings) $\Omega_\mathrm{strings}$. At the moment of misalignment, the energy density in such axions $\rho_\mathrm{mis}$ is given by
\begin{equation}
    \rho_\mathrm{mis}(t_{\rm osc})\simeq \frac{1}{2}f_a^2\langle\theta^2_\mathrm{mis}\rangle m_a(T_\mathrm{osc})^2,
\end{equation}
where $\langle\theta^2_\mathrm{mis}\rangle \simeq\pi^2/3$ is the mean square misalignment angle and $T_\mathrm{osc}$ is the temperature at which the the axion field starts to oscillate~\cite{Sikivie_2008}. The moment of oscillation $t_\mathrm{osc}$, corresponding to $T_\mathrm{osc}$, is found by solving $3H(T_\mathrm{osc})=m_a(T_\mathrm{osc})$. We detail how to compute the oscillation time in the early matter-dominated scenario scenario in Appendix~\ref{app:reheat_temp_in_md}.

We now compute $\Omega_\mathrm{PBH}/\Omega_\mathrm{mis}$ as
\begin{equation}
    \frac{\Omega_\mathrm{PBH}}{\Omega_\mathrm{mis}}=\frac{\rho_\mathrm{PBH}(t_c)}{\rho_\mathrm{mis}(t_\mathrm{osc})}\left[\frac{a(t_c)}{a(t_\mathrm{osc})}\right]^3 \frac{m_a(T_{\rm osc})}{m_a(T=0)},
\end{equation}
where the $[a(t_\mathrm{c})/a(t_\mathrm{osc})]^3$ factor accounts for dilution of misalignment DM up until the moment of PBH formation and $m_a(T_{\rm osc})/{m_a(T=0)}$ accounts for the increase in the axion mass of misalignment DM from $t_{\rm osc}$ until today. In the matter-dominated scenario, $[a(t_\mathrm{c})/a(t_\mathrm{osc})]^3=(t_\mathrm{c}/t_\mathrm{osc})^2$ while in the radiation dominated scenario, $[a(t_\mathrm{c})/a(t_\mathrm{osc})]^3=(t_\mathrm{c}/t_\mathrm{osc})^{3/2}$. After $t_{\mu/\sigma}$, where the whole string-wall network collapses, the relic axion DM particles and the PBHs both scale like matter for the rest of cosmic history, hence the fraction $\Omega_\mathrm{PBH}/\Omega_\mathrm{mis}$ stays fixed until the present day.

\begin{figure}
    \centering
    \includegraphics[width=0.8\textwidth,keepaspectratio]{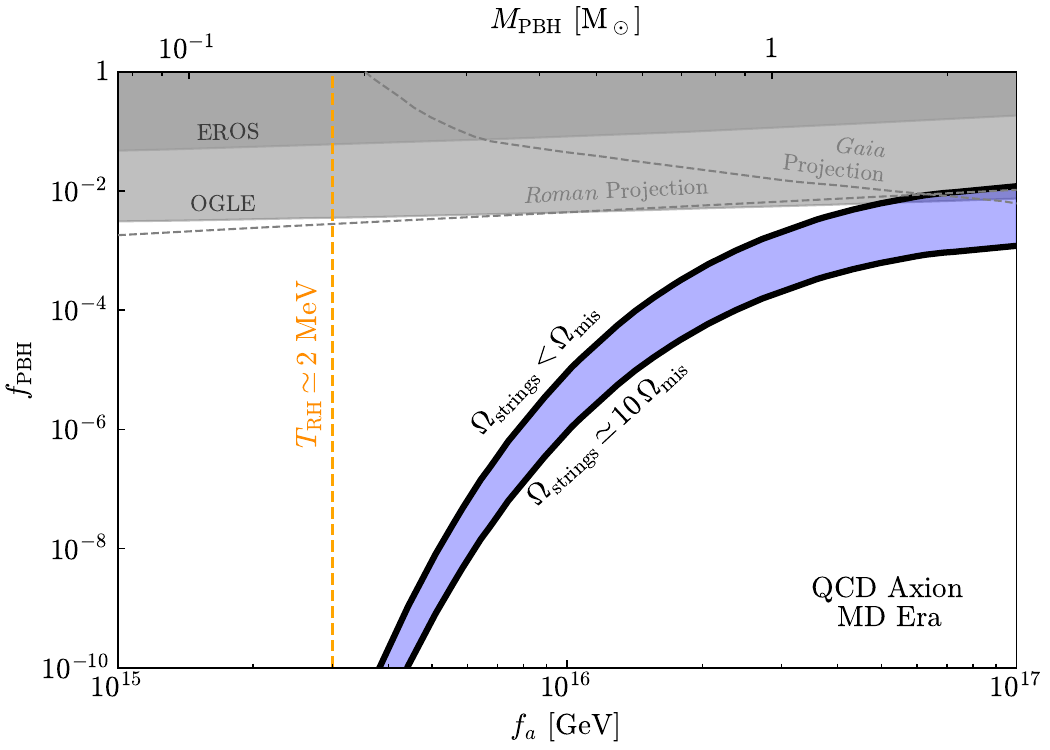}
\caption{The projected DM PBH fraction in an early matter-dominated cosmology. The bottom axis shows the QCD axion model $f_a$ corresponding to a given $f_\mathrm{PBH}$ abundance. The top axis shows the PBH mass $M_\mathrm{PBH}$ associated with each $f_a$. Note that the PBH spectrum for a given $f_a$ is nearly monochromatic, hence the one-to-one map from $f_a$ to $M_\mathrm{PBH}$. The top and bottom solid black lines show the $f_\mathrm{PBH}$ abundance from axion domain wall collapse in the case where the relic radiated string axion abundance $\Omega_\mathrm{strings}$ is $<\Omega_\mathrm{mis}$ and approximately $10\,\Omega_\mathrm{mis}$, respectively. $\Omega_\mathrm{strings}$ is entirely predicted by $f_a$ and the subject of extensive study, however its exact value remains uncertain due to simulation limitations. The two black lines correspond to edge cases for $\Omega_\mathrm{strings}$, meaning $f_\mathrm{PBH}$ may lie anywhere in the blue band. The gray solid regions show the PBH parameter space currently excluded by the EROS microlensing survey~\cite{Tisserand_2007} and by the recent OGLE-IV survey~\cite{Mroz:2024mse}. The gray dashed lines show projections for the PBH parameter space that will be reachable by \textit{Gaia} Data Release 4~\cite{Chen_2023} and by \textit{Roman}'s Galactic Bulge Time Domain Survey~\cite{derocco2023rogue,derocco2023new}. The dashed orange line shows where the reheat temperature $T_\mathrm{RH}$ equals the BBN temperature $T_\mathrm{BBN} \simeq 2$ MeV~\cite{Hasegawa:2019jsa} in the most minimal early matter-dominated cosmology when $\Omega_{\rm strings} \lesssim \Omega_{\rm mis}$~\cite{Nelson_2018}. Regions to the right of this line may overproduce axion DM so that consistent early matter-dominated cosmologies for the QCD axion require $f_{\rm PBH}$ to be below $\sim 10^{-10}$, although this constraint is easily avoided in ALP models. Projections for ALP models are shown in Figs.~\ref{fig:money_plot_alps},~\ref{fig:md_alps}.}
\label{fig:money_plot}
\end{figure}

Finally, $f_\mathrm{PBH}$, the fraction of present-day DM consisting of PBHs, is defined as
\begin{equation}
    f_\mathrm{PBH}=\frac{\Omega_\mathrm{PBH}}{\Omega_\mathrm{mis} + \Omega_\mathrm{strings} + \Omega_\mathrm{PBH}}\simeq \frac{\Omega_\mathrm{PBH}}{\Omega_\mathrm{mis} + \Omega_\mathrm{strings}},
\end{equation}
where $\Omega_\mathrm{strings}$ can be obtained from simulations. Despite being extensively studied, the number density of axions from string radiation is uncertain. As discussed in Sec.~\ref{sec:wallFormationTimes}, log corrections to the string density parameter $\xi$ could be slightly larger than $1.0$ by $T_\mathrm{QCD}$, but these log corrections are generally thought to be suppressed in an early MD cosmology~\cite{Visinelli_2010} due to the greater expansion rate of the universe compared to a RD cosmology. Fig.~\ref{fig:money_plot} shows $f_\mathrm{PBH}$ as a function of $f_a$  where the blue band indicates the uncertainty from string radiation. The gray solid regions  show current constraints from EROS~\cite{Tisserand_2007} and OGLE~\cite{Niikura_2019}, while the dashed gray regions show constraints projected to be achieved with the release of \textit{Gaia} Data Release 4~\cite{Chen_2023} and constraints projected to be achieved with the \textit{Roman Space Telescope}'s Galactic Bulge Time Domain Survey~\cite{derocco2023rogue,derocco2023new}. Note that these constraints from \textit{Roman} are somewhat optimistic and may fall up to an order of magnitude lower, depending on systematics~\cite{derocco2023rogue}.

Fig.~\ref{fig:money_plot} indicates that $f_\mathrm{PBH}$ plateaus at very high valaues of $f_a$ because in this part of the parameter space, even small, aspherical walls are energetic enough to successfully form PBHs upon collapsing. As $f_a$ is lowered below $f_a \lesssim$ a few times $10^{16}$ GeV, the walls become progressively less energetic, requiring their initial radius $R_0$ to be increasingly larger. However, the number density of enclosed walls with a very large initial radius is exponentially suppressed relative to small walls per Eq.~\eqref{eq:continuumf(R)}. This means that for $f_a\lesssim$ a few times $10^{15}$ GeV, $f_\mathrm{PBH}$ drops well below a range observable by gravitational lensing surveys in the foreseeable future. However, the projections suggest that near-future gravitational lensing surveys will be able to probe the very high $f_a$ region of the relevant $f_\mathrm{PBH}$ parameter space.

In the most minimal early MD cosmology, where a single particle species dominates the energy density of the universe at early times and eventually decays into SM radiation at $T_\mathrm{RH}$, obtaining the correct axion DM abundance at large $f_a$ requires low $T_\mathrm{RH}$. The region to the right of the orange vertical line of Fig.~\ref{fig:money_plot} indicates roughly where $T_{\rm RH}$ falls below the Big Bang Nucleosynthesis (BBN) temperature, $T_\mathrm{BBN}\simeq 2$ MeV \cite{Hasegawa:2019jsa} for $\Omega_{\rm strings} \lesssim \Omega_{\rm mis}$. For $T_{\rm RH} \lesssim T_\mathrm{BBN}$, which roughly corresponds to $f_a \gtrsim 3 \times 10^{15}$ GeV \cite{Nelson_2018}, the predicted abundance of light elements begins to observably differ from their measured values \cite{Hasegawa:2019jsa}. Moreover, for $f_a \gtrsim 10^{16}$ GeV, $T_{\rm PQ}$ also risks growing larger than the upper bound on the inflationary reheating temperature~\cite{Planck:2018jri}.
Hence, for the QCD axion, a consistent MD cosmology may limit $f_{\rm PBH}$ to lie on the exponentially sensitive tail of the distribution of large walls such that $f_{\rm PBH} \lesssim 10^{-10}$ -- out of reach of near-future lensing surveys. ALPs, which can form domain walls at times earlier than the QCD axion, can avoid this low $T_{\rm RH}$ bound and can thus have consistent cosmologies for $f_a > 3 \times 10^{15}$ GeV so that $f_{\rm PBH}$ is large and observable by next generation lensing surveys. We discuss this further in Appendix \ref{sec:ALPS}. 
\section{Conclusions}
\label{sec:conclusions}

Self-enclosed domain walls in axion models provide an exciting new avenue for PBH production, and these PBHs may potentially be observable by future gravitational lensing surveys. We study the number density and PBH formation efficiency of these closed domain walls in both radiation dominated and early matter-dominated cosmologies: the latter being of particular interest since it allows high values of the axion decay constant $f_a$.

First, we study the number density of self-enclosed domain walls near the moment the string-wall network collapses $t_{\mu/\sigma}$ using a simple geometric argument: namely, that the axion field must be completely uncorrelated on length scales larger than the string separation length $L$ and highly correlated on length scales shorter than $L$. Applying this argument to a discrete lattice representation of the early universe yields the distribution of self-enclosed domain walls of various sizes. Expanding the argument to a continuous random field furthermore yields the distribution of shapes of self-enclosed domain walls, which we take to be approximately spheroidal -- this is a valid approximation for studying PBH formation from domain wall collapse since the walls are smooth on scales $\sim L$, which is the scale at which the domain wall number density is peaked.

We then study the dynamics of these domain walls upon formation. At early times, the walls may stretch with the expansion of the universe until they enter the cosmic horizon. Numerically solving the low energy axion equations of motion in an FLRW background yields the radius of domain walls at the moment of horizon re-entry $R_\mathrm{RE}$ and the amount of (conformal) time it takes for walls to re-enter the horizon $\eta_\mathrm{RE}$ and to fully collapse $\eta_c$. Depending on the axion decay constant $f_a$, the wall may also become thinner and more energetic due to a higher wall tension as the axion mass continues to `turn on' during the dynamical stage of domain wall evolution.

Next, we simulate the subsequent stage of wall evolution: the inwards collapse, where the expansion of the universe may safely be neglected. We numerically solve the low energy axion equation of motion in a Minkowski spacetime and compute how much energy can be compressed into a small volume about the wall center of mass. The numerical solutions reveal that for a spherical domain wall, the fraction of total wall energy that can be compressed into a sphere of radius $r$ around the wall's center of mass is roughly unity for $r\gtrsim \delta_*$, but drops off $\propto r^3$ for $r\lesssim \delta_*$ where $\delta_*$ is the Lorentz contracted domain wall thickness. Importantly, the numerical solutions also show that the energy compressibility of aspherical walls is further power law suppressed relative to the spherical case. Realistic walls with asphericities can compress most of their energy into a radius $r \approx r_{\rm break}$ which is parametrically larger than the rest-frame wall thickness $\delta = 1/m_a$. 

We use these results to compute the initial domain wall size and shape necessary to form a PBH. For the $f_a$ parameter space of interest, there is a critical minimal initial domain wall radius $R_0$ necessary for successful PBH formation. This radius coincides with the domain wall being large enough that its Schwarzschild radius is somewhat greater than the wall thickness $R_{s} \approx r_{\rm break}$. We also account for angular momentum acquisition by the walls due to interactions with neighboring defects and gravitational attraction between domain walls and surrounding density perturbations. However, careful study reveals that these effects are marginal for the wall and PBH masses of interest.

Finally, we use all of these considerations to calculate the fraction of present-day DM that is composed of PBHs produced via axion domain walls collapse, $f_\mathrm{PBH}$, in an early matter-dominated cosmology, shown in Fig.~\ref{fig:money_plot}. We find that the present-day PBH abundance is sufficiently high in early matter-dominated scenarios with $f_a\gtrsim 10^{15}$ GeV to potentially be observable in the future. For $f_a \gtrsim 10^{16}$ GeV, the PBH abundance is large enough that current or near-future astrometric and photometric gravitational lensing surveys like \textit{Gaia} Data Release 4 and OGLE could potentially observe a PBH signal. However, for this $f_a$ range, the early matter-dominated reheat temperature $T_\mathrm{RH}$ comes into tension with the Big Bang Nucleosynthesis temperature $T_\mathrm{BBN}$ in the simplest cosmological scenario. For smaller values of $f_a$, this tension can be completely avoided, however the corresponding $f_\mathrm{PBH}$ is significantly smaller and will likely not be observable in the near future. Another way to avoid the potential tension with BBN, while still maintaining a near-future observable $f_\mathrm{PBH}$, is to consider axion-like particles, which we do in Appendix~\ref{sec:ALPS}.

In all of these scenarios, we have assumed that the string density parameter $\xi$ is roughly unity, however this could turn out to be inaccurate -- especially in the radiation dominated scenario discussed in Appendix~\ref{sec:ALPS} (see Sec.~\ref{sec:closed_dw_abundance} for a discussion of $\xi$). Should that be the case, with $\xi$ growing larger than roughly unity at the QCD phase transition, $\Omega_\mathrm{PBH}$ would be subject to a further exponential suppression due to the self-enclosed wall number density scaling in Eqns.~\eqref{eq:continuumf(V)}-\eqref{eq:continuumf(R)}. This would cause there to be fewer $\sim$horizon-sized walls, meaning exponentially fewer PBHs in the parameter space that relies on superhorizon growth to increase the domain wall mass and form PBHs. As can be seen from Fig.~\ref{fig:money_plot}, this is dominantly the region where $f_a \lesssim 10^{16}$ GeV. 

As a final note, we emphasize that the formalism developed here can be applied to other models that exhibit unstable domain walls approximately described by a Sine-Gordon equation of motion. We therefore hope to extend this work to other classes of models not involving the QCD axion in the future.

\section{Acknowledgements}
 We thank Robert Brandenberger, Malte Buschmann, Cyril Creque-Sarbinowski, Chris Dessert, Joshua W. Foster, Anson Hook, Junwu Huang, John March-Russell, Javier Redondo, Ken Van Tilburg, Giovanni Villadoro, and Neal Weiner for helpful discussions, Huangyu Xiao for clarification on low $T_\mathrm{RH}$ limits in axion matter-dominated cosmologies, and Robert Brandenberger, Cyril Creque-Sarbinowski, Chris Dessert, and Ken Van Tilburg for helpful comments on the manuscript. This work was supported in part through the NYU IT High Performance Computing resources, services, and staff expertise. DD is supported by the James Arthur Postdoctoral Fellowship and is grateful to the Galileo Galilei Institute and the CERN Theory Division where part of this work was performed. MK was funded by New York University's Carl P. Feinberg graduate fellowship in theoretical physics when this work was conducted. MK is also grateful for the hospitality of Perimeter Institute. Research at Perimeter Institute is supported in part by the Government of Canada through the Department of Innovation, Science and Economic Development Canada and by the Province of Ontario through the Ministry of Colleges and Universities. Both authors also thank the Center Computational Astrophysics at the Flatiron Institute for hosting us while we carried out part of this work. The Center for Computational Astrophysics at the Flatiron Institute is supported by the Simons Foundation. This research was supported in part by the Heising-Simons Foundation, the Simons Foundation, grant no. PHY-2210551 and grant no. PHY-2309135 to the Kavli Institute for Theoretical Physics (KITP). We made use of the software packages NumPy~\cite{numpy}, SciPy~\cite{Virtanen_2020}, and py-pde~\cite{py-pde}.

\appendix
\section{Reheat Temperature from an Early Matter-Dominated Era}
\label{app:reheat_temp_in_md}
In this section, we derive the necessary reheat temperature at the end of the early MD era, $T_{\rm RH}$, to give the correct axion DM abundance from the misalignment mechanism, $\Omega_{{\rm mis}, \rm MD}$, and from axion strings,  $\Omega_{\rm string, MD}$. This value of $T_{\rm RH}$ is self-consistently used to derive the temperature, $T_{\rm osc, \rm MD}$, when the axion begins oscillating in an early MD era. This is because the oscillation temperature $T_{\rm osc, MD}$ in a MD era depends on $T_{\rm RH}$ which enters into Hubble. Overall, the value of $T_{\rm osc, \rm MD}/{T_{\rm QCD}}$ is important since it controls whether the axion mass is still growing when walls form, which in turn affects the condition for PBH formation (see Sec.~\ref{sec:PBHConditions}). 

The wall formation temperature relative to $T_{\rm QCD}$ is
\begin{align}
    \label{eq:TOsc}
    \frac{T_{\rm osc}}{T_{\rm QCD}} &\simeq 
        \begin{dcases}
        \left(\frac{M_{\rm Pl}}{f_a} \right)^\frac{2}{4+n} \left(\frac{10}{8\pi^3 g_*(T_0)} \right)^\frac{1}{4+n} \quad \hfill &\text{RD era},
        \\
        \left(\frac{M_{\rm Pl}}{f_a} \right)^\frac{2}{8+n} 
         \left(\frac{T_{\rm RH}}{T_{\rm QCD}} \right)^\frac{4}{8+n} 
        \left(\frac{10}{8\pi^3 g_*(T_0)}\frac{g_*(T_{\rm RH})}{g_*(T_0)} \right)^\frac{1}{8+n} \quad \hfill &\text{MD (Non-Adiabatic) Era}.
        \end{dcases}
\end{align}
Here, $M_{\rm Pl} \simeq 1.22 \times 10^{19}$ GeV is the Planck constant and $g_*(T)$ are the relativistic degrees of freedom in the thermal bath at temperature $T$.  In a RD era, $T_{\rm osc}/T_{\rm QCD}$ depends dominantly on $f_a$, and to a much weaker extent, $g_*$. In a MD era, $T_{\rm osc}$ also depends on the temperature at which the universe returns to being radiation dominated, $T_{\rm RH}$. The lower $T_{\rm RH}$ is, the more the background axion abundance is diluted~\cite{Ramberg:2019dgi,Visinelli_2010,Nelson_2018}. 

The axion abundance from both misalignment and strings in a MD era is
\begin{align}
    \label{eq:OmegaMD}
    \Omega_{a, \rm MD}(f_a) = \underbrace{\Omega_{{\rm mis}, \rm RD}(f_a)\frac{Y_a(T_{\rm osc, MD})}{Y_a(T_{\rm osc, RD})}\frac{1}{D}}_{\Omega_{{\rm mis}, \rm MD}}\left(1+ \frac{\Omega_{\rm string, MD}}{\Omega_{{\rm mis}, \rm MD}}\right) \, .
\end{align}
The first term in Eq.~\eqref{eq:OmegaMD}, $\Omega_{{\rm mis}, \rm RD}$, is the energy density of axions, $\rho_a$, relative to the critical density of the universe, $\rho_c$, if the axions were generated solely by misalignment in the \textit{standard RD cosmology}
\begin{align}
    \Omega_{{\rm mis}, \rm RD} &= \frac{\rho_a(t_{\rm today})}{\rho_c(t_{\rm today})} \simeq \frac{\frac{1}{2} \langle \theta^2_\mathrm{mis} \rangle f_a^2 m_a(t_{\rm osc, RD}) m_a \frac{a(t_{\rm osc, RD})^3}{a(t_{\rm today})^3}}{\rho_c(t_{\rm today})} 
    \\
    &\simeq \Omega_{\rm DM} \left(\frac{f_a}{f_{a,\rm DM}} \right)^\frac{6+n}{4+n} \, .
\end{align}
Here,  $f_{a, \rm DM}$ is the decay constant that gives the correct DM abundance, $\Omega_{\rm DM}$, in a RD era from misalignment. For the QCD axion, $n \simeq 7$ so that~\cite{Adams:2022pbo} 
\begin{align}
\Omega_{{\rm mis}, \rm RD} \approx \Omega_{\rm DM} \left(\frac{f_a}{3 \times 10^{11} \, \rm GeV} \right)^{1.18}
\qquad \text{(QCD axion)} \, .
\end{align}
Note that in the post-inflation scenario considered in this paper, the misalignment angle is not a free parameter, but is instead the horizon averaged value of $\langle \theta^2 \rangle$. 
For ALPs, $f_{a, \rm DM}$ can be much larger, near $10^{16}$ GeV, since ALP mass can be much lighter than the QCD axion for a given $f_a$~\cite{Galanti:2022ijh}. 

The second term in Eq.~\eqref{eq:OmegaMD}, $Y_a(T_{\rm osc, MD})/Y_a(T_{\rm osc, RD})$, is the ratio of the redshift invariant axion yields when the axion begins oscillating at $T_{\rm osc, MD}$ ($T_{\rm osc, RD}$), in a MD (RD) era 
\begin{align}
    \frac{Y_a(T_{\rm osc, MD})}{Y_a(T_{\rm osc, RD})} &= \frac{n_a(T_{\rm osc, \rm MD})/s(T_{\rm osc, \rm MD})}{n_a(T_{\rm osc, \rm RD})/s(T_{\rm osc, \rm RD})}
    \\
    &= \frac{m_a(T_{\rm osc, MD})}{m_a(T_{\rm osc, RD})} \left(\frac{T_{\rm osc, RD}}{T_{\rm osc, MD}}\right)^3\left(\frac{g_{*s}(T_{\rm osc, RD})}{g_{*s}(T_{\rm osc, MD})}\right)
    \\
    &= \frac{\min 
    \left\{\left(\frac{T_{\rm QCD}}{T_{\rm osc, MD}}\right)^{n/2}, 1\right\}}
    {\min 
    \left\{\left(\frac{T_{\rm QCD}}{T_{\rm osc, RD}}\right)^{n/2}, 1\right\}} \left(\frac{T_{\rm osc, RD}}{T_{\rm osc, MD}}\right)^3\left(\frac{g_{*s}(T_{\rm osc, RD})}{g_{*s}(T_{\rm osc, MD})}\right).
\end{align}
Each yield corresponds to the conserved comoving number of axions once the axion begins oscillating. Since the temperature at which the axion begins oscillating in a  MD era is less than a RD era for the same $f_a$, these yields are different. The case where $n = 8$ and $T_{\rm osc, MD}, T_{\rm osc, RD} > T_{\rm QCD}$ corresponds to Eq.~(69) in ref.~\cite{Visinelli_2010}.

The third term in Eq.~\eqref{eq:OmegaMD}, $D$, is the dilution of the axion abundance by the entropy generated at the end of the early MD era cosmology when the universe returns to radiation domination. Specifically, the dilution is the increase in the entropy of the universe, $S = g_{*s} a^3 T^3$, at the end of the MD era (at temperature $T_{\rm RH}$) compared to the entropy when the axion begins oscillating (at $T_{\rm osc, MD}$),
\begin{align}
    \label{eq:dilutionFactor}
    D &= \frac{S(T_{\rm RH})}{S(T_{\rm osc, MD})} = \frac{g_{*s}(T_{\rm RH}) a(T_{\rm RH})^3 T_{\rm RH}^3}{g_{*s}(T_{\rm osc, MD}) a(T_{\rm osc, MD})^3 T_{\rm osc, MD}^3}
    \\
    &= 
    \begin{dcases}
        \frac{g_{*s}(T_{\rm RH})}{g_{*s}(T_{\rm osc, MD})} \left[\left(\frac{g_{*}(T_{\rm osc, MD})}{g_{*}(T_{\rm RH})}\right)^{2/3}\left(\frac{T_{\rm osc, MD}}{T_{\rm RH}}\right)^{8/3}\right]^3  \left(\frac{T_{\rm RH}}{T_{\rm osc, MD}}\right)^3   \qquad &(T_{\rm osc, MD} < T_{\rm NA})
        \\
        \simeq \frac{g_{*}(T_{\rm osc, MD})}{g_{*}(T_{\rm RH})}\left(\frac{T_{\rm osc, MD}}{T_{\rm RH}}\right)^5 
        \\
        \frac{g_{*s}(T_{\rm RH})}{g_{*s}(T_{\rm NA})} \left[\left(\frac{g_{*}(T_{\rm NA})}{g_{*}(T_{\rm RH})}\right)^{2/3}\left(\frac{T_{\rm NA}}{T_{\rm RH}}\right)^{8/3}\right]^3  \left(\frac{T_{\rm RH}}{T_{\rm NA}}\right)^3  \qquad &(T_{\rm osc, MD} > T_{\rm NA})
        \\
        \simeq  \frac{g_{*}(T_{\rm NA})}{g_{*}(T_{\rm RH})}\left(\frac{T_{\rm NA}}{T_{\rm RH}}\right)^5 \simeq\frac{T_{\rm MD}}{T_{\rm RH}}
    \end{dcases}
\end{align}
\begin{figure}
    \centering
    \includegraphics[width=1.0\textwidth]{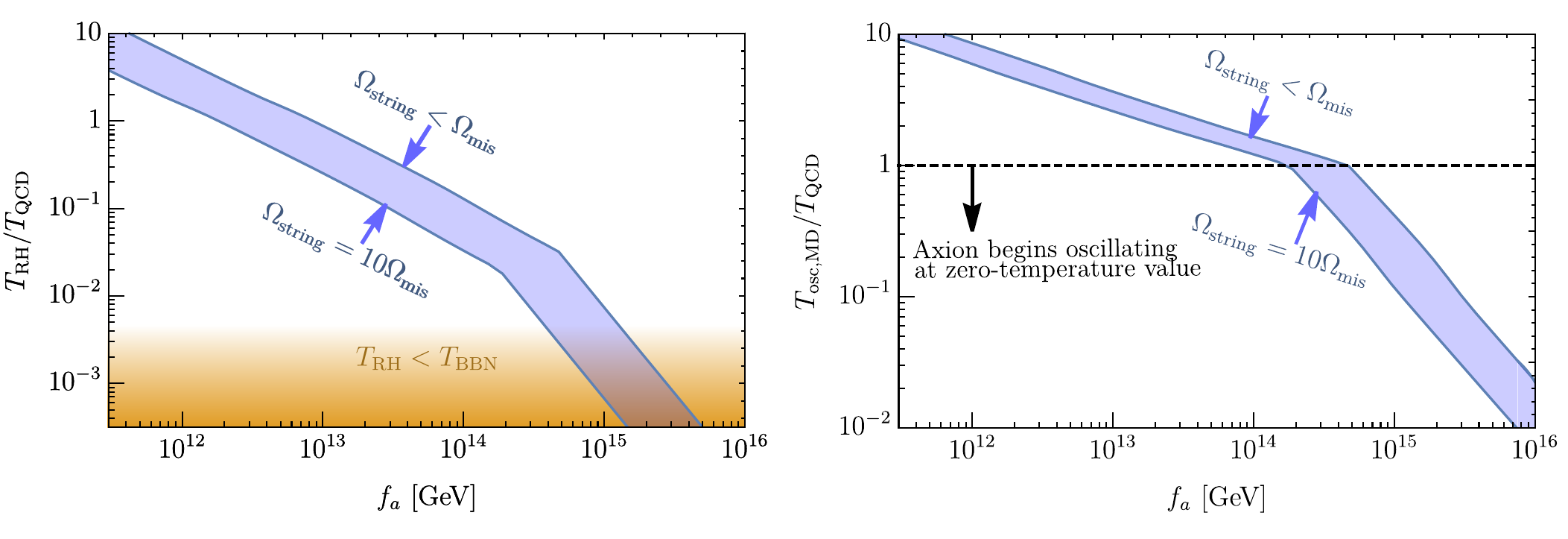}
\caption{\textit{Left}: The reheat temperature $T_\mathrm{RH}$ over the QCD temperature $T_\mathrm{QCD}$ as a function of $f_a$. The blue band indicates the uncertainty in $T_\mathrm{RH}$ due to the string radiated axion energy density $\Omega_\mathrm{string}$, which alters what $T_\mathrm{RH}$ is needed to obtain the correct present-day axion DM abundance. The orange shaded region indicates the part of parameter space where $T_\mathrm{RH}$ is driven below the Big Bang Nucleosynthesis temperature $T_\mathrm{BBN}$ in the most minimal early matter-dominated cosmology. \textit{Right}: The axion field oscillation temperature $T_\mathrm{osc,MD}$ over the QCD temperature $T_\mathrm{QCD}$ as a function of $f_a$. The oscillation temperature controls when a majority of misalignment axion DM is produced. The blue band again indicates uncertainty in $T_\mathrm{osc,MD}$ due to $\Omega_{\mathrm{string}}$ not being accurately known. The black dotted line indicates the boundary between the region of parameter space where the axion has reached its zero temperature value at $T=T_\mathrm{osc,MD}$ and where the mass is still turning on at $T=T_\mathrm{osc,MD}$.}
    \label{fig:TMDQCDAxion}
\end{figure}
Here, we use the fact that the entropy of the universe during the MD era is unchanged until the energy density of the relativistic decay products of the field responsible for matter domination overtakes the primordial radiation, which marks the start of the non-adiabatic regime of matter domination and occurs at temperature $T_{\rm NA} \simeq (T_{\rm RH}^4 T_{\rm MD})^{1/5}$ where $T_{\rm MD}$ is the temperature at which the early MD era begins~\cite{Co:2015pka}. That is, between $T_{\rm MD}$ and $T_{\rm NA}$, the scale factor vs. temperature relation of the universe is the usual $a \propto g_{*s}^{-1/3} T^{-1}$, while between $T_{\rm NA}$ and $T_{\rm RH}$, the expansion is non-adiabatic with $a \propto g_{*s}^{-2/3} T^{-8/3}$~\cite{Kolb:1990vq}. As can be seen from the two cases in Eq.~\eqref{eq:dilutionFactor}, the amount of dilution depends on whether the axion starts oscillating after or before the universe enters the non-adiabatic regime of the MD era. The first case is equivalent to the reciprocal of Eq.~(70) in ref.~\cite{Visinelli_2010}. Note the largest possible dilution occurs when $T_{\rm osc, MD}$ occurs during the non-adiabatic regime of the early MD era.

Finally, the last term in Eq.~\eqref{eq:OmegaMD} is the contribution to $\Omega_a$ from axion string decay. The ratio of axion DM from strings compared to misalignment, $\Omega_{\rm string}/\Omega_{{\rm mis}}$, depends on the number of strings per horizon, $\xi$, and how relativistic the axions emitted by the strings are. As discussed in Sec.~\ref{sec:wallFormationTimes}, although $\xi$ is expected to be $\mathcal{O}(1)$, there is large uncertainty from simulations due to extrapolation of $\xi$ from $\sim 1$ to $\mathcal{O}(10)$ which could make $\Omega_{\rm string, RD} \sim 10 \Omega_{{\rm mis}, \rm RD}$ in a RD era. In a MD era, however, the faster expansion rate of the universe is expected to make $\xi$ much smaller even with logarithmic extrapolation of simulations~\cite{Visinelli_2010}.

The left panel of Fig.~\ref{fig:TMDQCDAxion} shows the necessary $T_{\rm RH}/T_{\rm QCD}$ to give the observed axion DM abundance from axion misalignment and string decay (i.e. $\Omega_{a, \rm MD} = \Omega_{\rm DM}$) as a function of $f_a$ for the QCD axion. The thickness of the blue line corresponds to when $\Omega_{\rm string,MD} < \Omega_{{\rm mis}, \rm MD}$ (the top contour) to $\Omega_{\rm string,MD} \approx 10\Omega_{{\rm mis}, \rm MD}$ (the bottom contour). In the orange shaded region, the reheat temperature after the MD ends is below $T \approx 4$ MeV and is excluded by BBN~\cite{deSalas:2015glj}. This limits the largest $f_a$ consistent with the most minimal early MD cosmology to be $f_a \sim$ a few times $10^{15}$ GeV.

Similarly, the right panel of Fig.~\ref{fig:TMDQCDAxion} shows the corresponding $T_{\rm osc, MD}/T_{\rm QCD}$ using the correct $T_{\rm RH}$ from the left panel to give the correct axion DM abundance. For $f_a \gtrsim 5 \times 10^{14}$ GeV, the axion is already at its zero temperature value by the time the earliest axion walls form, as shown by the dashed horizontal line. Consequently, for $f_a$ larger than this, the rest-frame wall thickness of the walls is always unchanged in time. 
\section{Superhorizon Evolution of Aspherical Domain Walls}
\label{app:superhorizon_asphericities}
\begin{figure}
    \centering
    \includegraphics[width=1.0\textwidth,height=0.7\textheight,keepaspectratio]{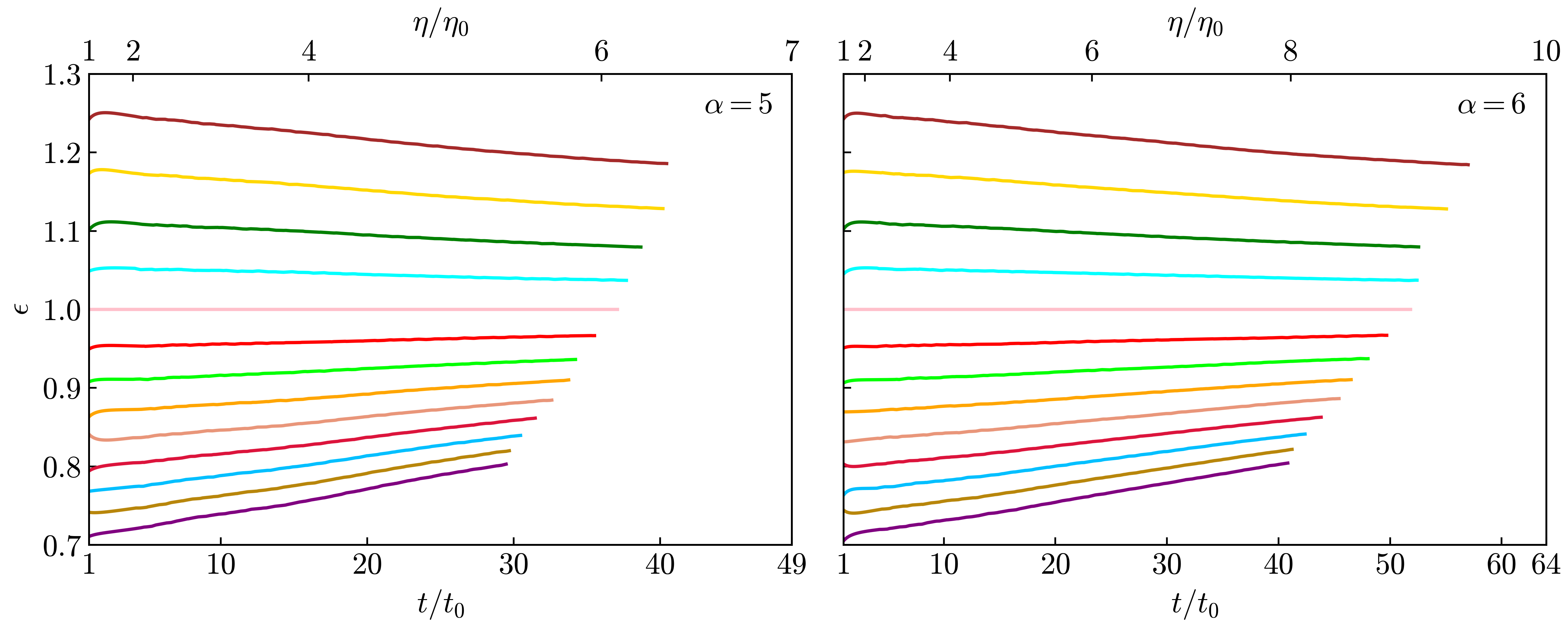}
\caption{The eccentricity evolution of ellipsoidal domain walls up until the moment of wall re-entry. \textit{Left:} The eccentricity parameter $\epsilon$ as a function of real time $t$ and conformal time $\eta$ for $\alpha=5$ walls assuming radiation domination. The distinct colored curves correspond to distinct walls with the same $\alpha \equiv R_0/t_0$, but distinct ellipsoid parameters initialized at $t/t_0=\eta/\eta_0=1$. Notice that walls with $\epsilon<1$ re-enter the horizon faster than walls with $\epsilon>1$, and that walls tend to become less elliptical before re-entry. \textit{Right}: The same as left, but with $\alpha=6$. Note that larger walls re-enter the horizon later.}
    \label{fig:ellipse_hubble_evol}
\end{figure}

Here, we expand our analysis of the superhorizon evolution of domain walls described in Sec.~\ref{sec:SupHorizonExpansion} to aspherical self-enclosed domain walls. We use Eq.~\eqref{eq:ellipsoid_eqn} to paramaterize the surface of the aspherical wall and Eq.~\eqref{eq:aspherical_field_config} as our initial field configuration. We then study how this field configuration evolves via Eq.~\eqref{eq:axion_hubble_eom}, completely analogous to the spherical case in Sec.~\ref{sec:SupHorizonExpansion}. As argued in Sec.~\ref{sec:closed_dw_abundance}, we approximate any aspherical wall formed with a characteristic mean radius of curvature $R_0$ as an ellipsoid with semi-major axis $R_0 \epsilon_1$ and semi-minor axis $R_0 \epsilon_2$, which parameterize the asphericity of the domain wall. We simulate the superhorizon evolution of a host of these ellipsoidal walls of different $\alpha \equiv R_0/t_0$ and different $\epsilon_1$ and $\epsilon_2$. The primary takeaway from these simulations is that during the superhorizon stretching phase, the elliptical walls become more spherical, with $(\epsilon(t_{0}) - \epsilon(t_{\rm RE}))/\epsilon(t_0)\simeq 0.1-0.2$. Fig.~\ref{fig:ellipse_hubble_evol} shows several examples of how a single eccentricity parameter $\epsilon = \epsilon_1$ (with $\epsilon_2 = 1$) evolves with time up until horizon re-entry. Since the asphericity becomes mildly less severe, we safely neglect this superhorizon correction. However, we emphasize that asphericities are of monumental importance in subhorizon collapse dynamics, primarily due to how they prevent the compression of energy. We account for this in details in Sec.~\ref{sec:SubHorizonCollapse} and Sec.~\ref{sec:PBHConditions}.
\section{Domain Wall Self-Gravitation}
\label{app:self_gravitation}
\begin{figure}
    \centering    \includegraphics[width=0.7\textwidth,height=0.7\textheight,keepaspectratio]{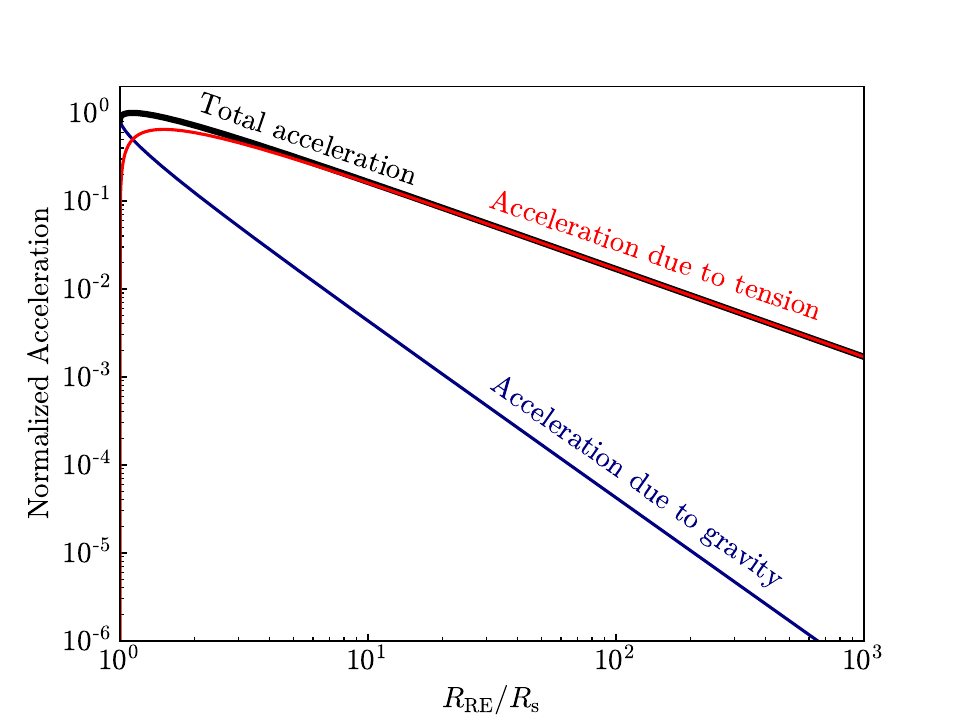}
\caption{The instantaneous acceleration of a perfectly spherical domain wall normalized to its peak acceleration as a function of the wall re-entry radius to its Schwarzschild radius $R_{\rm RE}/R_s$. In red, the inward acceleration of the domain wall due to tension alone. In blue, the inward acceleration of the domain wall due to gravity alone. In black, the total inward acceleration of the domain wall. The total energy of the wall is taken to be $E=4\pi f_a^2 m_a R_{\rm RE}^2$ with $f_a=10^{14}$ GeV. Note that the wall acceleration due to gravity only becomes significant when the wall radius is of order the Schwarzschild radius. For axion domain walls, $R_s\ll R_{\rm RE}$, meaning walls will typically have reached a contraction speed near the speed of light by the time the wall radius is close to the Schwarzschild radius. This makes the enhanced acceleration due to gravity negligible when computing the total collapse time.}
\label{fig:gravity_acceleration}
\end{figure}

It was pointed out by refs.~\cite{PhysRevD.23.852,PhysRevD.30.712} that domain walls are gravitationally repulsive. For spherical walls close to their collapse radius, where gravity will act the most strongly, we may neglect the Hubble expansion of the universe and simply consider the axion wall field configuration on an asymptotically flat spacetime background. Here, this `repulsion' is simply the statement that an observer exterior to the wall must accelerate inward in order to stay a fixed height above the wall surface. For a spherical axion domain wall, Birkhoff's theorem implies that the spacetime region exterior to the wall is described by a Schwarzschild metric~\cite{PhysRev.164.1776}. Inside the domain wall, the spacetime is entirely flat, so an observer inside the wall must also accelerate inward to stay a fixed distance above the wall surface. We note that this is in contrast to static planar domain walls, where a genuine repulsion is experienced by an observer near any side of the wall. Solving the Einstein field equations for a timelike spherical shell of equal energy density and tension $\sigma$ yields the equation ~\cite{PhysRevD.30.712}
\begin{equation}
\label{eq:collapse_w_gravity}
        (\mathcal{A}+\mathcal{B})\ddot{R}=-2\frac{\mathcal{A} G M}{R^2}-2\frac{\mathcal{A}\mathcal{B} (\mathcal{A}+\mathcal{B})}{R},
\end{equation}
where $G$ is the gravitational constant, $M=4\pi\sigma \mathcal{R}^2(1-2\pi\sigma G \mathcal{R})$ is the mass of the wall as measured by an observer far away and
\begin{equation}
    \mathcal{A}=(1+\dot{R}^2)^{1/2},\hspace{0.5in}
    \mathcal{B}=\bigg[1-\frac{2 G M}{R}+\dot{R}^2\bigg]^{1/2}.
\end{equation}
where the dot denotes a time derivative with respect to an observer instantaneously comoving with the wall~\cite{PhysRevD.30.712}. Note that gravity aids the domain wall collapse via the first term in Eq.~\eqref{eq:collapse_w_gravity}. However, the gravitational contribution is negligible compared to the pure tension term unless $R\lesssim \sigma R_{\rm RE}^2/M_{\text{Pl}}^2\sim m_a f_a^2 R_{\rm RE}^2/M_{\text{Pl}}^2\sim R_{s}$, where $R_s$ is the Schwarzschild radius. Thus, the gravitational contraction is completely negligible until the last moments of collapse, at which point the wall will already have accelerated to nearly the speed of light. Therefore, the attractive gravitational force will not yield a significant speedup or change in the wall momentum in the collapse and we ignore it in the text. For visualization, Fig.~\ref{fig:gravity_acceleration} compares how the acceleration due to gravity is largely subdominant to the wall self-tension except when the wall has already contracted very close to the Schwarzschild radius. 
\section{PBH Abundance in ALP Models}
\label{sec:ALPS}
\begin{figure}[t]
    \centering
    \includegraphics[width=0.7\textwidth,keepaspectratio]{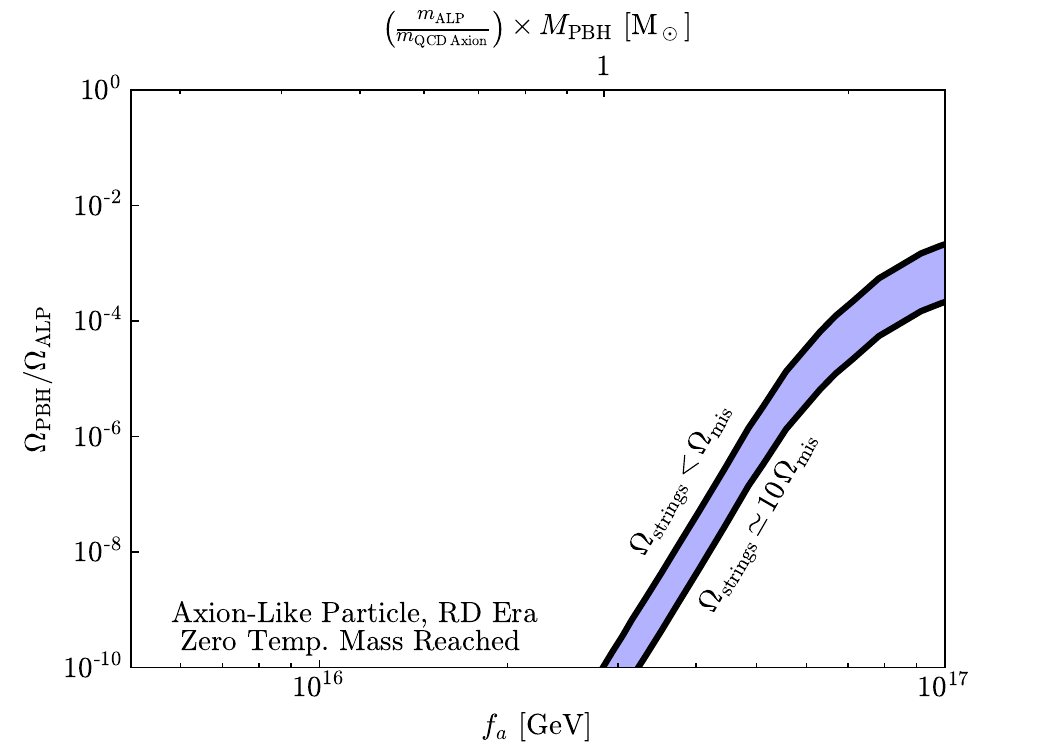}
\caption{The projected fraction of ALP DM consisting of PBHs assuming radiation domination and a static ALP mass at wall formation. The bottom axis shows the ALP model $f_a$ corresponding to a given $f_\mathrm{PBH}$ abundance. The top axis shows the PBH mass $M_\mathrm{PBH}$ associated with each $f_a$. Note that the PBH spectrum for a given $f_a$ is nearly monochromatic, hence the one-to-one map from $f_a$ to $M_\mathrm{PBH}$. The top and bottom solid black lines show the $f_\mathrm{PBH}$ abundance from ALP domain wall collapse in the case where the relic radiated string axion abundance $\Omega_\mathrm{strings}$ is less than $\Omega_\mathrm{mis}$ and approximately $10\,\Omega_\mathrm{mis}$, respectively. $\Omega_\mathrm{strings}$ is entirely predicted by $f_a$ and the subject of extensive study, however its exact value remains uncertain due to simulation limitations. The two black lines correspond to edge cases for $\Omega_\mathrm{strings}$, meaning $f_\mathrm{PBH}$ may lie anywhere in the blue band. Note that this figure assumes that the ALP has reached its zero temperature mass at wall formation. Also note that $m_a$ is now a free parameter which must be picked to obtain a PBH mass corresponding to each $f_a$ (however, this $m_a$ is fixed by cosmology if one assumes that all of DM comes from ALP misalignment).}
\label{fig:money_plot_alps}
\end{figure}
Here we compute the PBH abundance in axion-like particle (ALP) models with temperature-independent masses $(n = 0$ per Eq.~\eqref{eq:axion_mass_temp}). ALP models accommodate a more general set of cosmologies than QCD axions because $f_a$ and $m_a$ are independent parameters. First, we consider an ALP model in a RD cosmology. Unlike in the QCD axion case, ALPs with high values of $f_a$ are still cosmologically viable if the ALP mass is sufficiently lowered. We find that in a RD era, the PBH abundance in ALP models is nevertheless smaller than the PBH abundance in the QCD axion + MD cosmology considered in the main text, as shown in Fig.~\ref{fig:money_plot_alps}. This is mainly because of the slower expansion rate of the universe during a RD era, which means that domain walls do not get stretched as much before entering the horizon. The time interval between distinct ALP DM/PBH production stages (formation, misalignment, collapse) is also less favorable than in the MD case unless the ALP possesses a temperature dependent mass like the QCD axion.
Fig~\ref{fig:money_plot_alps} shows the ratio of $\Omega_{\rm PBH}/\Omega_{\rm mis}$ for ALPs in a RD era, assuming that the ALPs are at their zero temperature mass before wall formation and the string density is at $\xi \simeq 1$. The ratio here is true regardless if the ALP misalignment abundance is the DM or if it is overproduced and later diluted since the PBHs and misalignment axions are produced nearly concurrently so that $\Omega_{\rm PBH}/\Omega_{\rm mis}$ is redshift invariant. Note that since string scaling in the RD scenario is more susceptible to log corrections, the $\Omega_\mathrm{PBH}$ result presented here may be subject to exponential corrections if $\xi$ is not $\mathcal{O}(1)$. This is in contrast to Fig.~\ref{fig:money_plot} (the main result of this work) where systematics are under better control due to the MD cosmology (although the physical value of $\xi$ has still not been fully determined, even though it's believed to be close to unity~\cite{Visinelli_2010}).

\begin{figure}
    \centering
    \includegraphics[width=0.7\textwidth,keepaspectratio]{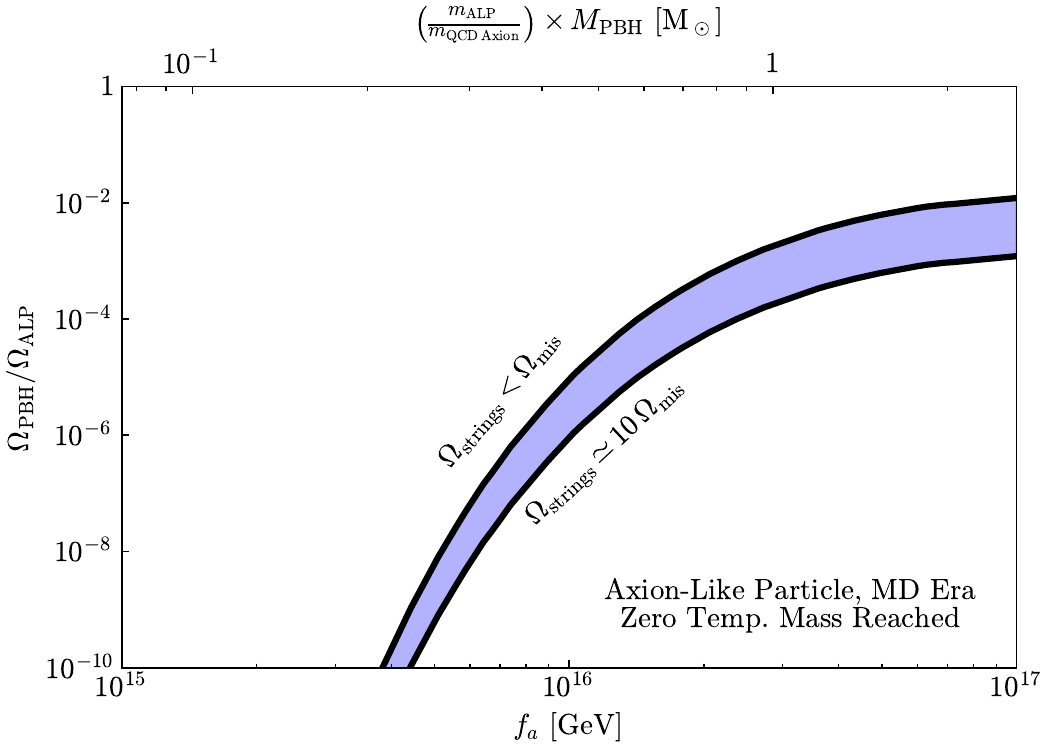}
\caption{The projected fraction of ALP DM consisting of PBHs assuming matter domination and a static ALP mass at wall formation. The bottom axis shows the ALP model $f_a$ corresponding to a given $f_\mathrm{PBH}$ abundance. The top axis shows the PBH mass $M_\mathrm{PBH}$ associated with each $f_a$. Note that the PBH spectrum for a given $f_a$ is nearly monochromatic, hence the one-to-one map from $f_a$ to $M_\mathrm{PBH}$. The top and bottom solid black lines show the $f_\mathrm{PBH}$ abundance from ALP domain wall collapse in the case where the relic radiated string axion abundance $\Omega_\mathrm{strings}$ is less than $\Omega_\mathrm{mis}$ and approximately $10\,\Omega_\mathrm{mis}$, respectively. $\Omega_\mathrm{strings}$ is entirely predicted by $f_a$ and the subject of extensive study, however its exact value remains uncertain due to simulation limitations. The two black lines correspond to edge cases for $\Omega_\mathrm{strings}$, meaning $f_\mathrm{PBH}$ may lie anywhere in the blue band. Note that this figure assumes that the ALP has reached its zero temperature mass at wall formation. Also note that $m_a$ is now a free parameter which must be picked to obtain a PBH mass corresponding to each $f_a$ (however, this $m_a$ is fixed by cosmology if one assumes that all of DM comes from ALP misalignment). This plot is identical to Fig.~\ref{fig:money_plot}, except the PBH mass map differs, and the $T_\mathrm{RH}$ constraint of Fig.~\ref{fig:money_plot} can easily be overcome by a suitable choice of $m_\mathrm{ALP}$.}
\label{fig:md_alps}
\end{figure}

ALP domain wall formation in a MD cosmology is also a possible cosmological scenario. We show the ALP MD $f_\mathrm{PBH}$ in Fig.~\ref{fig:md_alps}. This projection is nearly identical to Fig.~\ref{fig:money_plot}, except the ALP mass is now a free parameter, which in turn makes $m_\mathrm{PBH}$ a free parameter. Note that in contrast to Fig.~\ref{fig:money_plot}, the ALP scenario easily avoids coming into tension with $T_\mathrm{BBN}$ if the ALP mass is slightly offset from the QCD axion line. For high values of $f_a$, where the PBH yield is largest, one needs $m_\mathrm{ALP}>m_\mathrm{QCD\; Axion}$. This, in turn, makes the PBHs produced slightly lighter.

\clearpage
\bibliographystyle{JHEP}
\bibliography{Bib_Axion_PBH}
\end{document}